\documentclass[useAMS,usenatbib]{mn2e}

\title[Spectral variability in faint high frequency peakers]{Spectral variability in faint high frequency peakers}
\author[M. Orienti et al.]
  {M. Orienti$^{1,2}$\thanks{E-mail: orienti@ira.inaf.it},
D. Dallacasa$^{1,2}$, C. Stanghellini$^{2}$\\
$^{1}$Dipartimento di Astronomia, Universit\`a di Bologna, via Ranzani 1,
I-40127, Bologna, Italy \\
$^{2}$Istituto di Radioastronomia - INAF, Via P. Gobetti 101, I-40129
Bologna, Italy\\}
\date{Received \today; accepted ?}

\pagerange{\pageref{firstpage}--\pageref{lastpage}} \pubyear{2002}

\def\LaTeX{L\kern-.36em\raise.3ex\hbox{a}\kern-.15em
    T\kern-.1667em\lower.7ex\hbox{E}\kern-.125emX}

\begin{document}

\label{firstpage}

\maketitle

\begin{abstract}
We present the analysis of simultaneous multi-frequency Very Large
Array (VLA)
observations of 57 out of 61 sources from the ``faint'' high frequency
peaker (HFP) sample
carried out in various epochs. Sloan Digital Sky Survey SDSS 
data have been
used to identify the optical counterpart of each radio source. From
the analysis of the multi-epoch spectra we find that 24 sources 
do not show evidence of spectral variability, while
12 objects do not possess a peaked spectrum anymore at least in one of the
observing epochs. Among the remaining 21 sources showing some degree
of variability, we find that in 8 objects the spectral properties
change consistently with the expectation for a 
radio source undergoing adiabatic expansion. The comparison between the
variability and the optical identification suggests that
the majority of 
radio sources hosted in galaxies likely represent the young radio
source population, whereas the majority of those associated with
quasars are part of a different population similar to flat-spectrum
objects, which possess peaked spectra during short intervals of their life,
as found in other samples of high-frequency peaking objects. 
The analysis of the optical images from the SDSS
points out the presence of
companions around 6 HFP hosted in galaxies, 
suggesting that young radio sources resides in groups.
\end{abstract}


\section{Introduction}

Our knowledge of the first stages of the evolution of powerful
  radio sources is based
on the study of the population of high frequency peaking radio
sources. In the framework of models explaining the evolution of
individual radio sources, the spectral peak of
young radio sources occurs at high frequencies. Given their small size,
in these sources the synchrotron self-absorption (SSA) is a
very effective mechanism. As the source grows, the peak frequency is
expected to shift towards lower frequencies as a consequence of adiabatic
expansion. An alternative explanation suggests that the spectral
  peak is due to free-free absorption from a ionized medium enshrouding
the radio emission \citep{bicknell97}.
Both scenarios are supported by the empirical
anti-correlation found by \citet{odea97} from the study of samples of
compact steep spectrum (CSS) and gigahertz-peaked spectrum (GPS) radio
sources. The former have peak frequencies around a few hundred MHz,
typical sizes of a few kpc and ages of 10$^{5}$ - 10$^{6}$ years, 
whereas the latter have spectra peaking
around 1 GHz, typical sizes of about 1 kpc or less and ages of 10$^{3}$
- 10$^{4}$ years. However, it is worth noting that the
  consistency between the source size and the spectral peak often
  found in the most compact sources strongly
  support the synchrotron self-absorption scenario \citep{momo,
    tingay03}.\\
High frequency peakers \citep[HFP;][]{dd00}, with a spectral peak
occurring at frequencies above a few GHz, are thus the best candidates
to be newly born radio sources, with ages between 10$^{2}$ - 10$^{3}$
years.\\
The radio properties of HFPs have been derived by the analysis of the
``bright'' HFP sample \citep{dd00,tinti05,mo06a,mo07,mo08a}. 
In particular, from the
multi-epoch analysis of their radio spectra it has been found that the
sample is composed of two different populations. 
One population consists of radio
sources that maintain the convex spectrum without showing variability,
whereas the other comprises radio sources that change their spectral
shape, becoming also flat-spectrum objects, and possessing
substantial flux density variability. The different spectral
properties shown by the two populations suggest that the former
represent young radio sources still in an early stage of their
evolution, while the latter are beamed objects. The
analysis of the multi-epoch spectral behaviour has proved to be a
powerful tool to discriminate between the two populations
\citep{torniainen05,tinti05,mo07}. In fact, beamed radio sources,
although usually characterized by a flat and variable spectrum, may be
selected in samples of high-frequency peaking objects when their
emission is dominated by a flaring knot in a jet. On the other
hand, young radio sources are known to be the least variable class of
extragalactic radio sources \citep{odea98}. However, it must be
mentioned that in the youngest objects, substantial variability in the
optically-thick part of the spectrum is expected as a consequence of
either the source growth/evolution
\citep[e.g. J1459+3337,][]{mo08}, or changes in the possible
  absorber screen, or a combination of both \citep{tingay03}.\\

In this paper we present a multi-epoch analysis based on simultaneous
multi-frequency VLA data of the radio spectra of 57 high frequency
peakers from the ``faint'' HFP sample \citep{cs09}. This
sample was selected as the ``bright'' HFP sample \citep{dd00} by
cross-correlating the 87GB survey 
at 4.9 GHz with the NVSS at 1.4 GHz, and
including only sources fainter than 300 mJy at 4.9 GHz within a
restricted area around the northern galactic cap (for details on
the sample see Stanghellini et al. 2009). The study of the radio
properties of a sample of faint HFPs is the first step in
understanding the first stages of the radio source evolution. So far,
spectral studies have been carried out for the bright HFP sources
only, and an extension to fainter objects is necessary. The
evolution models developed so far
\citep[e.g.][]{cf95,snellen00,kaiser97} predict that in the earliest
stage the radio luminosity progressively increases, implying that the
youngest objects are likely to be found among faint
sources. Furthermore, in faint HFPs, boosting effects should be less
relevant, making the contamination from blazars less severe than what
found in samples of brighter objects.\\

Throughout this paper we assume the following cosmology: $H_{0} = 71
{\rm km \; s^{-1} \; Mpc^{-1}}$, $\Omega_{\rm M} = 0.27$, and
$\Omega_{\Lambda} = 0.73$ in a flat Universe. The spectral index is
defined as $S{\rm (\nu)} \propto \nu^{- \alpha}$.\\

\begin{table}
\caption{VLA observations and configurations}
\begin{center}
\begin{tabular}{cccc}
\hline
Date&Conf&Proj.ID&Code\\
\hline
&&&\\
Sep. 2003&AnB&AD488&a\\
Jan.2004&BnC&AD494&b\\
Nov.2006& C &AO210&c\\
Apr.2007& D &AO210&d\\
&&&\\
\hline
\end{tabular}
\label{vla_oss}
\end{center}
\end{table}

\section{VLA observations and data reduction} 

Simultaneous multi-frequency VLA observations of 57 out of the 61
sources from the ``faint'' HFP sample \citep{cs09} were carried out
during different runs between September 2003 and April 2007
(Table \ref{vla_oss}). Observations were performed in L band (with the two
IFs centered at 1.415 and 1.665 GHz), C band (with the two IFs
centered at 4.565 and 4.935 GHz), X band (with the two IFs centered at
8.085 and 8.465 GHz), U band (14.940 GHz), K band (22.460 GHz), and
in Q band (43.340 GHz). At each frequency, the target
sources were observed for about 1 minute, cycling through
frequencies. During each run, the primary flux density calibrator
either 3C\,286 or 3C\,48 was observed for about 3 minutes at each
frequency. Secondary calibrators were chosen on the basis of their
distance from the targets in order to minimize the telescope slewing
time, and they were observed for 1.5 min at each frequency, every 20
min.\\
The data reduction was carried out following the standard procedures
for the VLA implemented in the NRAO AIPS package. The flux density at
each frequency was measured on the final image produced after a few
phase-only self-calibration iterations. 
In the L band it was generally necessary to
image a few confusing sources falling within the primary beam. All the
target sources appeared unresolved at any frequency. During the
observations of a few sources, strong RFI at 1.420 and 1.665 GHz
was present, and in those cases the measurements of the flux density were
not possible.\\
Uncertainties on the determination of the absolute flux density scale 
are dominated by amplitude errors.
Based on the variations of antenna complex gains during the various
observations, we can conservatively estimate an uncertainty of
$\sim$3\% in L, C, and X bands,
$\sim$5\% in U band, and $\sim$10\% in K and Q bands. The rms noise
level on the image plane is about 0.1 mJy/beam in C and X bands, and
about 0.2-0.4 mJy/beam in L, U, and K bands. In the Q band it accounts
for 0.4-0.5 mJy/beam, becoming comparable to the amplitude calibration
errors for sources fainter than $\sim$20 mJy.
Results are presented in Section 4.\\

\begin{figure*}
\begin{center}
\includegraphics{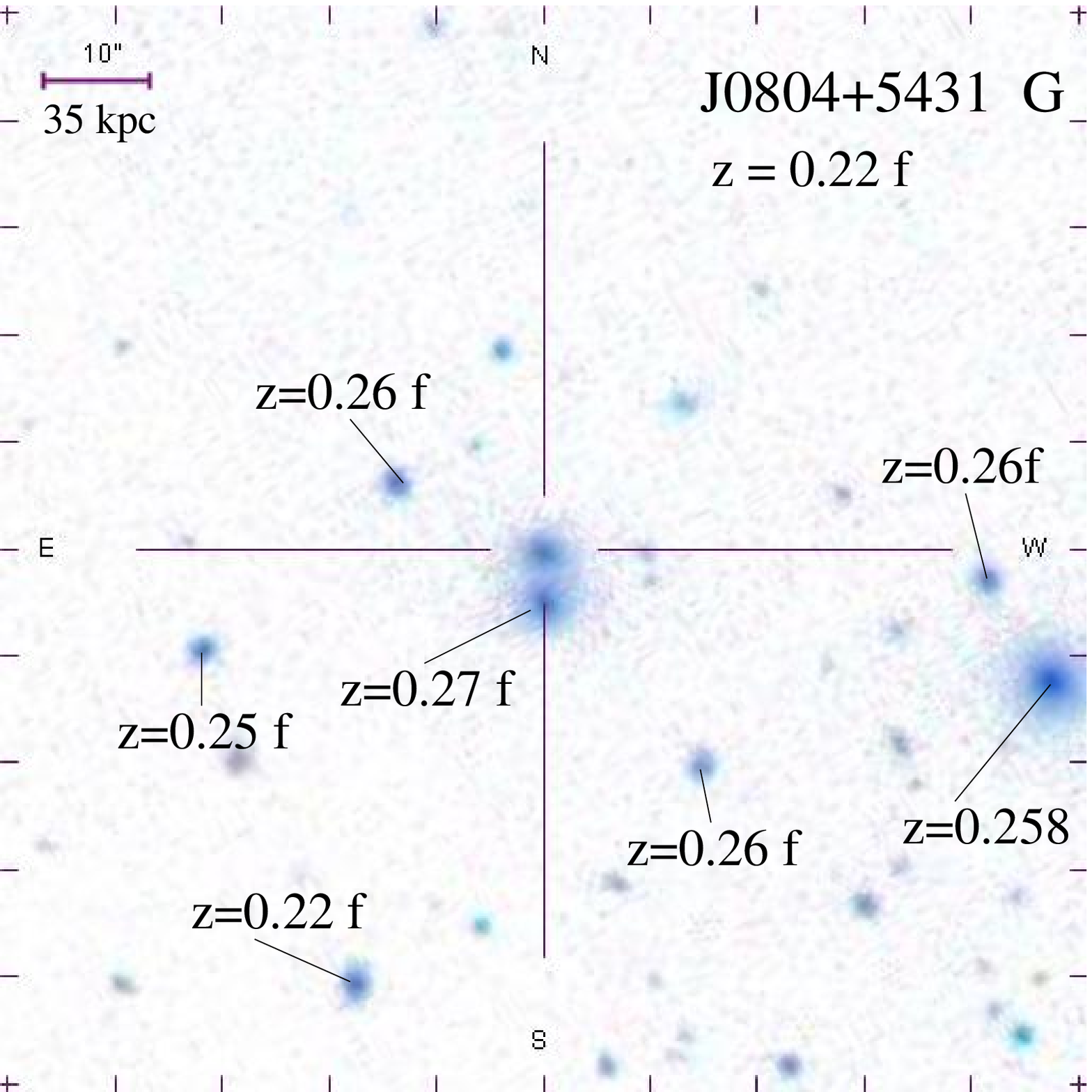}
\includegraphics{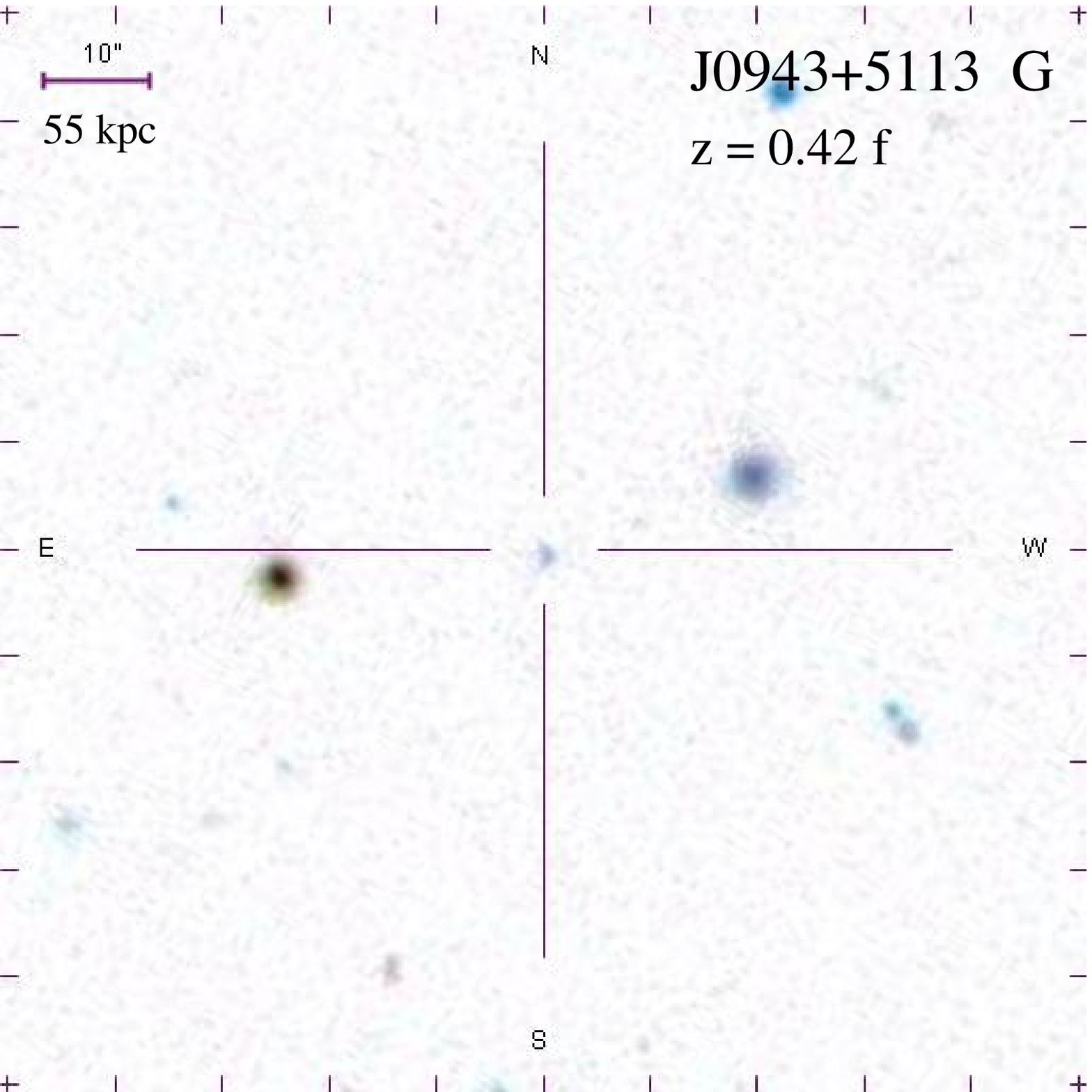}
\includegraphics{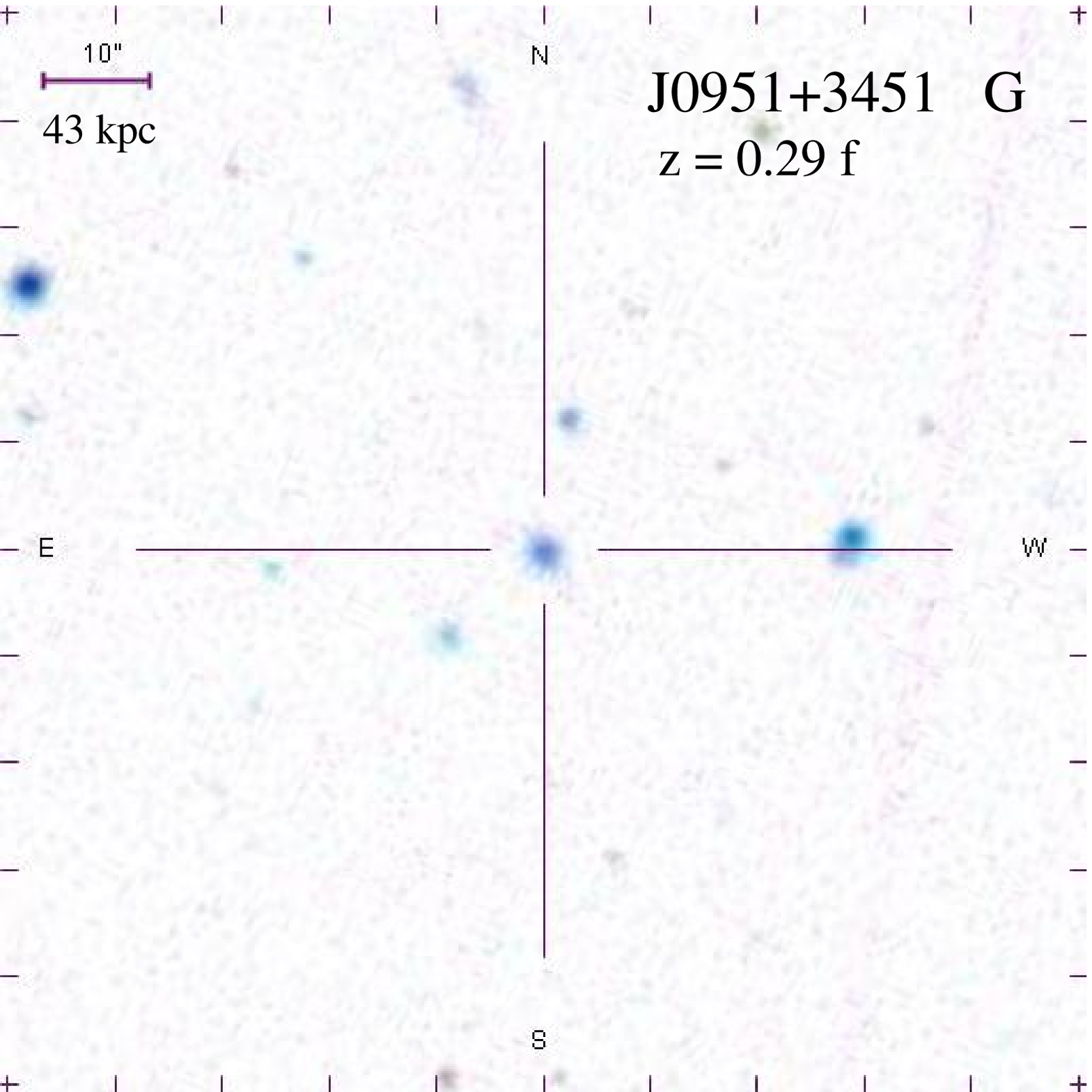}
\includegraphics{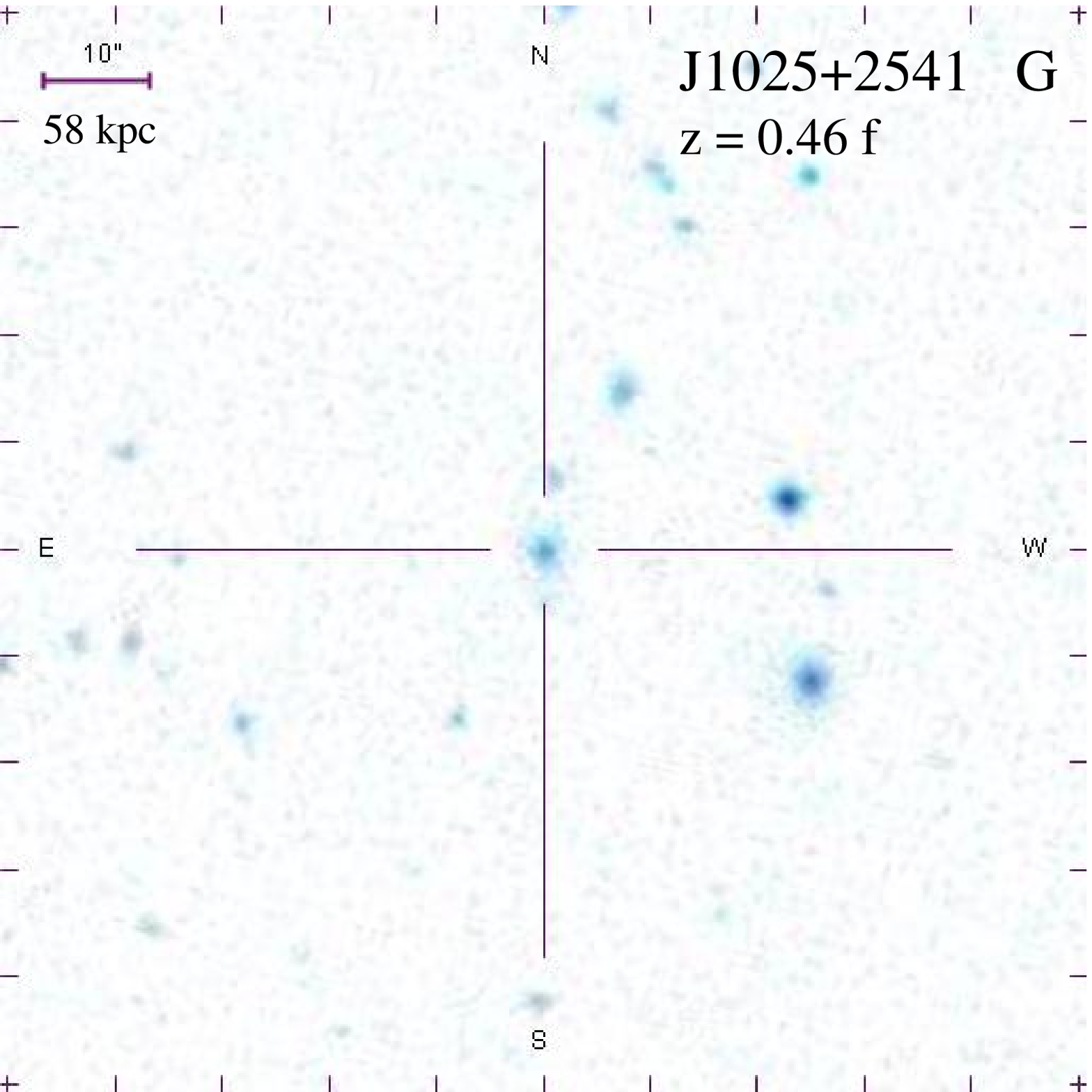}
\includegraphics{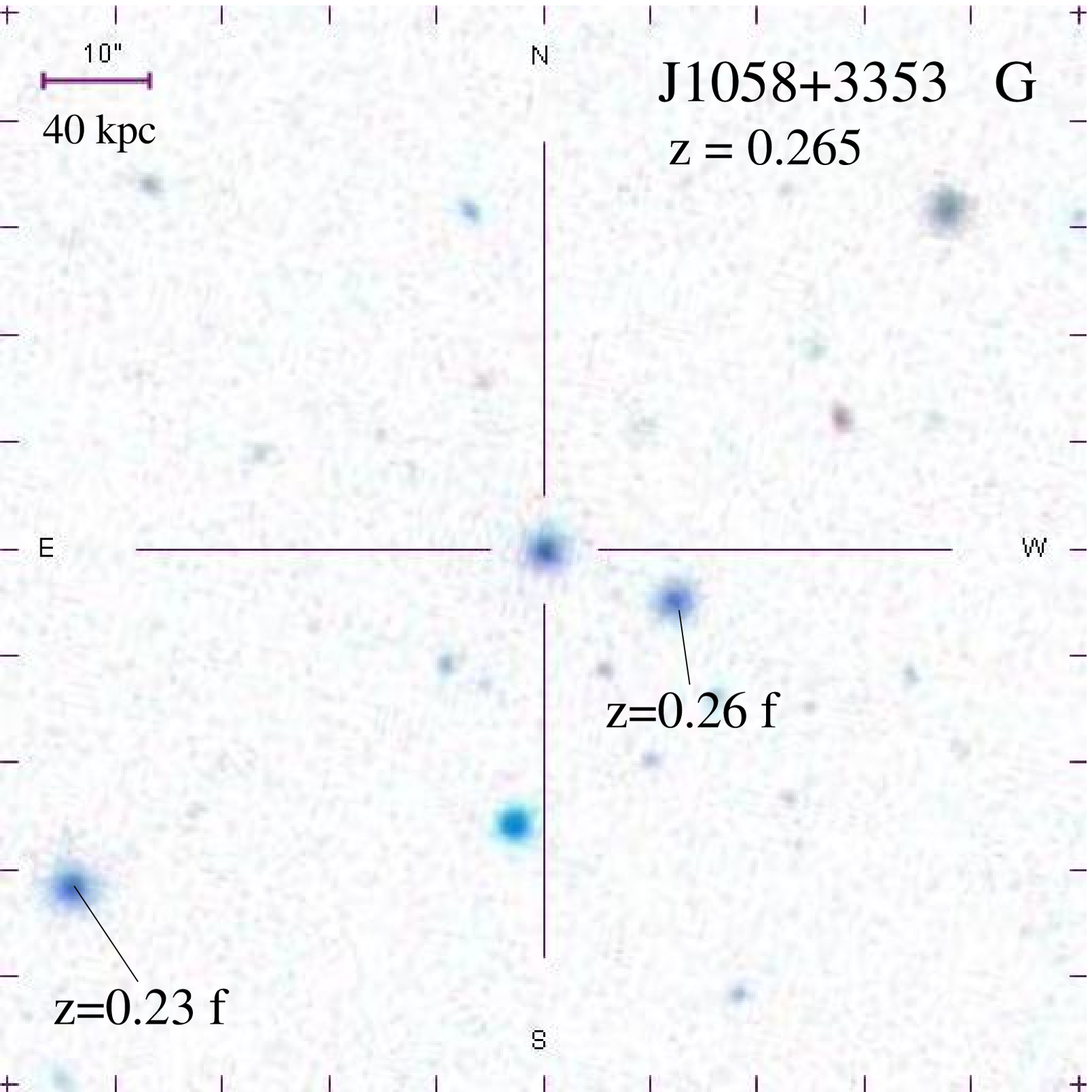}
\includegraphics{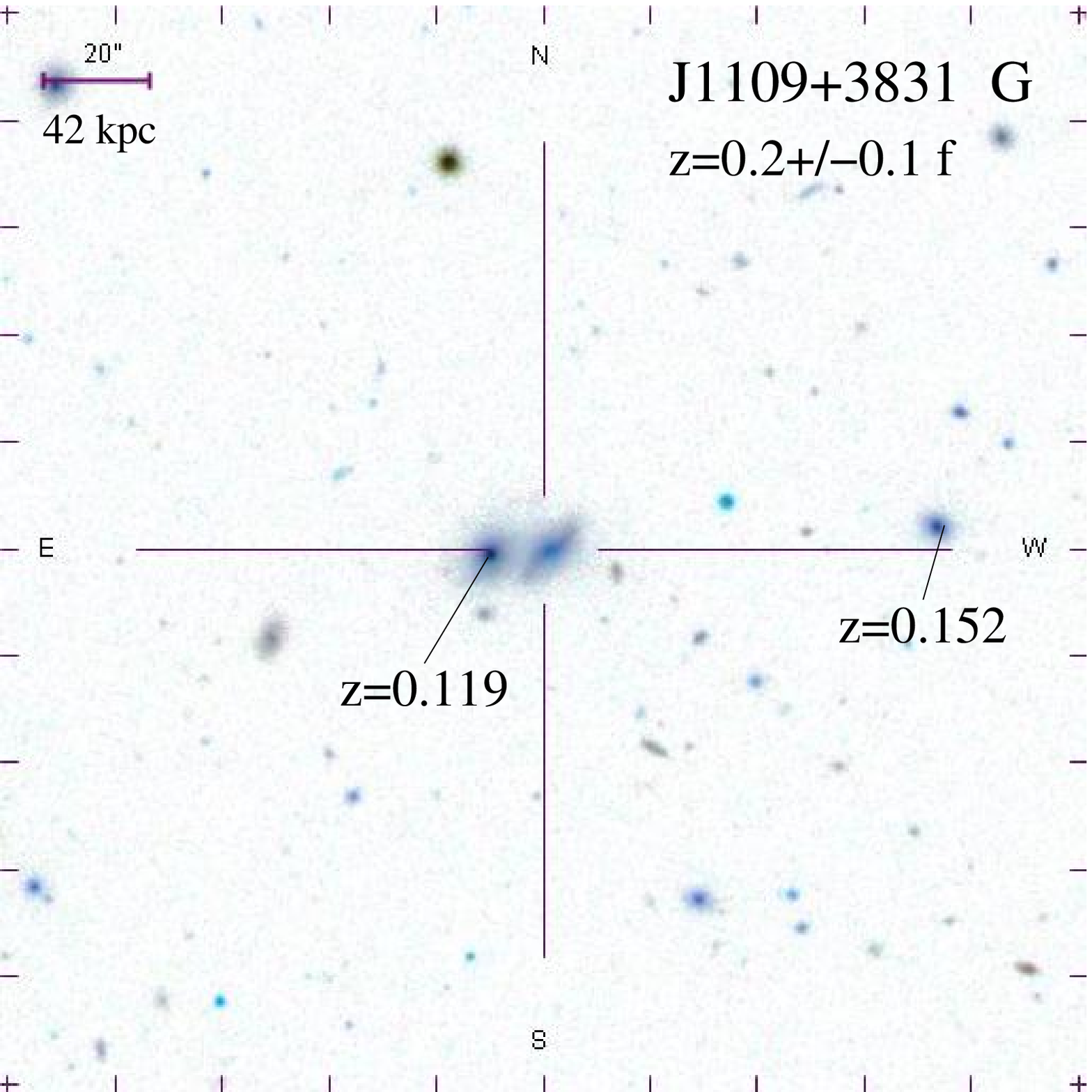}
\includegraphics{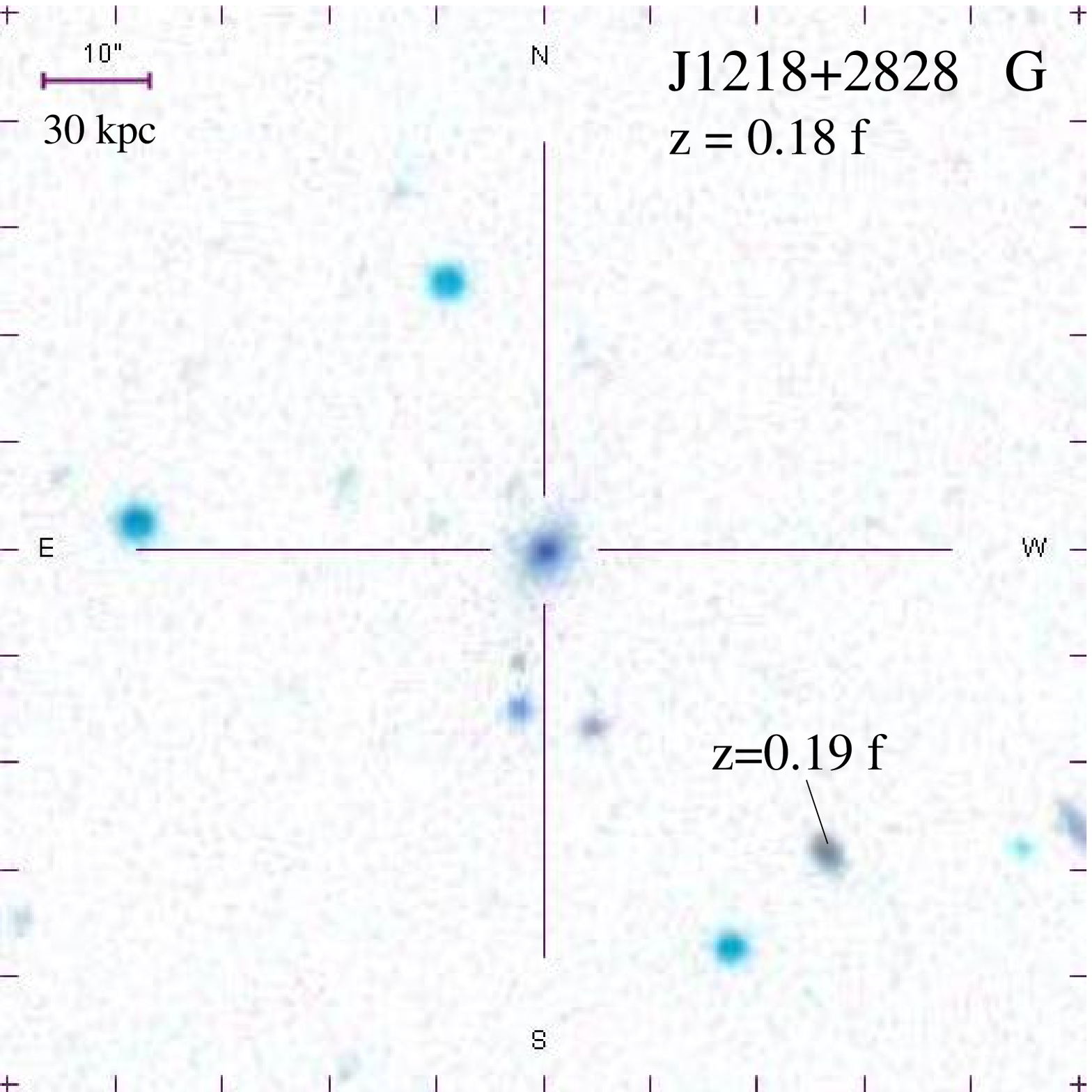}
\includegraphics{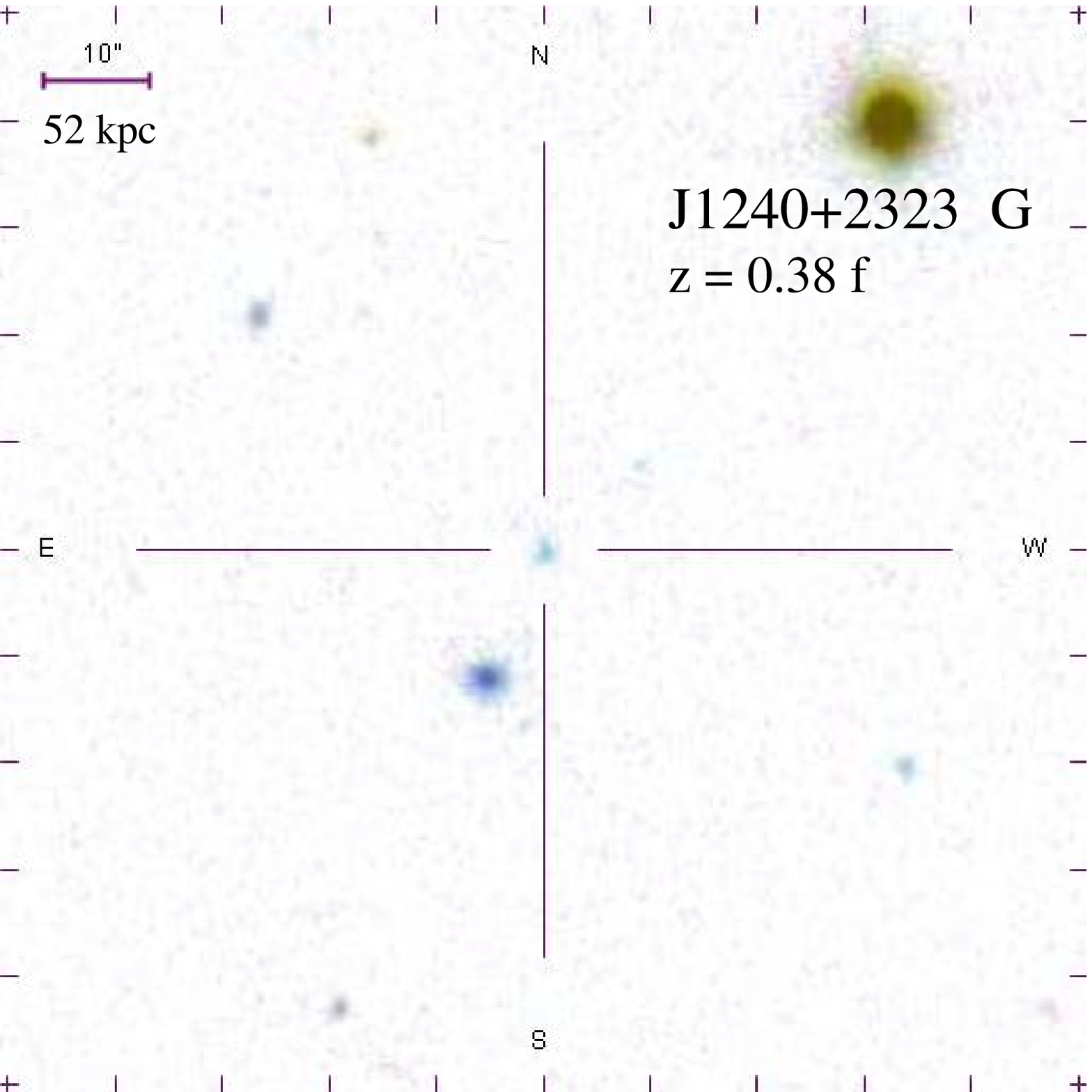}
\includegraphics{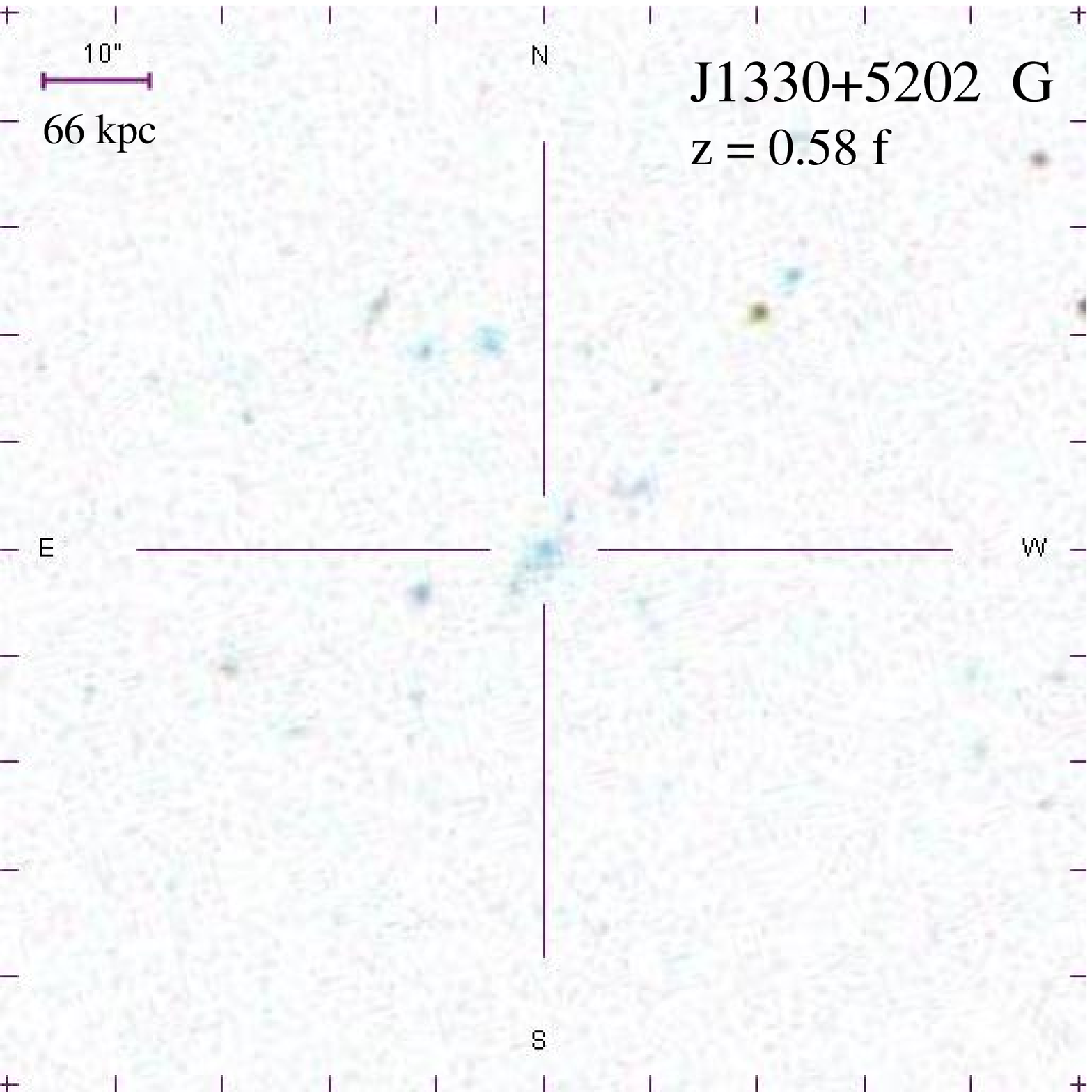}
\vspace{19cm}
\caption{Optical images from the SDSS DR7 of the 12 HFP radio sources 
  identified with a galaxy. On the images we report the source name,
  the redshift (an ``f'' indicates a photometric redshift). If the
  source forms a group, we report the redshift of the companion
  galaxies, when available. The field width has been chosen to show a
  region of about 250 kpc around the galaxy hosting the HFP radio
  source.}
\label{sdss_images}
\end{center}
\end{figure*}

\addtocounter{figure}{-1}
\begin{figure*}
\begin{center}
\includegraphics{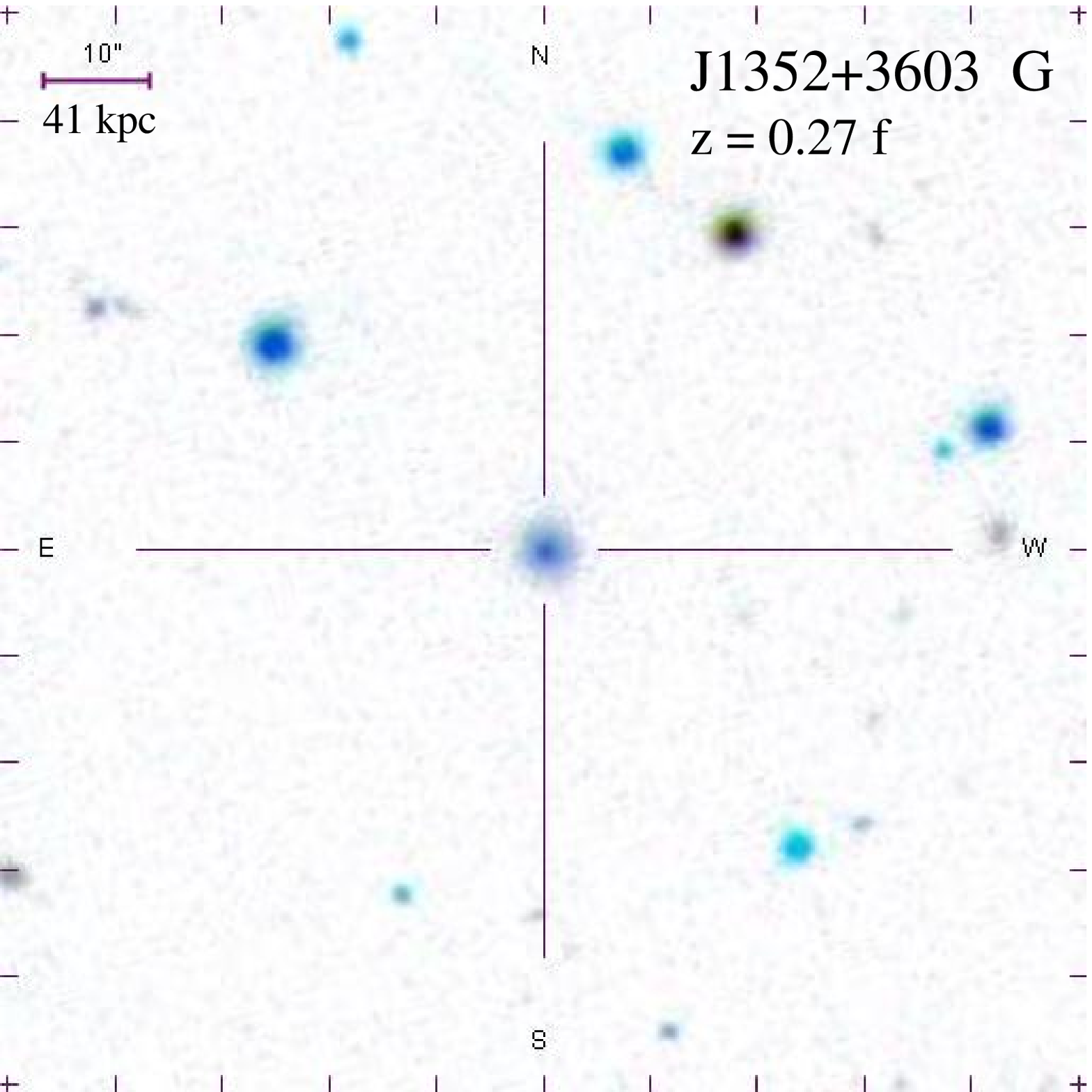}
\includegraphics{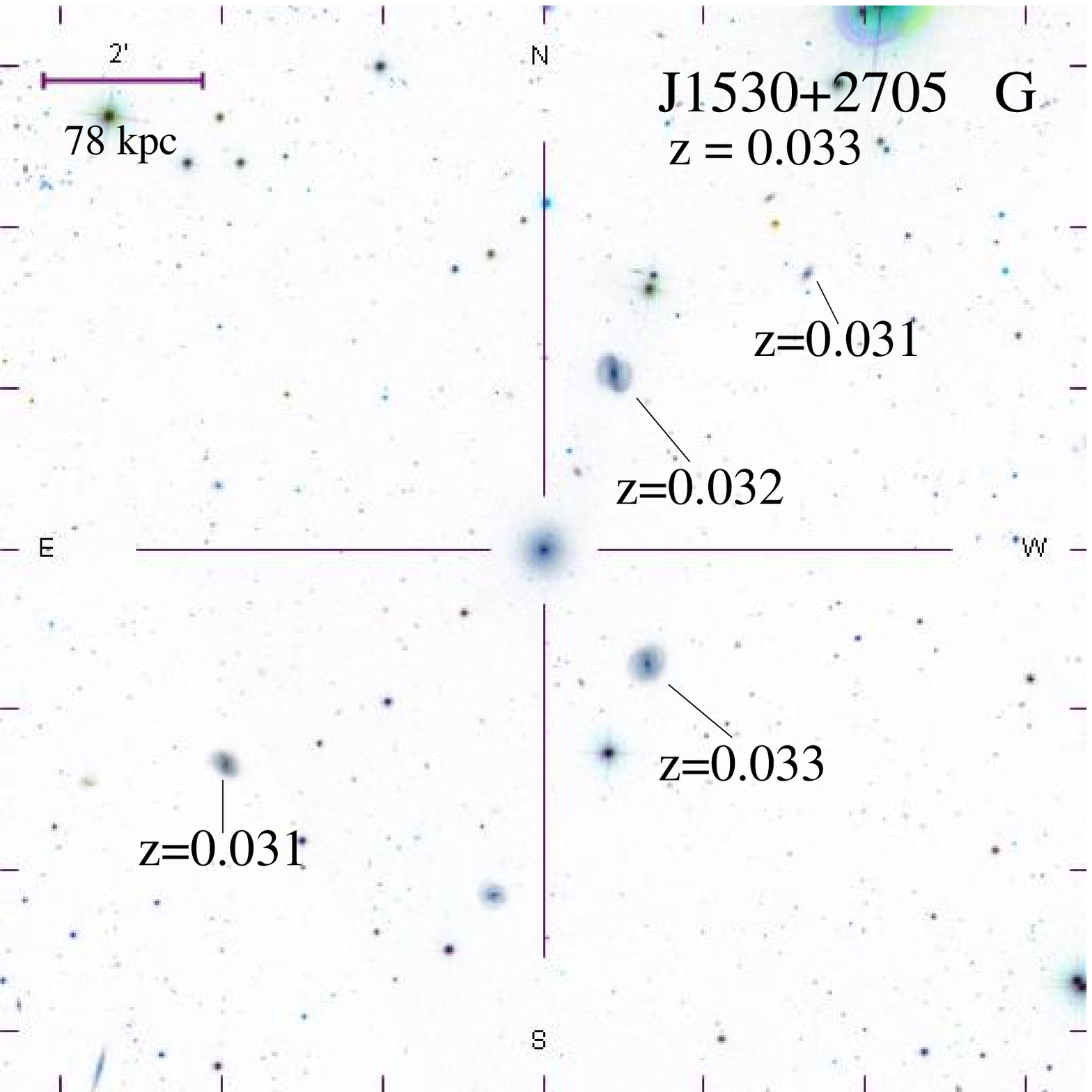}
\includegraphics{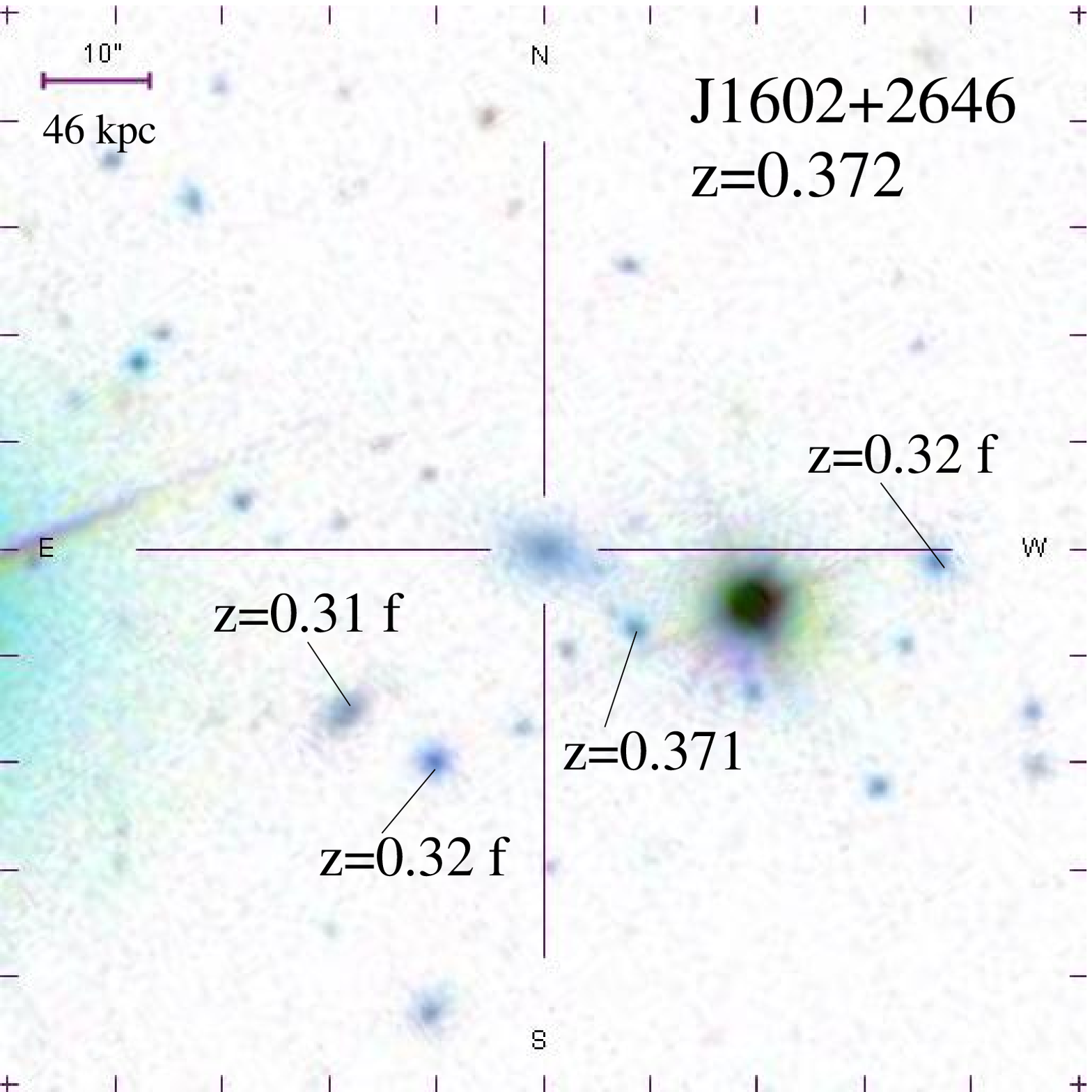}
\vspace{7cm}
\caption{Continued.}
\end{center}
\end{figure*}

\begin{figure}
\begin{center}
\includegraphics{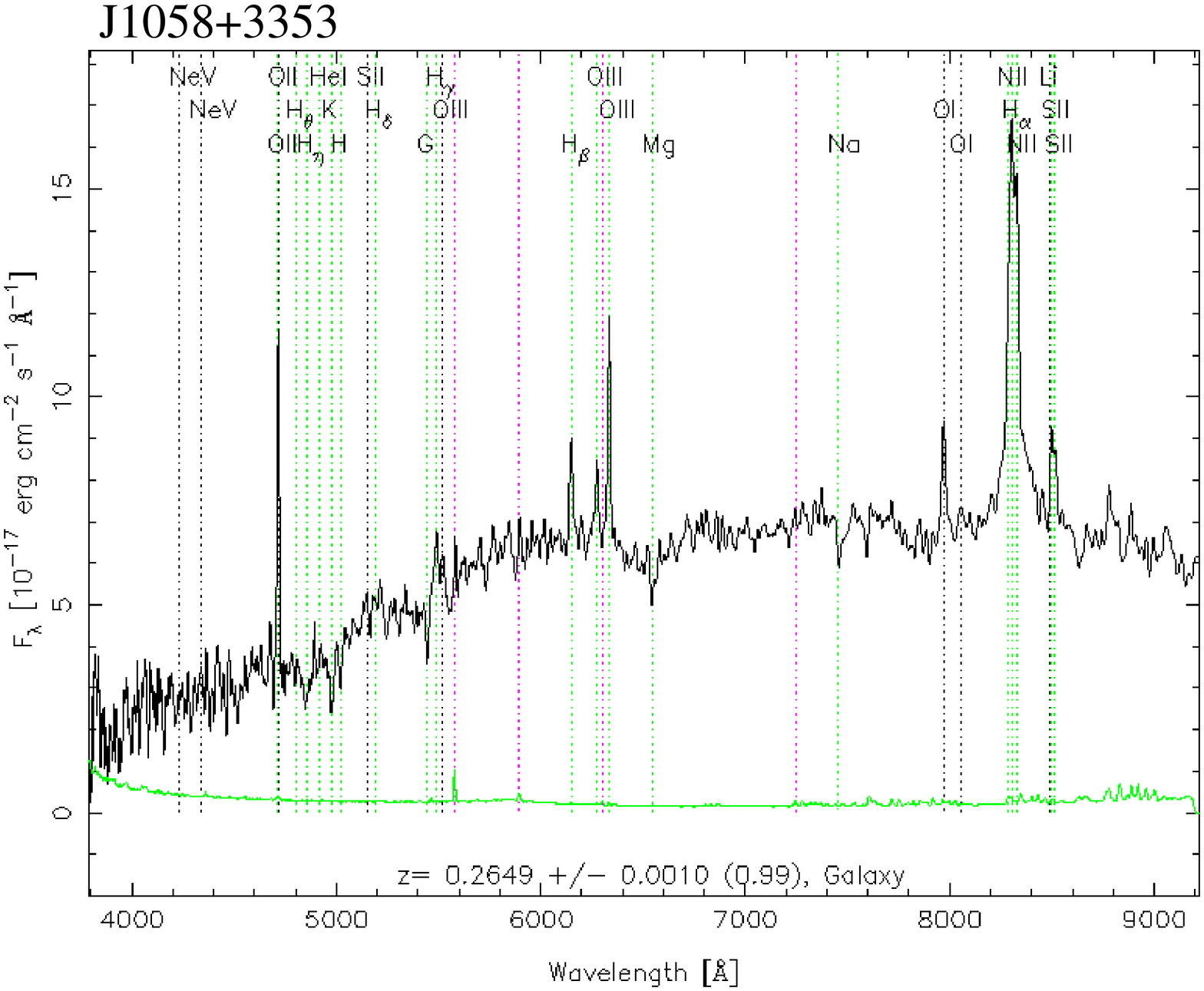}
\includegraphics{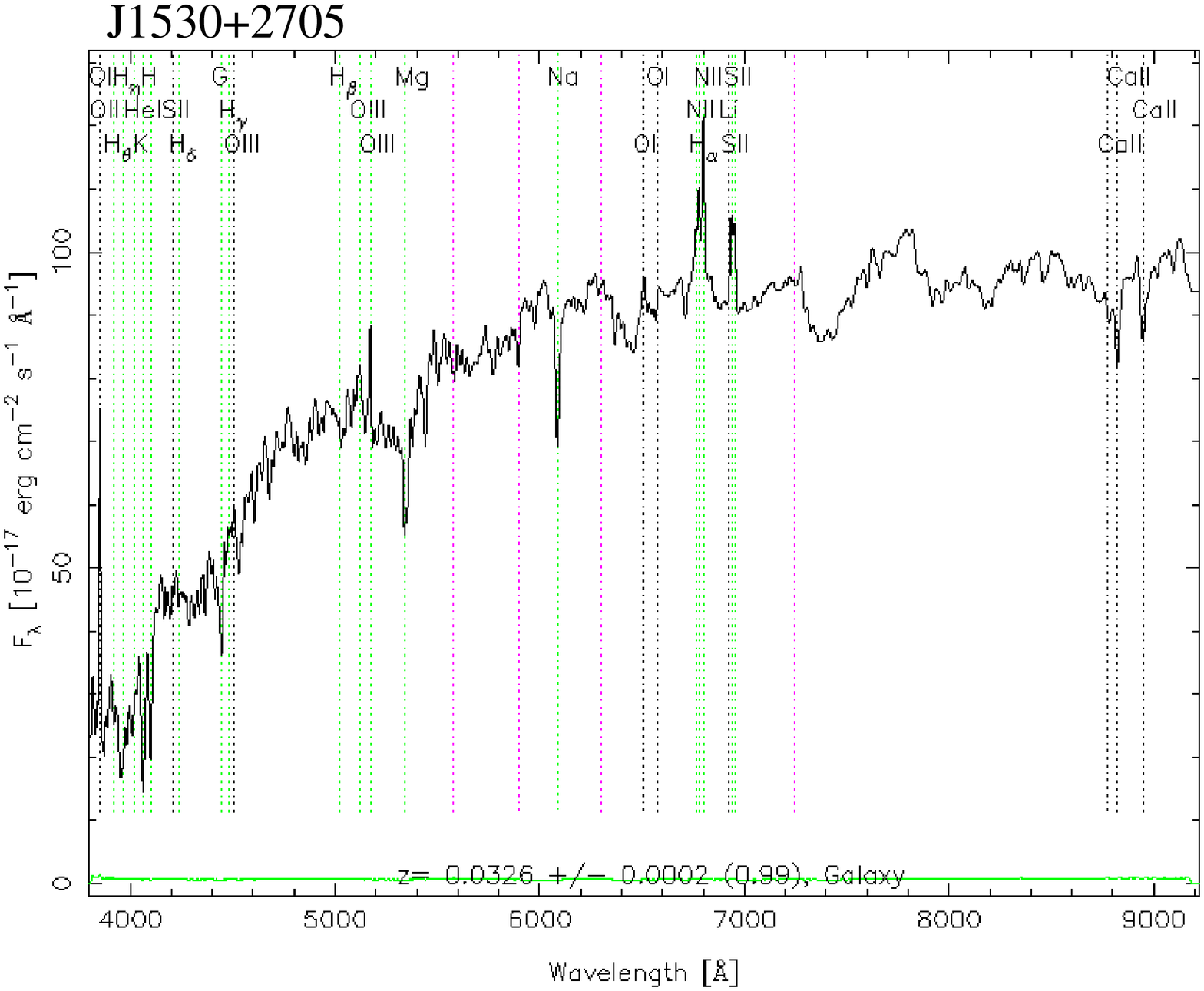}
\includegraphics{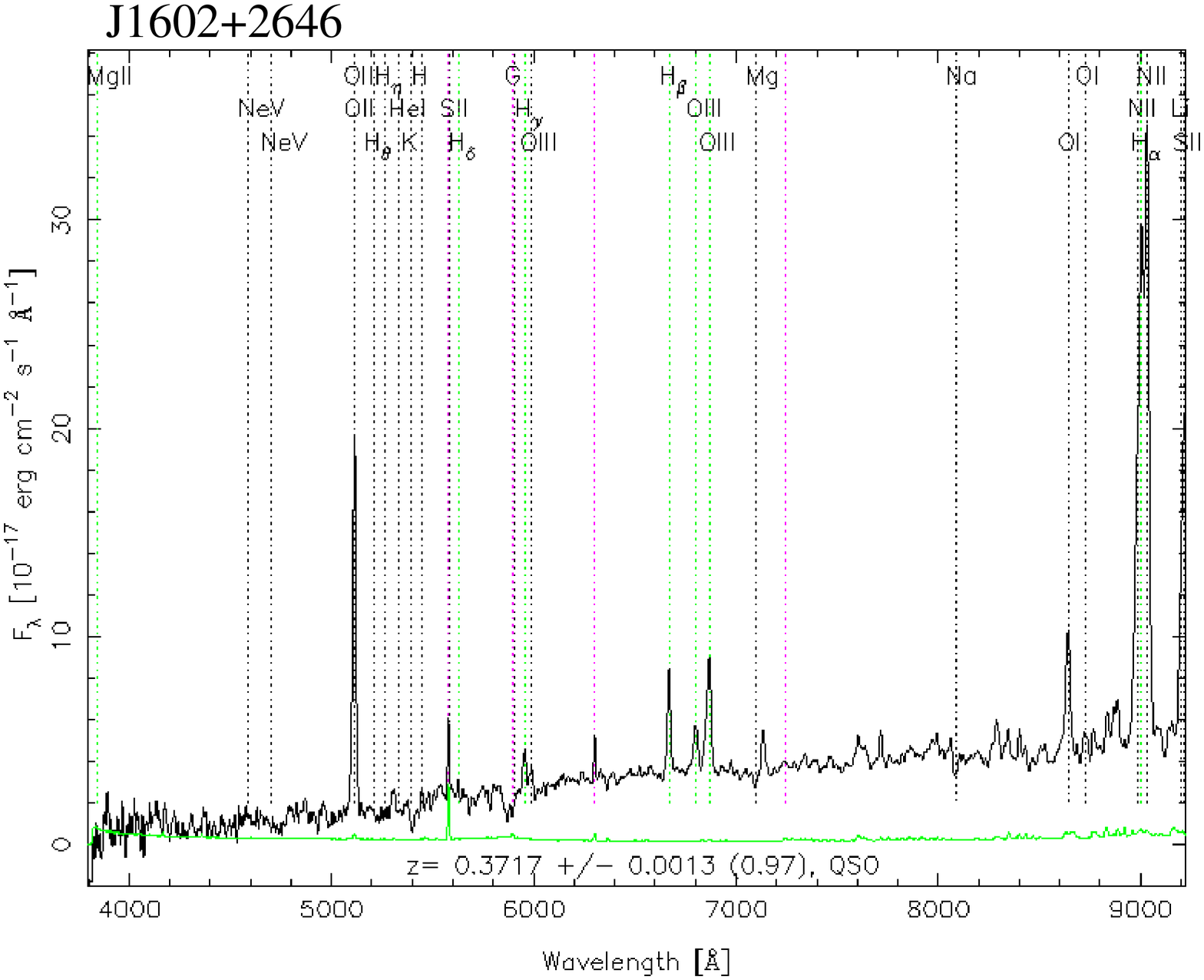}
\vspace{23cm}
\caption{Optical spectra from the SDSS DR7 
of the galaxies J1058+3353 ({\it top}),
  J1530+2705 ({\it center}), J1602+2642 ({\it bottom}).}
\label{spettro_sdss}
\end{center}
\end{figure}

\begin{figure}
\begin{center}
\includegraphics{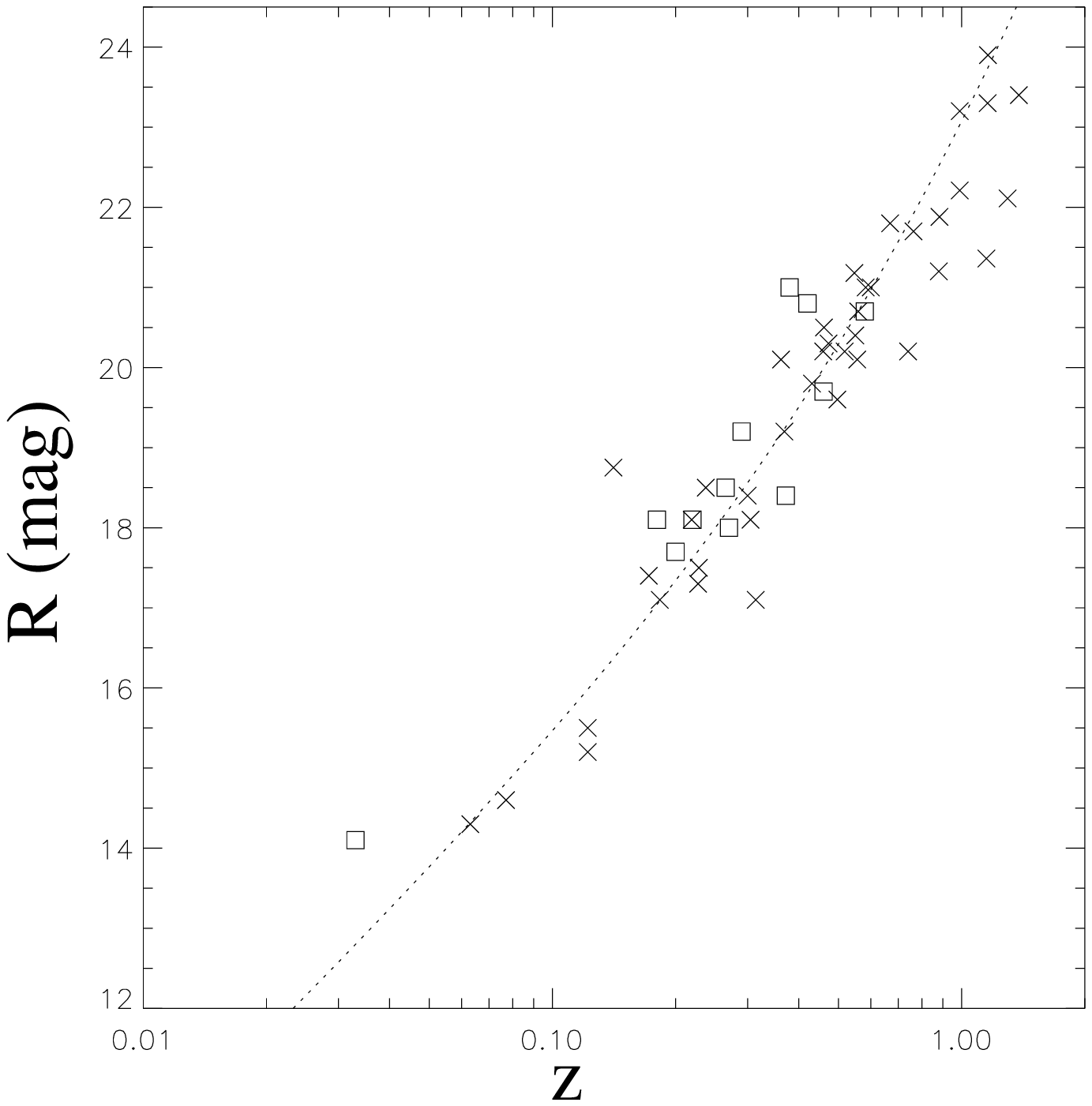}
\vspace{6cm}
\caption{The Hubble diagram of HFP galaxies ({\it squares}) and GPS
  galaxies ({\it crosses}, Snellen et al. 2002). The dotted line
  represents the Hubble relation for GPS galaxies as found by \citet{sn96}.}
\label{hubble}
\end{center}
\end{figure}

\section{Optical data}

To determine the optical properties of the sources in the faint HFP
sample, we complemented the information available in the literature
with that provided by the SDSS DR7 \citep{sdss7}. The optical
properties of each object (like source extension and magnitude) have
been carefully inspected beyond the automated procedures in the SDSS, 
in order to unambiguously identify the host
(i.e. quasar, galaxy, or empty field) of each radio source.\\
Of the 57 sources considered in this paper, 12 are identified with 
galaxies with redshift between 0.03 and 0.6; 33 are quasars with a
higher redshift, typically in the range from 0.6 to 3.0, 
while 12 sources still lack an
optical counterpart (labelled as empty field (EF) in Table
\ref{flussi_vla}). \\
Images and optical information have been retrieved by means of the SDSS
DR7 Finding Chart Tool. In Table \ref{flussi_vla} we report the R
magnitude, converted in the Johnson-Kron-Cousins BVRI system, 
and the spectroscopic redshift when available. An ``f''
indicates a photometric redshift.\\
In Fig. \ref{sdss_images} we present the optical images from the SDSS
DR7 \citep{sdss7} of the sources hosted in galaxies. \\
For the 3 galaxies with available spectroscopic data we show
the spectrum in Fig. \ref{spettro_sdss} and we summarize the information on
the main lines in Table \ref{line_sdss}.\\

\subsection{Optical properties}
A characteristic arising from Fig. \ref{sdss_images} is
the presence of companions within a projected distance of about 150 -
200 kpc 
around 6 HFP galaxies. Although in
J0804+5431, J1058+3353, J1109+3831, and J1218+2828 this evidence comes
from photometric information only, in the case of J1530+2705 and
J1602+2646 the association is confirmed by spectroscopic redshifts,
supporting the idea that young radio sources reside in groups, as
found in other works \citep{mo06b,sn02,cs93}. The relatively
small redshift of J1530+2705 allows us to identify the spiral morphology of
the brightest companions where also a bar is clearly visible.\\ 
The galaxies hosting the HFPs are usually the brightest elliptical at
the group centre. In the case of J0804+5431 the radio source is hosted
by an elliptical galaxy that is at a projected distance of about 160 
kpc to the north-east  
from the brightest galaxy at the centre of the group and it is at a
projected distance of about 20 kpc to the north of another
elliptical. \\
An intriguing case is represented by
J1109+3831 whose hosting galaxy is a spiral that is located at a
projected distance of about 20 kpc from an elliptical. A possible
identification error between optical and radio images has been
excluded by the analysis of the
optical spectrum of the companion, which lacks the typical lines
displayed by active galaxies. \\ 
Among the HFPs identified with galaxies, 3 objects (J1058+3353,
J1530+2705, and J1602+2646) have an optical spectrum in the SDSS
DR7. For J1602+2646, 
the optical spectrum seems to be well fitted by a QSO template, since
a large fraction of the light comes from the nuclear region. However,
both Fig. \ref{sdss_images} and 
the analysis of the diagnostics O[II]/H$_{\beta}$ - O[III]/H$_{\beta}$
clearly indicate that this source is hosted in a galaxy.\\
The emission lines detected in these objects (Table
\ref{line_sdss}) are those typical for radio galaxies, showing [O\,II]
$\lambda$3727, [O\,III] $\lambda \lambda$4959, 5007,
H$_{\alpha}$/N\,II, H$_{\beta}$ lines, and [O\,I] $\lambda$6300 in
J1058+3353 and in J1602+2646. The [O\,II]/[O\,III] line ratios
indicate low ionization, as also found in the optical spectra of faint
GPS sources \citep{snellen99}.\\
In the spectrum of each galaxy, the absorption lines associated with
stellar populations (4300-\AA \,G band, 5175-\AA \,Mg, and 5900-\AA \,Na D)
are clearly visible (Fig. \ref{spettro_sdss}), in particular in J1530+2705,
where the relatively small redshift allows the detection of the Ca\,II
lines. The spectra of the companion galaxies (not shown here) 
have absorption features
similar to those found in the HFPs, but without prominent [O\,II],
[O\,III], and H$_{\beta}$ emission lines, as expected in non-active
objects.\\
To check whether the galaxies of the faint HFP sample follow the
Hubble relation found by \citet{sn96}, we added the 12 galaxies with
spectroscopic and photometric redshift in the GPS R-band Hubble
diagram of \citet{sn02}. 
The Hubble diagram, where
the HFP galaxies ({\it squares}) have been added to the GPS galaxies
({\it crosses}) from \citet{sn02}, is presented in Fig. \ref{hubble},
and it indicates that HFP galaxies have a tight distribution in the
apparent magnitude-redshift relation, as also found in GPS galaxies
\citep{sn96, sn02}.\\

\begin{table*}
\caption{Spectral lines of the HFP galaxies with available optical
  spectrum from the SDSS DR7. Column 1: source name (J2000); Col. 2:
  line; Col. 3: line frequency in the observer's frame; Col. 4: line
  flux density in the observer's frame; Col. 5: equivalent width in
  the rest frame.}
\begin{center}
\begin{tabular}{ccccc}
\hline
Source&Line&$\lambda_{\rm oss}$&$S_{\rm line, obs}$&EW$_{\rm rest}$\\
 & &\AA &10$^{-16}$ erg s$^{-1}$ cm $^{-2}$& \AA \\
\hline
&&&&\\
J1058+3353&     [O\,I]&7915.6$\pm$1.0&  2.1$\pm$0.9 &  5.4$\pm$0.7\\
          &    [O\,II]&4713.7$\pm$5.3&  6.3$\pm$1.0 & 10.0$\pm$0.7\\
          &   [O\,III]&6273.6$\pm$0.8&  0.9$\pm$0.3 &  3.2$\pm$0.4\\
          &   [O\,III]&6334.2$\pm$0.3&  9.9$\pm$1.0 & 10.0$\pm$0.4\\
          & H$_{\beta}$&6149.1$\pm$0.9&  2.0$\pm$0.4 &  5.8$\pm$0.5\\
          &H$_{\alpha}$&8304.2$\pm$6.2& 34.3$\pm$1.0 & 26.0$\pm$0.8\\ 
J1530+2705&     [O\,I]&6507.5$\pm$0.6&  0.5$\pm$0.1 &  0.6$\pm$0.1\\
          &    [O\,II]&3848.3$\pm$4.1& 18.3$\pm$3.3 &  5.5$\pm$0.3\\
          &   [O\,III]&5120.8$\pm$5.1&  1.1$\pm$0.2 &  1.6$\pm$0.2\\
          &   [O\,III]&5171.2$\pm$0.1&  6.9$\pm$0.7 &  2.2$\pm$0.1\\
          &H$_{\alpha}$&6779.5$\pm$5.0&  6.8$\pm$0.3 &  2.3$\pm$0.1\\
J1602+2646&     [O\,I]&8643.8$\pm$2.4& 24.7$\pm$5.2 & 26.4$\pm$2.6\\
          &    [O\,II]&5111.8$\pm$0.4&239.7$\pm$15.8&115.6$\pm$2.9\\
          &   [O\,III]&6800.7$\pm$0.8&  4.7$\pm$0.7 & 13.2$\pm$0.9\\
          &   [O\,III]&6865.6$\pm$0.5& 23.5$\pm$2.1 & 29.6$\pm$1.1\\
          &  H$_{\beta}$&6669.1$\pm$0.3& 19.0$\pm$1.6& 21.9$\pm$0.8\\
          &H$_{\alpha}$&9004.7$\pm$22.5&519.6$\pm$10.5&164.7$\pm$3.3\\
&&&&\\
\hline     
\end{tabular}
\label{line_sdss}
\end{center}
\end{table*}

\begin{table*}
\caption{Multi-frequency VLA flux density of 57 sources from the faint
HFP sample. Column 1: source name (J2000); Col. 2: optical
identification from the SDSS DR7: G=galaxy, Q=quasar; EF=empty field;
Col. 3: R magnitude; Col. 4: redshift. An ``f'' indicates a
photometric redshift from the SDSS DR7; Col. 5: the observing code
from Table \ref{vla_oss}; Cols. 6 - 14: flux density at 1.4, 1.7,
4.5, 5.0, 8.1, 8.4, 15.3, 22.2, and 43.2 GHz respectively; Cols. 15 and 16:
the spectral index computed below and above the spectral peak,
respectively.}
\begin{center}
\begin{tabular}{cccccccccccccccc}
\hline
Source&ID&mag&z&code&$S_{1.4}$&$S_{1.7}$&$S_{4.5}$&$S_{5.0}$&$S_{8.1}$&$S_{8.4}$&$S_{15.3}$&$S_{22.2}$&$S_{43.2}$&$\alpha_{\rm
b}$&$\alpha_{\rm a}$\\
 & &R& & &mJy&mJy&mJy&mJy&mJy&mJy&mJy&mJy&mJy& & \\
(1)&(2)&(3)&(4)&(5)&(6)&(7)&(8)&(9)&(10)&(11)&(12)&(13)&(14)&(15)&(16)\\
\hline
&&&&&&&&&&&&&&&\\
J0736+4744&Q &20.3& -   &b&39&45&60&60&49&47&29&18&9&-0.3&0.8\\
          &  &    &     &c&30& -&53&54&46&48& - &19& - &-0.5&0.7\\
J0754+3033&Q &17.3&0.769&a&69&77&170&173&182&180&151&130& - &-0.6&0.3\\
          &  &    &     &b&65&70&161&165&174&173&150&116&76&-0.6&0.4\\
          &  &    &     &c&- &- &150&154&166&166&- &95&- & -&0.6\\
J0804+5431&G &18.1&0.22f&c&37&- &82&81&73&72&- &37&- &-0.7&0.4\\
J0819+3823&Q &21.6&-    &a&18&25&115&120&101&96&48&17&- &-1.4&1.3\\
             &    &  &   &b&16&23&113&116&97&93&45&25&10&-1.6&1.0\\
             &    &  &   &c&14&- &120&127&115&112&- &26&-&-1.7&1.1\\
J0821+3107&Q &16.9&2.625&c&93&- &95&92&75&74&- &31&- &- &0.7\\
J0905+3742&EF&-   &-    &a&54&64&102&100&72&68&29&11&- &-0.5&1.3\\
          &  &    &     &b&51&74&102&99&71&68&35&18&8&-0.6&1.1\\
J0943+5113&G &20.8&0.42f&a&77&90&160&147&69&64&16&9&- &-0.6&1.8\\
             &    &   &  &c&72&- &163&152&79&75&- &17&- &-0.6&1.4\\
J0951+3451&G &19.2&0.29f&a&19&29&62&62&56&55&37&26&- &-1.0&0.6\\
             &    &   &  &c&23&- &63&64&59&58&- &24&- &-0.8&0.6\\
J0955+3335&Q &17.3&2.491&a&45&58&104&106&95&92&60&36&- &-0.6&0.7\\
             &    &   &  &c&49&- &94&94&79&76&- &29&- &-0.5&0.8\\
J1002+5701&EF&-   &-    &a&28&39&130&129&72&65&15&- &- &-1.3&1.8\\
J1004+4328&EF&-   &-    &b&11&15&36&37&29&29&10&16&10&-1.0&0.6\\
             &    &   &  &c&12&- &36&37&33&32&- &19&- &-1.0&0.4\\
J1008+2533&Q &18.3&1.960&b&49&66&113&112&100&99&130&161&135&- &- \\
          &  &    &     &c&50&- &116&116&107&108&- &150& - &- &- \\
J1020+2910&EF&-   &-    &b&24&27&16&16&15&14&13&10&6&- &0.4\\
          &  &    &     &c&16&- &15&15&14&14&- &7&- &- &0.3\\
J1020+4320&Q &18.8&1.964&b&118&157&253&247&191&183&116&80&31&-0.6&0.7\\
J1025+2541&G &19.7&0.46f&b&24&35&47&47&30&29&13&9 &5 &-0.6&1.0\\
J1035+4230&Q &19.1&2.440&b&28&28&90&95&98&96&68&44&14&-0.7&0.7\\
J1037+3646&EF&-   &-    &b&70&94&146&141&99&94&55&33&11&-0.6&0.9\\
J1044+2959&Q &18.9&2.983&b&- &- &144&177&163&159&120&96&65&- &0.6\\
J1046+2600&EF&-   &     &b&13&- &38&38&31&30&15&7&4&-0.8&1.0\\
J1047+3945&Q &20.0&-    &b&41&- &39&39&30&30&20&12&9&- &0.4\\
J1052+3355&Q &16.9&1.407&b&- &- &38&36&22&20&11&5 &5 &- &1.2\\
          &  &    &     &c&14&- &35&32&20&18&- &6 &- &-0.8&1.1\\
J1053+4610&EF&-   &-    &b&11&- &36*&38&42&42&54&59&42&-0.6&- \\
          &  &    &     &c&22&- &39&40&60&62&- &75&- &-0.4&- \\
J1054+5058&Q &22.0&-    &c&12&- &20&21&31&32&- &40&- &-0.4&- \\
J1058+3353&G &18.5&0.265&a&20&26&39&40&41&42&51&51&- &-0.4&- \\
J1107+3421&EF&-   & -   &a&21&32&73&72&52&50&21&6&- &-1.1&1.6\\
          &  &    &     &b&25&38&75&73&52&48&25&14&6&-1.0&1.1\\
          &  &    &     &c&28&- &73&72&51&48&- &9 &- &-0.8&1.3\\
J1109+3831&G &17.7&0.2f &a&12&14&53&59&90&91&77&56&- &-1.1&0.4\\
          &  &    &     &b&13&15&50&55&88&89*&81&59&23&-1.0&0.4\\
          &  &    &     &c&14&- &50&55&95&98&- &53&- &-1.1&0.6\\
J1135+3624&EF& -  &-    &a&28&37&58&58&46&45&21&11&- &-0.6&1.0\\
          &  &    &     &c&28&- &59&59&50&49&- &13&-&-0.6&1.0\\
J1137+3441&Q &18.6&0.835&a&25&34&78&83&85&121&157&169&- &-0.7&- \\
J1203+4803&Q &16.2&0.817&a&187&- &529&562&721&734&788&761&- &-0.6&- \\
          &  &    &     &c&218&252&416&425&452&458&- &384&-
&-0.4&0.2\\
J1218+2828&G &18.1&0.18f&a&25&28&92&96&98&97&69&58&- &-0.7&0.5\\
J1239+3705&Q &21.4&-    &a&- &- &93&101&129&130&112&86&- &-0.5&0.4\\
          &  &    &     &c&13&17&96&107&145&146&- &100&- &-1.3&0.4\\
&&&&&&&&&&&&&&&\\
\hline
\end{tabular}
\label{flussi_vla}
\end{center}
\end{table*}

\addtocounter{table}{-1}
\begin{table*}
\caption{Continued}
\begin{center}
\begin{tabular}{cccccccccccccccc}
\hline
Source&ID&mag&z&code&$S_{1.4}$&$S_{1.7}$&$S_{4.5}$&$S_{5.0}$&$S_{8.1}$&$S_{8.4}$&$S_{15.3}$&$S_{22.2}$&$S_{43.2}$&$\alpha_{\rm
b}$&$\alpha_{\rm a}$\\
 & &R& & &mJy&mJy&mJy&mJy&mJy&mJy&mJy&mJy&mJy& & \\
(1)&(2)&(3)&(4)&(5)&(6)&(7)&(8)&(9)&(10)&(11)&(12)&(13)&(14)&(15)&(16)\\
\hline
&&&&&&&&&&&&&&&\\
J1240+2323& G&21.0&0.38f &a& 27& 28& 52& 53& 59& 60& 54& 49& - &-0.6& 0.2\\
          &  &    &      &c& 24& - & 58& 60& 65& 65& - & 56& - &-0.6&-\\
J1240+2425& Q&16.9& 0.831&a& 78& - & 60& 58& 45& 44& 32& 27& - &  - &0.4\\
          &  &    &      &c& 57& - & 58& 56& 45& 44& - & 17& - &  - &0.5\\
J1241+3844& Q&21.5&  -   &c& 20& 21& 23& 25& 20& 20& - & 13& - &-0.2&0.4\\
J1251+4317& Q&18.7& 1.453&a& - & - & 54& 57& 80& 82&112&116& - &-0.4&-\\
          &  &    &      &c& 25& - & 72& 80&138&142& - &105& - &-1.0&0.3\\
J1258+2820& Q&19.2& -    &a& 25& 32& 51& 52& 51& 50& 43& 38& - &-0.6&0.2\\
          &  &    &      &c& 23& 34& 45& 45& 54& 54& - & 54& - &-0.3&-\\
J1300+4352& Q&19.6& -    &c&140&155&116&115& 97& 95& - & 81& - &  - &0.2\\
J1309+4047& Q&18.9& 2.910&a& 37& - &139&131&118&115& 76& 40& - &-1.1&0.8\\
          &  &    &      &c& 34& 51&132&133&113&110& - & 33& - &-1.1&0.9\\
J1319+4951& Q&19.1&  -   &a& 25& - & 49& 49& 39& 39& 25& 21& - &-0.5&0.5\\ 
          &  &    &      &c& 22& 30& 50& 49& 43& 43& - & 20& - &-0.7&0.6\\
J1321+4406& Q&21.2&  -   &a& 62& - & 73& 74& 71& 70& 58& 50& - &-0.1&0.3\\
          &  &    &      &c& 65& 73& 78& 77& 75& 74& - & 38& - &-0.1&0.4\\ 
J1322+3912& Q&17.5& 2.985&a&117& - &227&223&181&176&118& 86& - &-0.6&0.6\\
          &  &    &      &c&116&135&200&196&150&146& - & 56& - &-0.5&0.8\\
J1330+5202& G&20.7& 0.58f&c& 91&103&176&180&180&177& - &156& - &-0.4&0.1\\
J1336+4735& Q&19.7&  -   &a& - & - & 61& 61& 52& 51& 26& 22& - & -  &0.7\\
          &  &    &      &d& 26& - & 67& 67& 57& 56& - & 30& - & -  &0.5\\
J1352+3603& G&18.0& 0.27f&a& 65& 70&102&103& 95& 93& 65& 37& - &-0.4&0.7\\
          &  &    &      &d& 59& - & 97& 97& 84& 82& - & 34& - &-0.4&0.7\\
J1420+2704& Q&20.3&  -   &a& 14& - & 55& 57& 55& 53& 34& 25& - &-1.1&0.6\\
          &  &    &      &d& 14& - & 62& 63& 58& 56& - & 26& - &-1.2&0.6\\
J1436+4820&EF& -  &  -   &a& 20& - & 72& 72& 61& 58& 32& 13& - &-1.0&1.1\\
J1459+3337& Q&16.6& 0.645&a& 22& 30&195&221&403&415&470&435& - &-1.3&-\\
J1530+2705& G&14.1& 0.033&a& 13& 14& 43& 46& 45& 43& 17& 13& - &-1.0&0.8\\
          &  &    &      &d& 11& 12& 48& 49& 42& 41& - & 23& - &-1.1&0.5\\
J1547+3518& Q&21.2&  -   &a& - & - & 53& 57& 67& 67& 74& 72& - &-0.3& -\\
          &  &    &      &d& 12& - & 49& 51& 57& 58& - & 74& - &-0.6& -\\
J1602+2647& G&18.4& 0.372&a& 39& - &131&148&238&244&259&224& - &-0.8& -\\
J1613+4223& Q&19.8&  -   &b& 42& - &206&201&119&110& 39& 14& - &-1.4&1.7\\
          &  &    &      &d& 40& - &201&194&114&105& - & 13& - &-1.4&1.7\\
J1616+4632& Q&19.3& 0.950&d& 88& - &141&144&148&147& - &129& - &-0.3&0.1\\
J1617+3801& Q&19.0& 1.607&a& 23& - & 71& 75& 99&100& 95& 67& - &-0.8&0.4\\
J1624+2748&EF&  - &   -  &a& 20& - &109&118&175&177&198&173& - &-1.0& -\\
J1651+3417&EF&  - &   -  &b& - & - & 46& 48& 58& 58& 49& 22& 16& -  &0.9\\
J1702+2643& Q&17.2&   -  &b& 35& 39& 41& 42& 49& 50& 61& 65& 73&-0.2&-\\
J1719+4804& Q&15.3& 1.084&a& 75& - &126&135&158&157&115& 85& - &-0.4&0.6\\
          &  &    &      &b& 66& - &122&127&150&148&113& 73& 30&-0.5&0.7\\
          &  &    &      &d& 60& - &112&114&100& 97& - & 43& - &-0.5&0.6\\
&&&&&&&&&&&&&&&\\
\hline
\end{tabular}
%\label{flussi_vla}
\end{center}
\end{table*}

\section{Results}

\subsection{Spectral properties}

Simultaneous multi-frequency observations carried out at different
epochs are necessary to monitor the spectral behaviour and variability
of high-frequency peaking radio sources. Variations in the spectral
properties, like peak frequency, spectral shape and flux density, are
strong indicators of the true nature of the source \citep{mo07,
    tingay03}. 
In young radio sources the spectral properties are not
expected to change, while spectral variability is a typical 
characteristic of beamed radio sources.\\
To determine a possible variation of the spectral peak,
for each source we fitted the simultaneous radio spectrum at each epoch
with a pure analytical function:

\begin{displaymath}
{\rm Log} S = a + {\rm Log} \nu \cdot (b + c \,{\rm Log} \nu)
\end{displaymath}

\noindent where $S$ is the flux density, $\nu$ the frequency, and {\it
a, b} and {\it c} are numeric parameters without any direct physical
meaning. We prefer to adopt this function instead of that used by
\citet{cs09} because it better represents the data, providing more
accurate values for the peak parameters. The best fits to the spectra
are shown in Fig. \ref{radio_spettri}, and the derived peak frequencies
at the various epochs are reported in Table
\ref{variability}. Statistical errors derived from the fit are not
representative of the real uncertainty on the estimate of the peak
frequency. 
For this reason we prefer to
assume a conservative uncertainty on the peak frequency of 10\%.
The position of the spectral peak is well constrained when
the peak occurs at a frequency well sampled by the observations, becoming
less accurate when the frequency coverage is not as appropriate. For
example, in the case a source has a spectrum peaking
at the edges of the frequency coverage (i.e. L or K/Q bands), the fit
provides parameters that are less constrained than in the case of
sources with the spectral peak occurring around 5-10 GHz, where both
the optically thin and thick emission are properly sampled. Spectra
poorly constrained are also those lacking observations at some
frequencies, as in the case of 
J1044+4328, J1052+3355, and J1547+3518 where the lack of data
either at 1.4/1.7 or 15.3 GHz precluded a reliable determination of
their peak frequencies.    \\
By comparing the distribution of the peak frequency of all the sources 
at the various epochs, we do not
find remarkable differences; the changes are 
usually within the uncertainties. The median value of the peak
frequency of the whole sample at each epoch has not changed
significantly: $\nu_{\rm p} = 5.8$ GHz in the 1998-1999 epoch
\citep[namely the observations of the HFP candidates that have been
  used for selecting the faint HFP sample;][]{cs09}; 
$\nu_{\rm p} = 6.2$ GHz in the 2003-2004 epoch; and
$\nu_{\rm p} = 5.4$ GHz in the 2006-2007 epoch. Among the sources
studied here, a
few cases show significant variation: in 7 sources
(J1053+4610, J1058+3353, J1203+4803, J1251+4317, J1258+2820,
J1300+4352, and J1616+4632) the peak measured in the most recent
epoch has substantially shifted towards
either higher or lower frequencies. This result indicates that these
sources are part of the blazar population, since in young radio
sources the shift of the peak towards lower frequencies is not expected to
be so remarkable at least on this short time scale (see Section
5).\\
We computed the spectral
index both below ($\alpha_{\rm b}$) and above ($\alpha_{\rm a}$) the
peak frequency, by fitting a straight line in the optically-thick and
-thin part of the spectrum, respectively, following the approach by \citet{torniainen05} and \citet{mo07}. We considered ``flat'' those
sources with both $\alpha_{\rm b} > -0.5$ and $\alpha_{\rm a} <
0.5$. In a few sources, depending on the peak frequency, we could fit
either $\alpha_{\rm b}$ or $\alpha_{\rm a}$ only, in order to avoid
the flattening near the peak. 
We find that 12 sources (2 galaxies, 9
quasars, and 1 empty field) show a flat spectrum during at least one
of the observing epochs, implying that they are part of the blazar
population. The fitted optically thick and thin
spectral indices are reported in Table \ref{flussi_vla}. \\
A case that is worth discussion in detail is the quasar
J1008+2533. The radio
spectrum shown by this source during two of the observing runs
presented here turned out to be a composition of two different
spectra: convex at frequencies below 8.4 GHz, and inverted at higher
frequencies (see Fig. \ref{radio_spettri}). This shape is similar to
that shown by the bright HFP J0927+3902. In J0927+3902 the
two-component spectrum is explained by its core-jet structure
\citep[see e.g.][]{mo06a}: at frequencies below 1 GHz the spectrum is
dominated by the emission from the jet, while at higher
frequencies the contribution from the self-absorbed core becomes more
important, becoming the dominant emission above $\sim$ 10 GHz. 
Such a scenario is well supported by pc-scale morphological information by
multi-frequency VLBI data \citep{alberdi00}. 
A similar
explanation may apply to the case of the faint HFP
J1008+2533. Another possibility is that in our new epochs a flare
from a 
self-absorbed knot in the jet occurred, causing an increase of the flux density
at high frequencies, as also found in blazar objects. 
The lack of pc-scale morphology information does not allow
us to unambiguously determine the origin of this complex spectrum.\\
 
\begin{figure*}
\begin{center}
\includegraphics{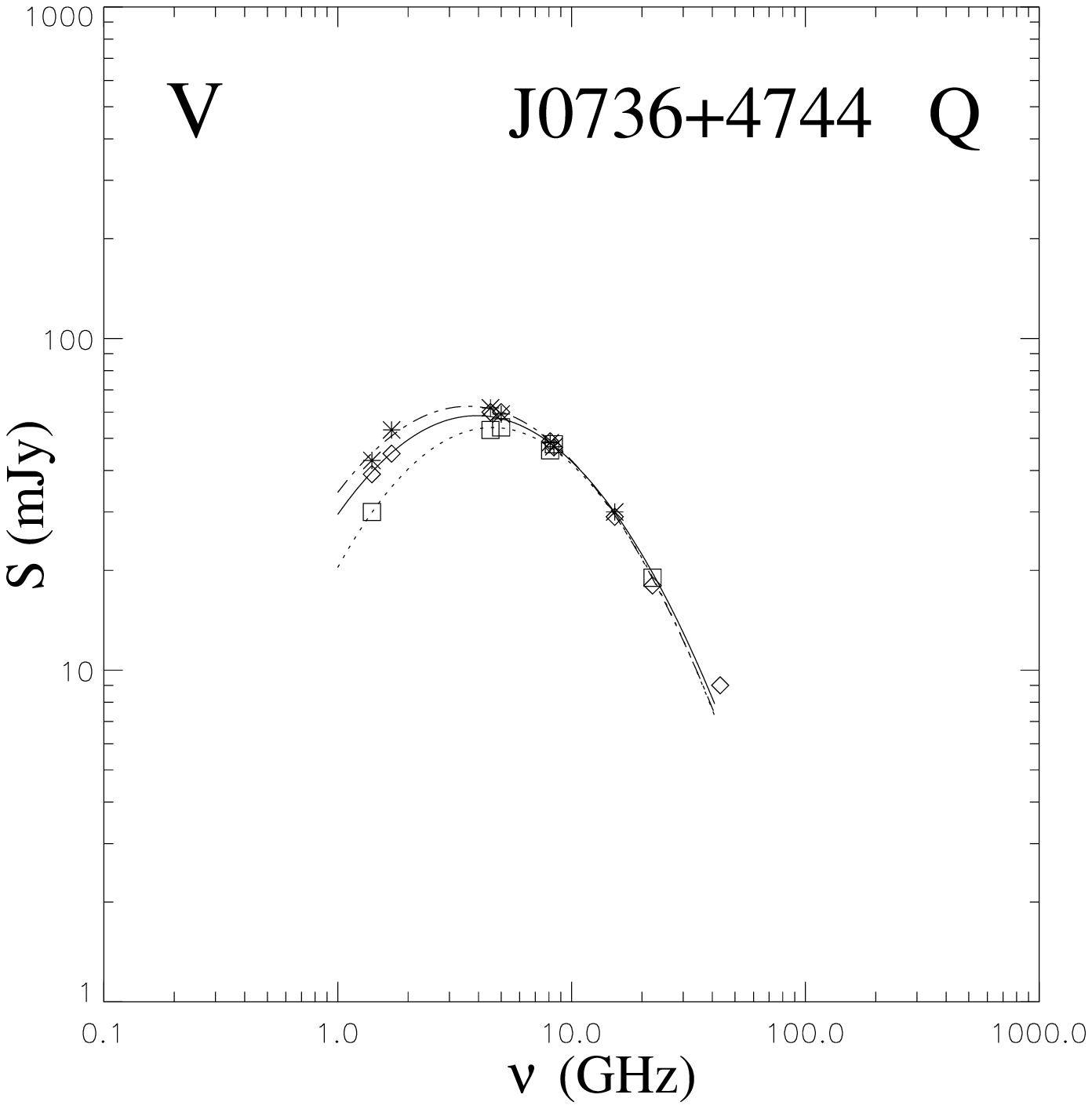}
\includegraphics{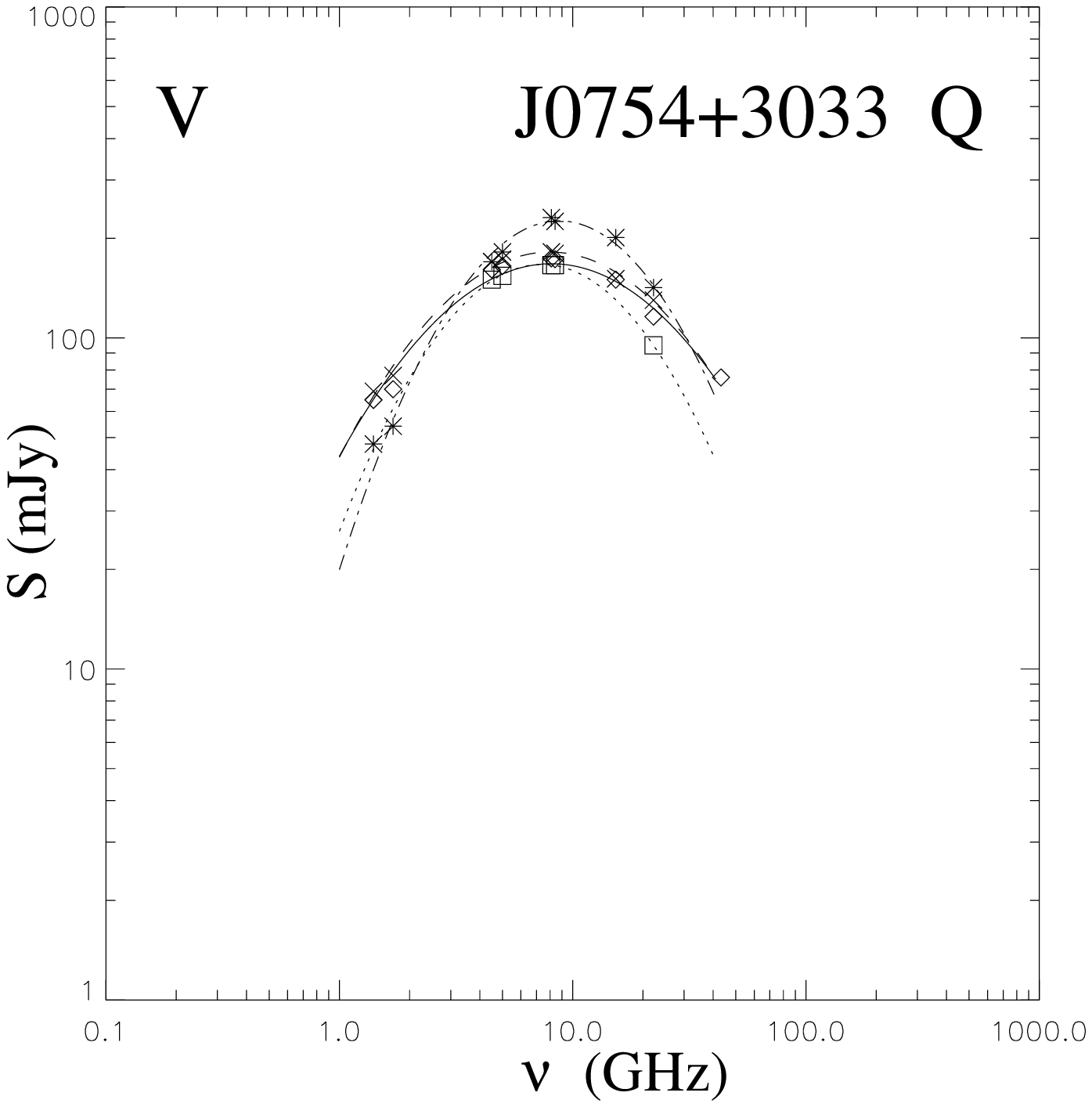}
\includegraphics{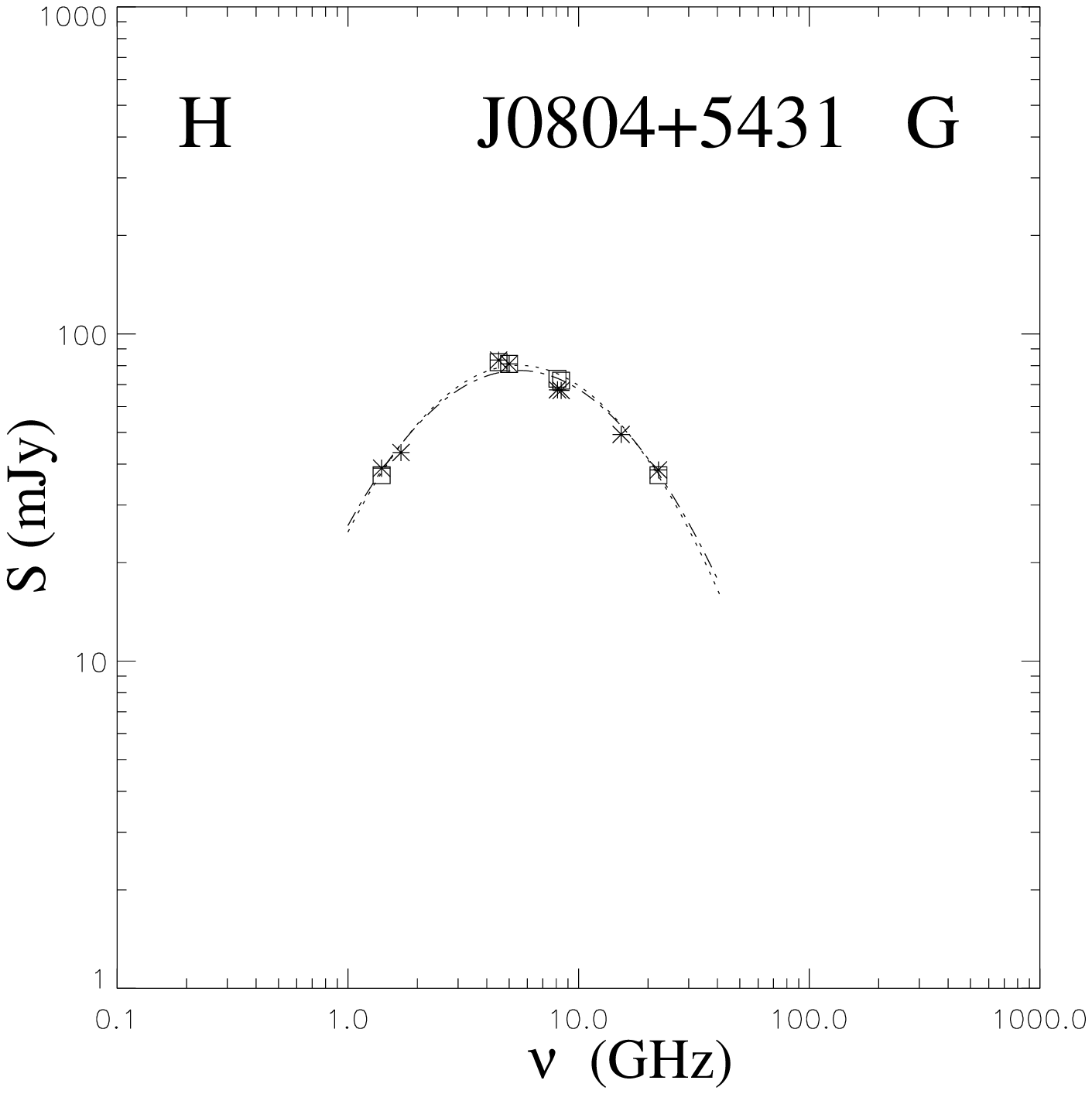}
\includegraphics{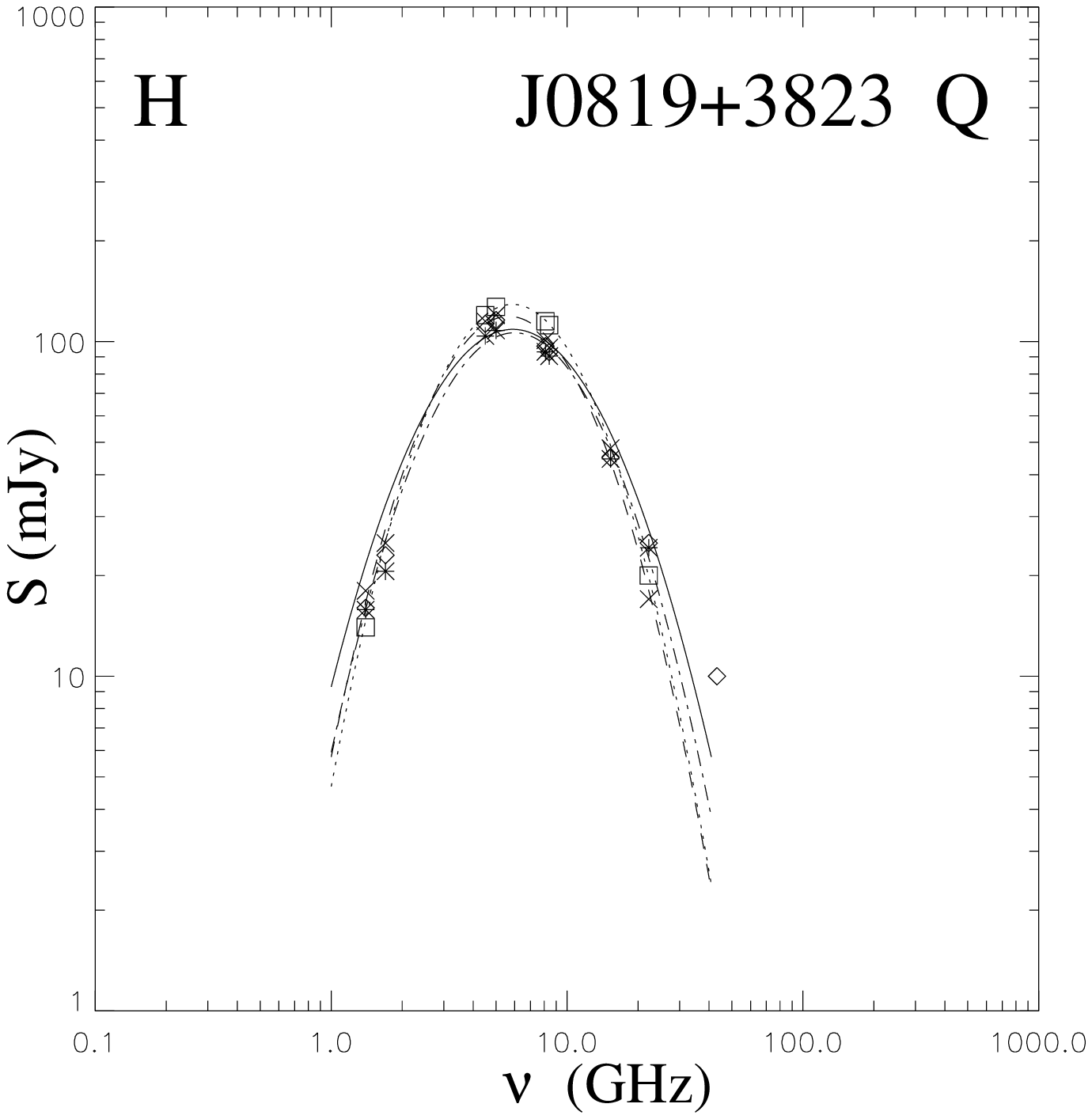}
\includegraphics{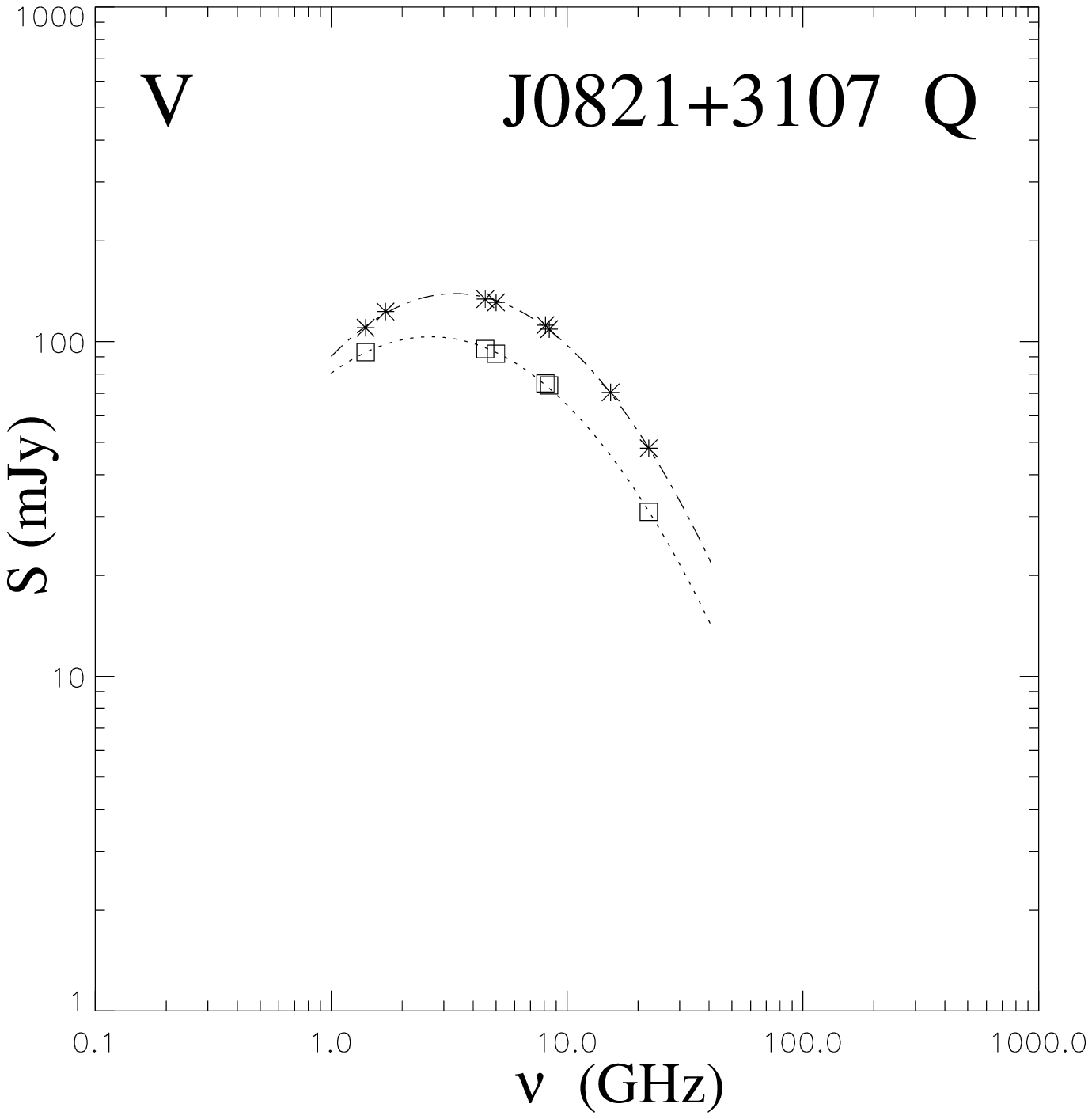}
\includegraphics{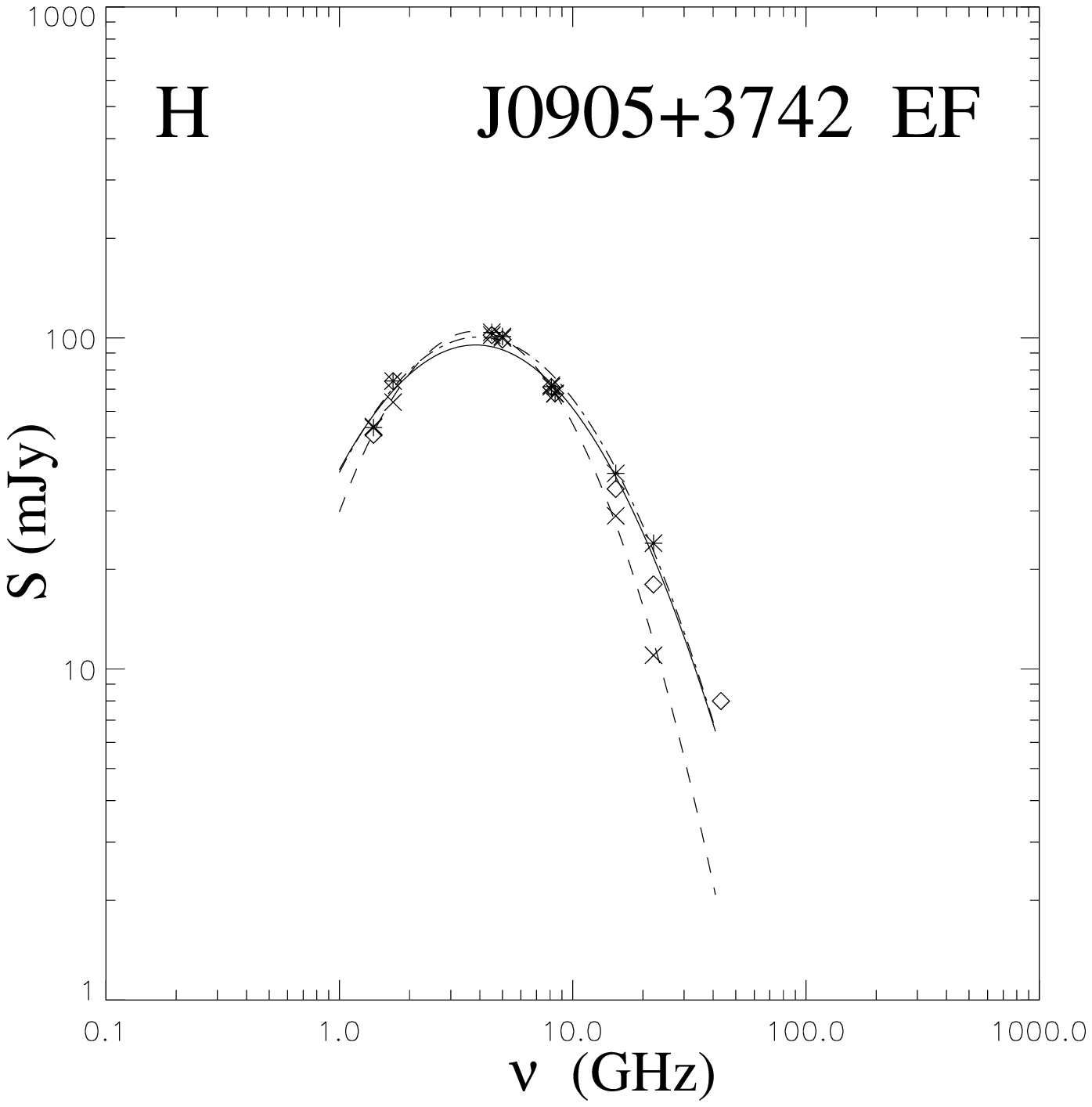}
\includegraphics{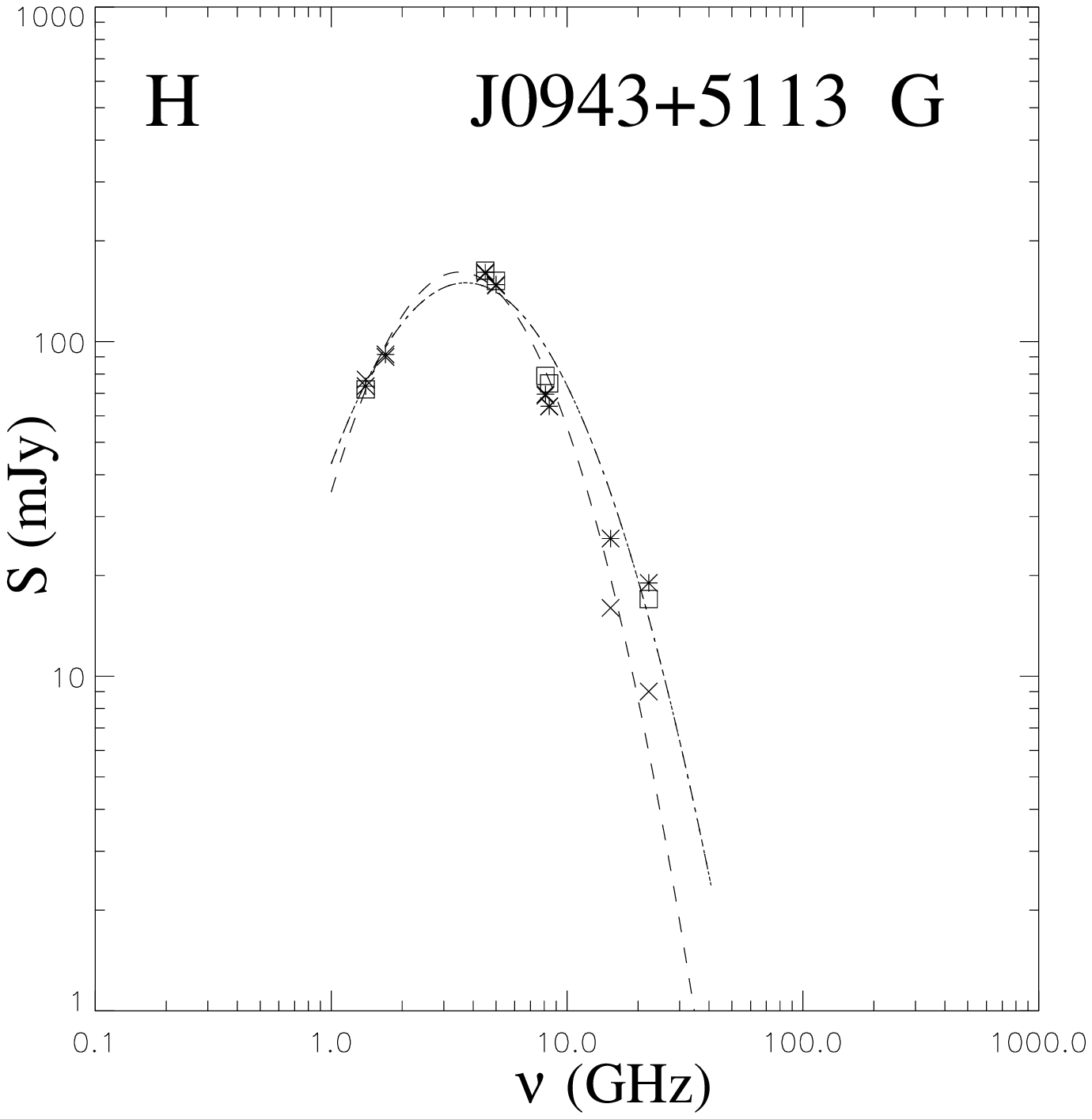}
\includegraphics{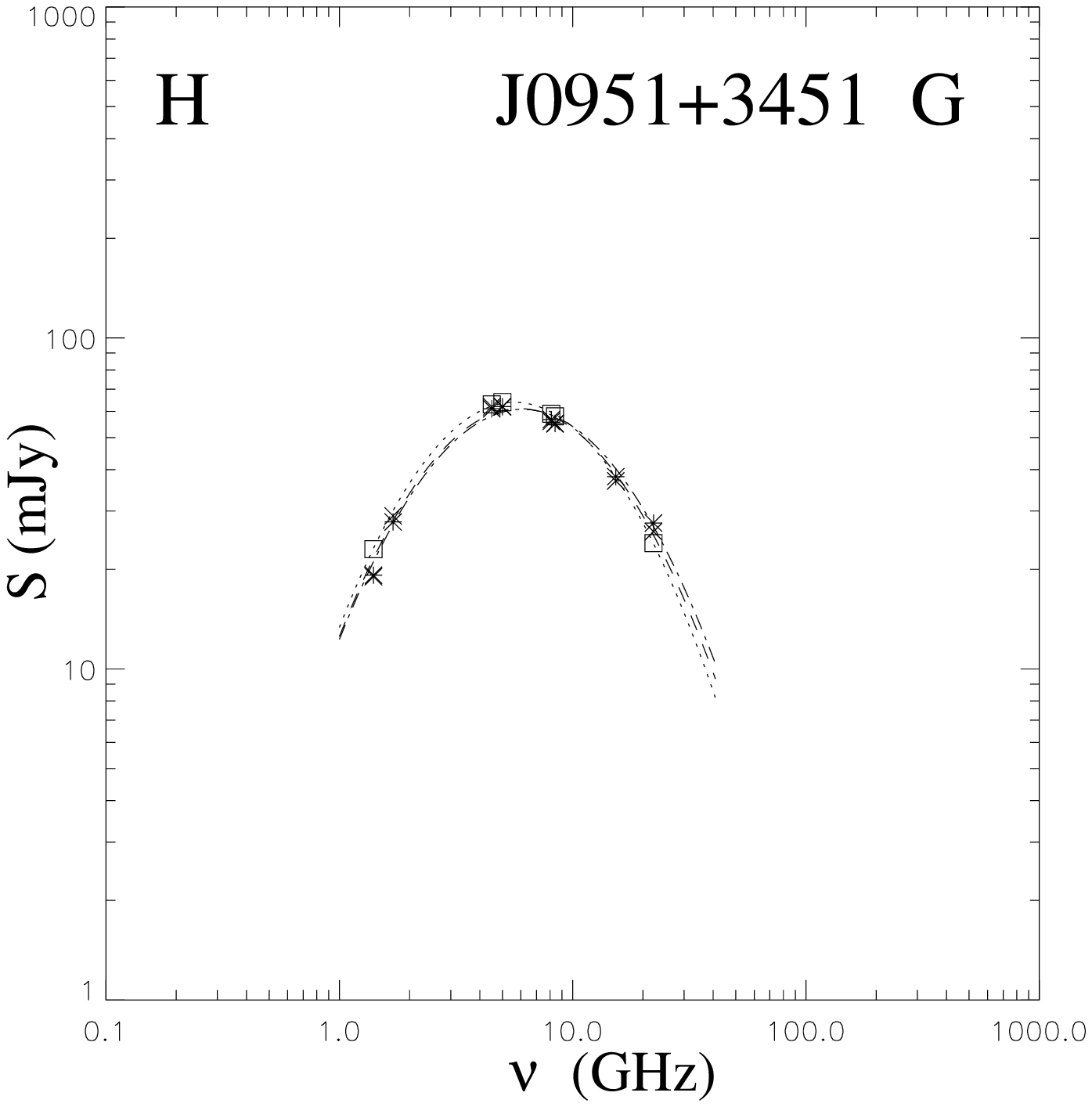}
\includegraphics{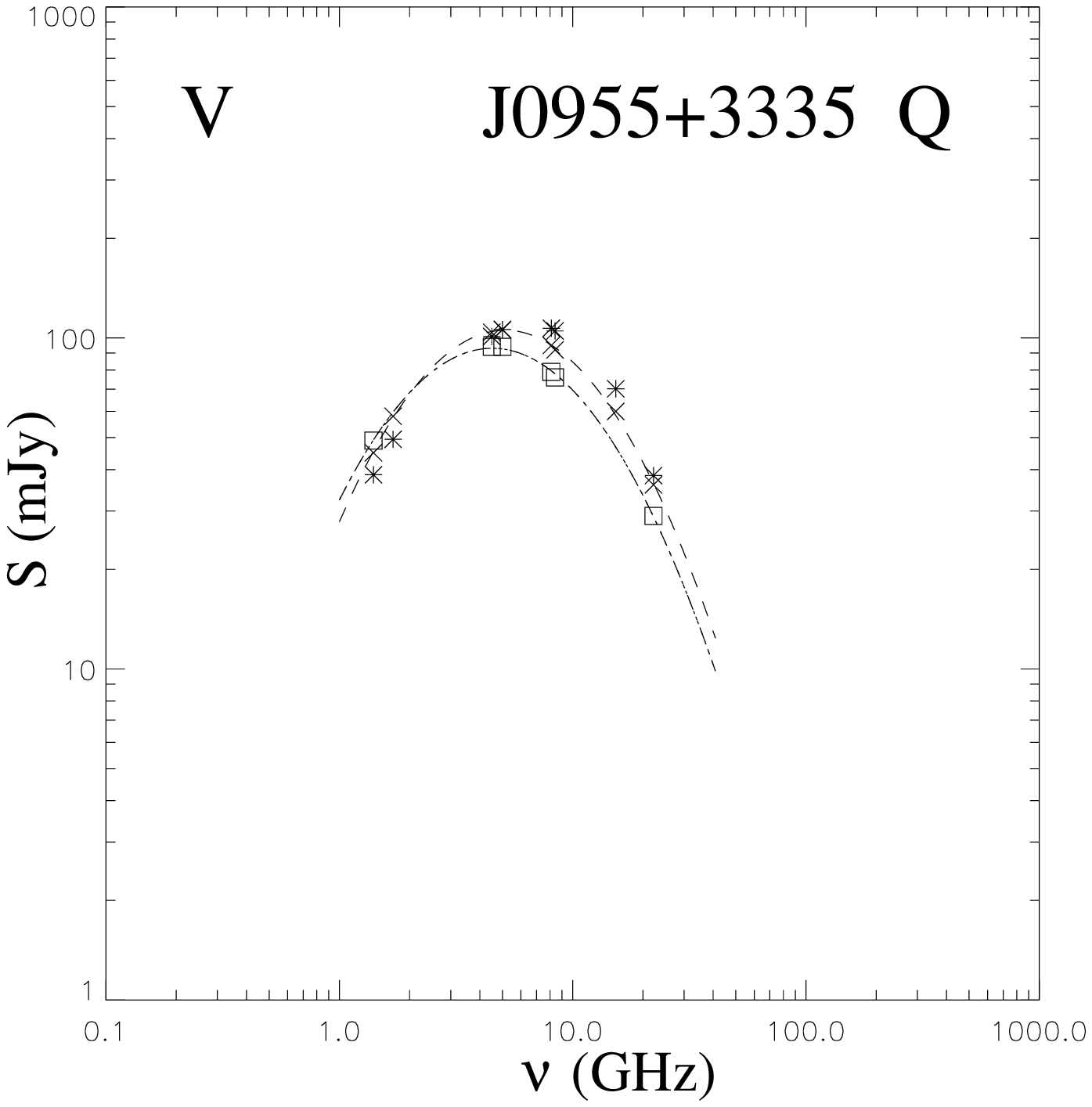}
\includegraphics{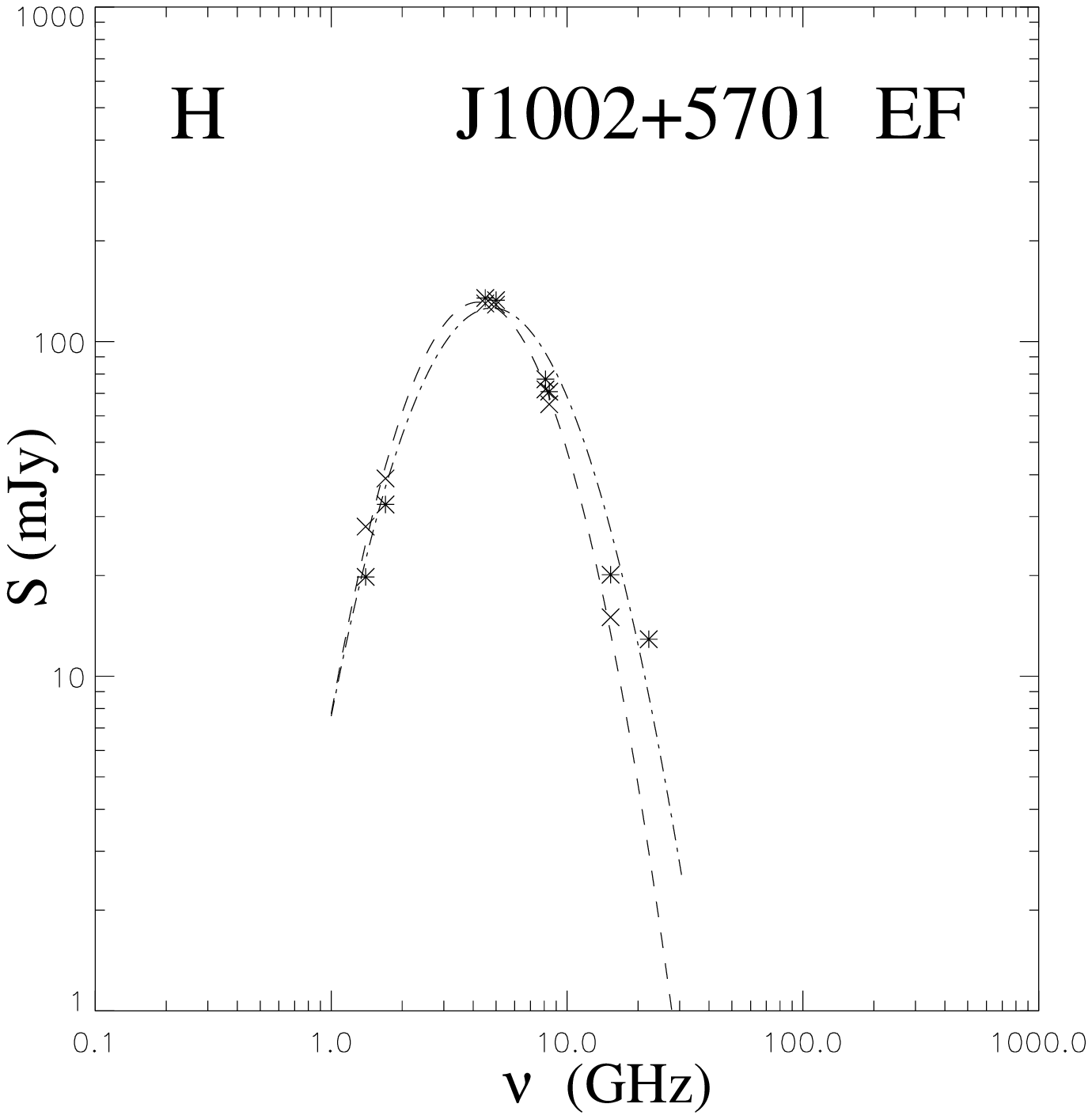}
\includegraphics{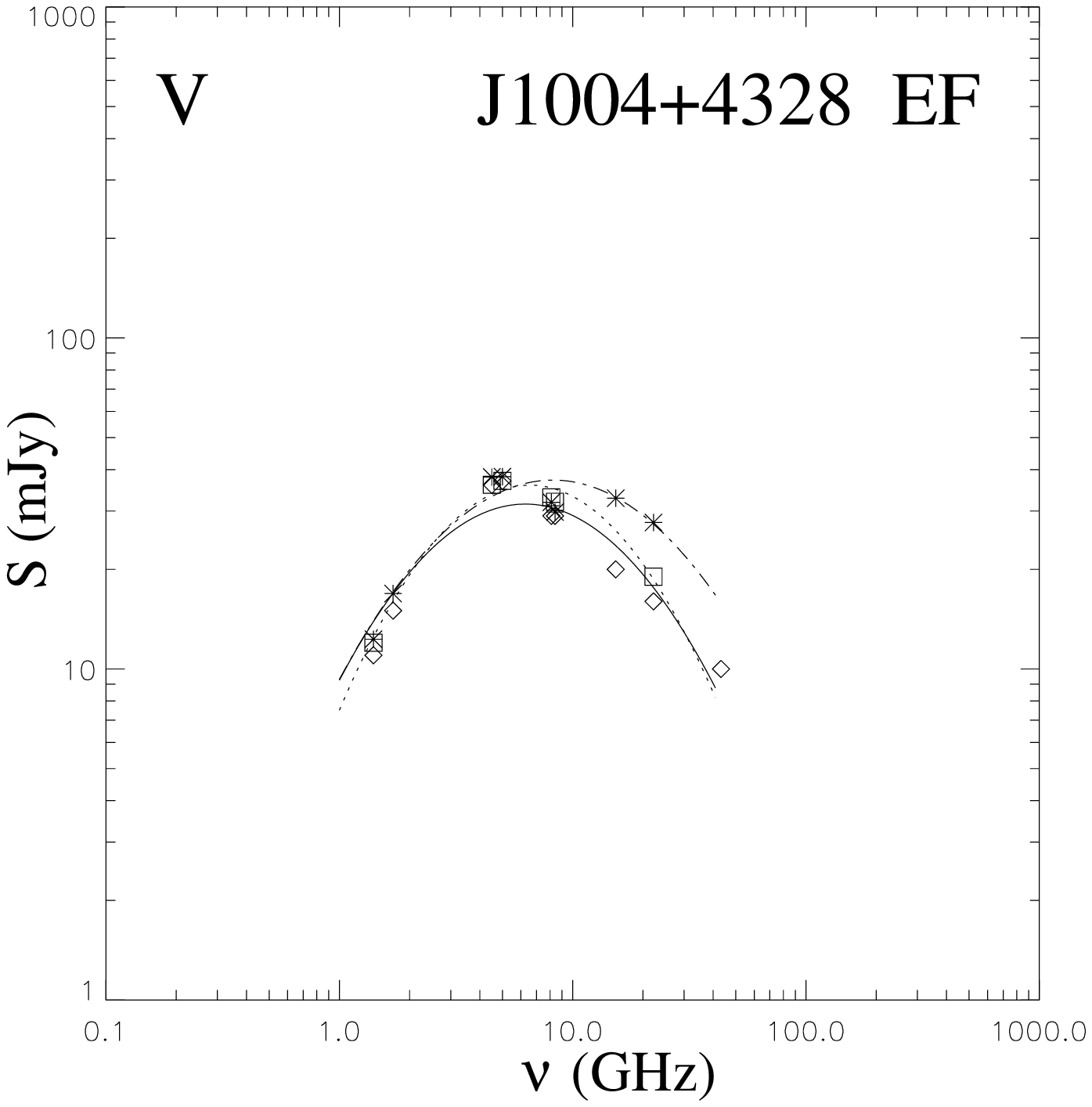}
\includegraphics{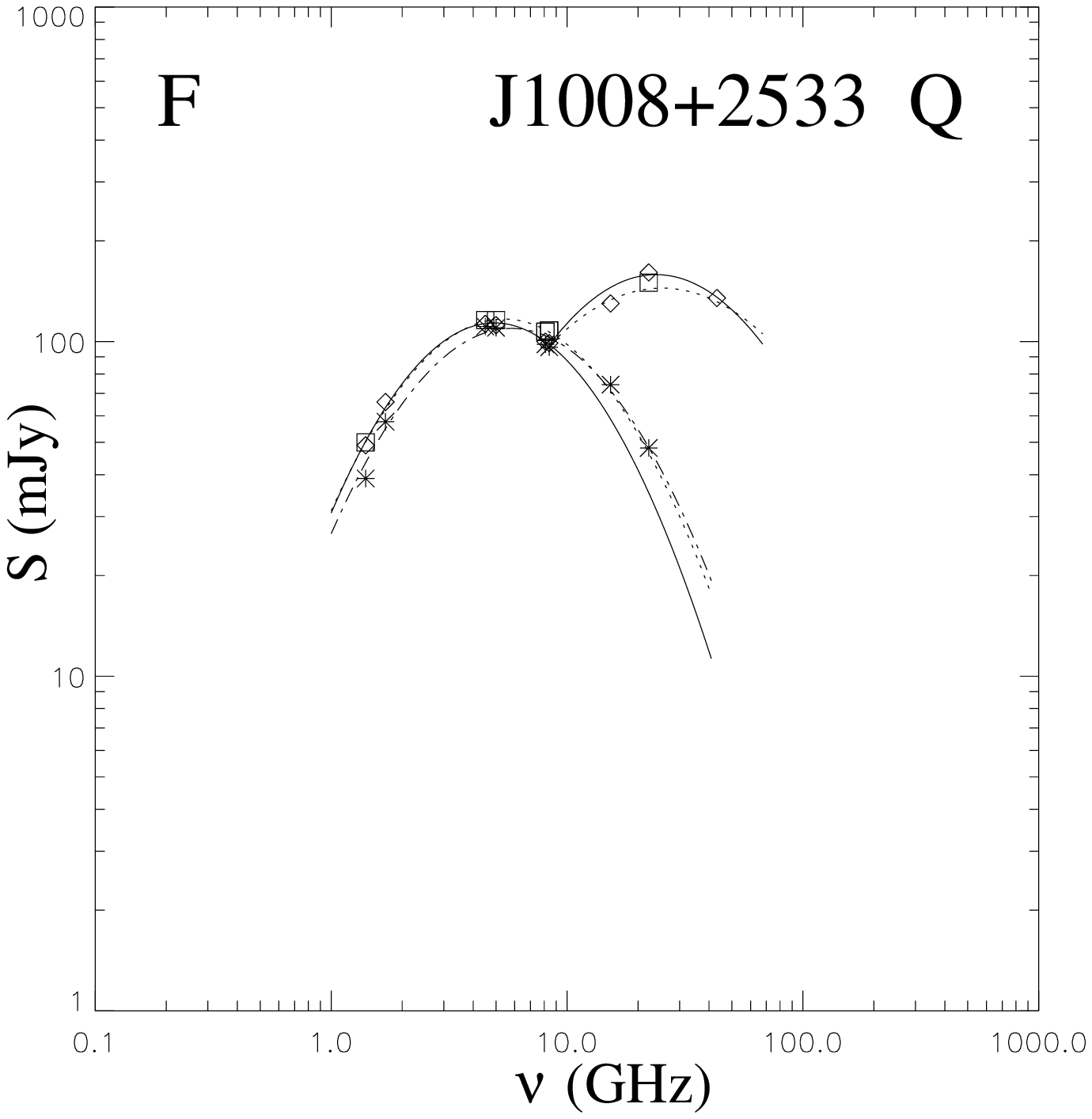}
\includegraphics{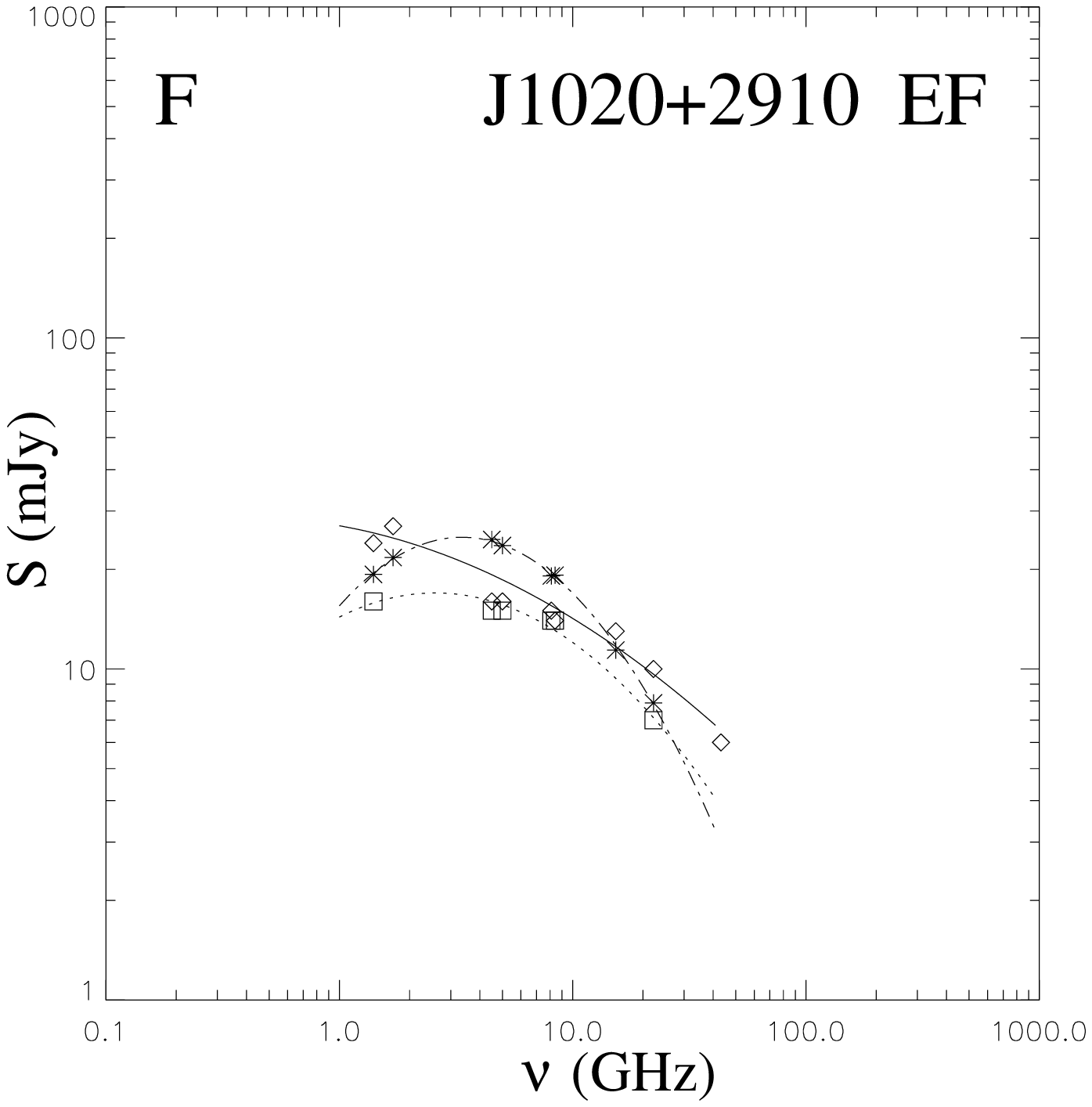}
\includegraphics{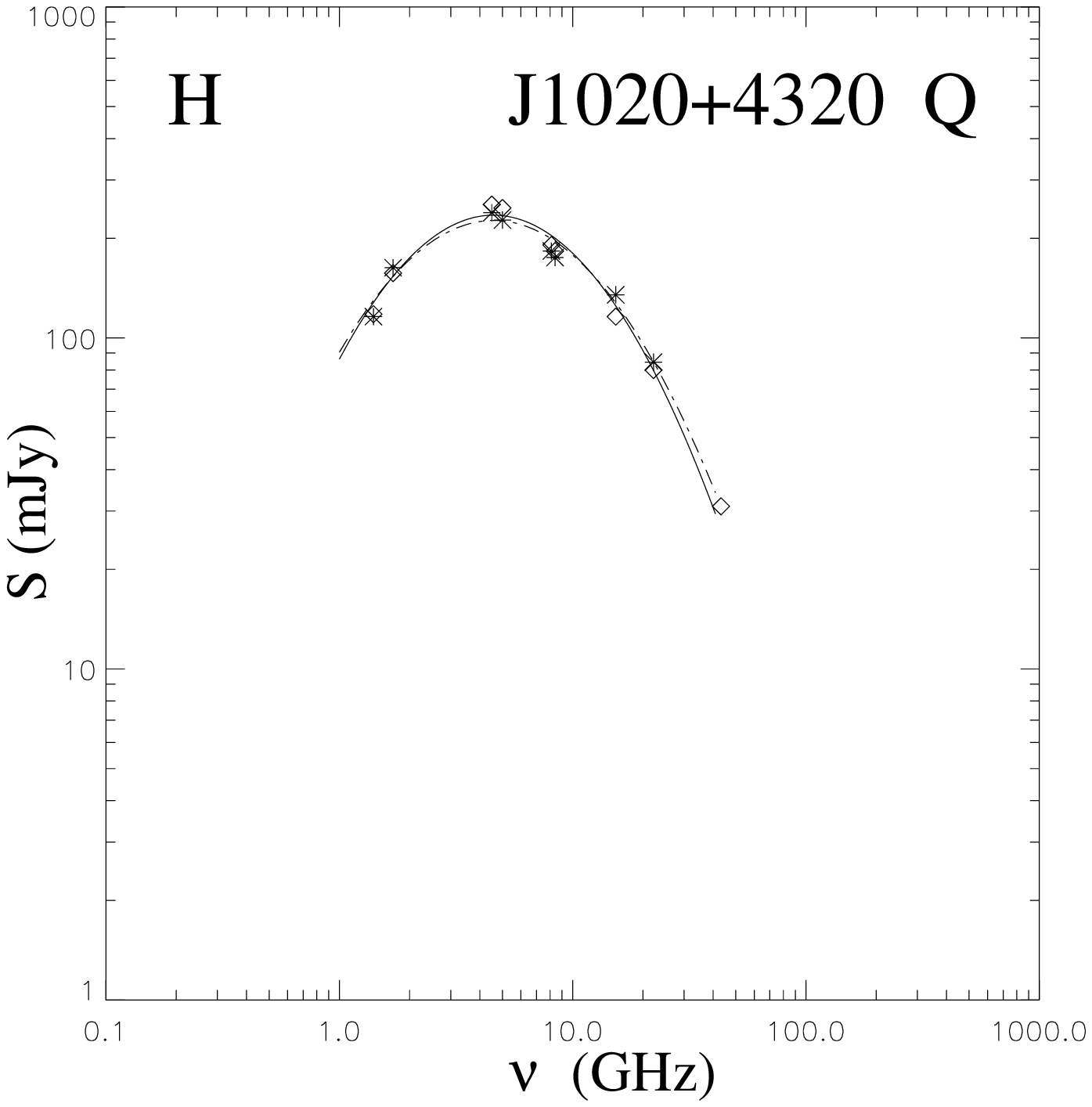}
\includegraphics{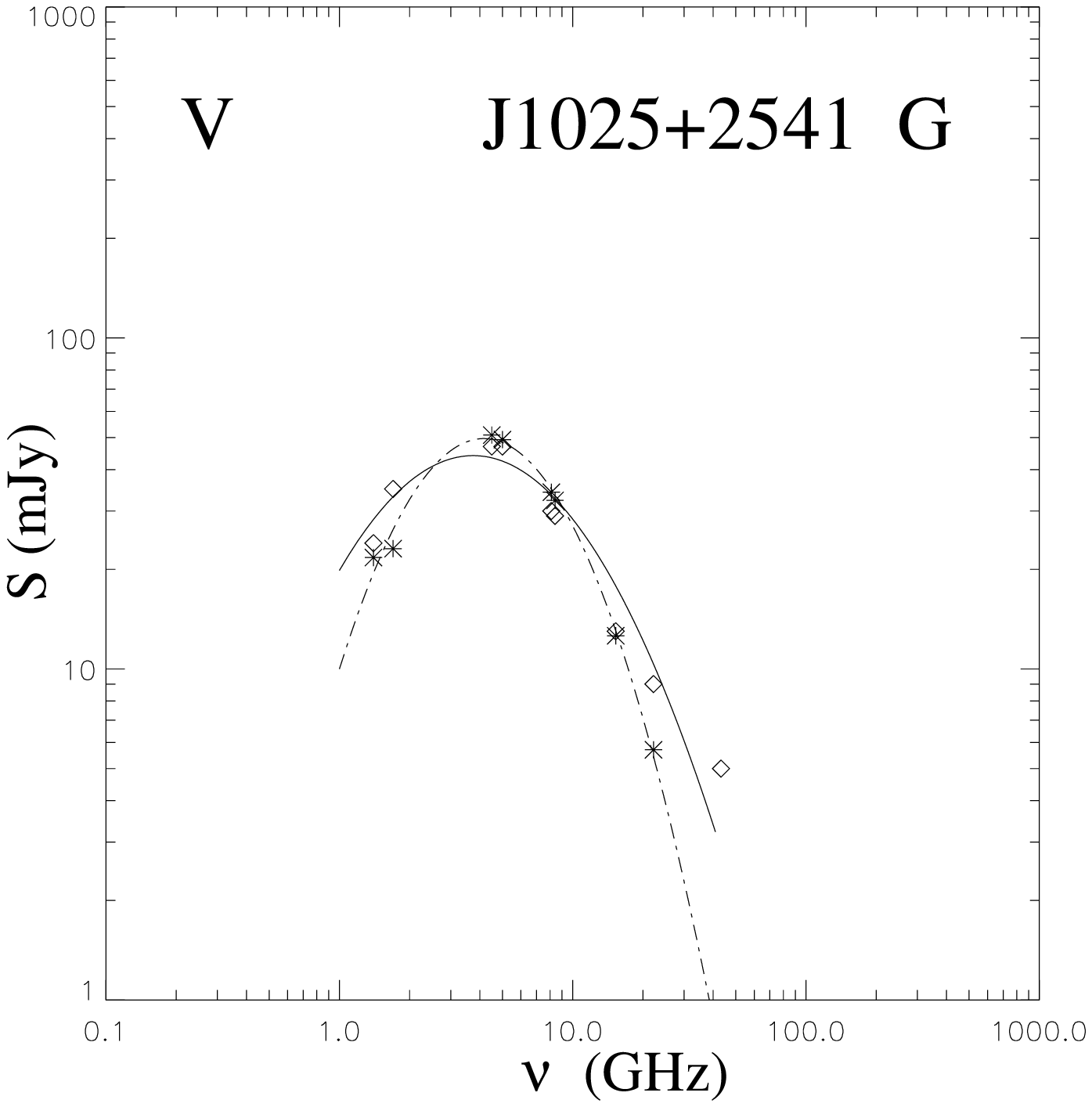}
\includegraphics{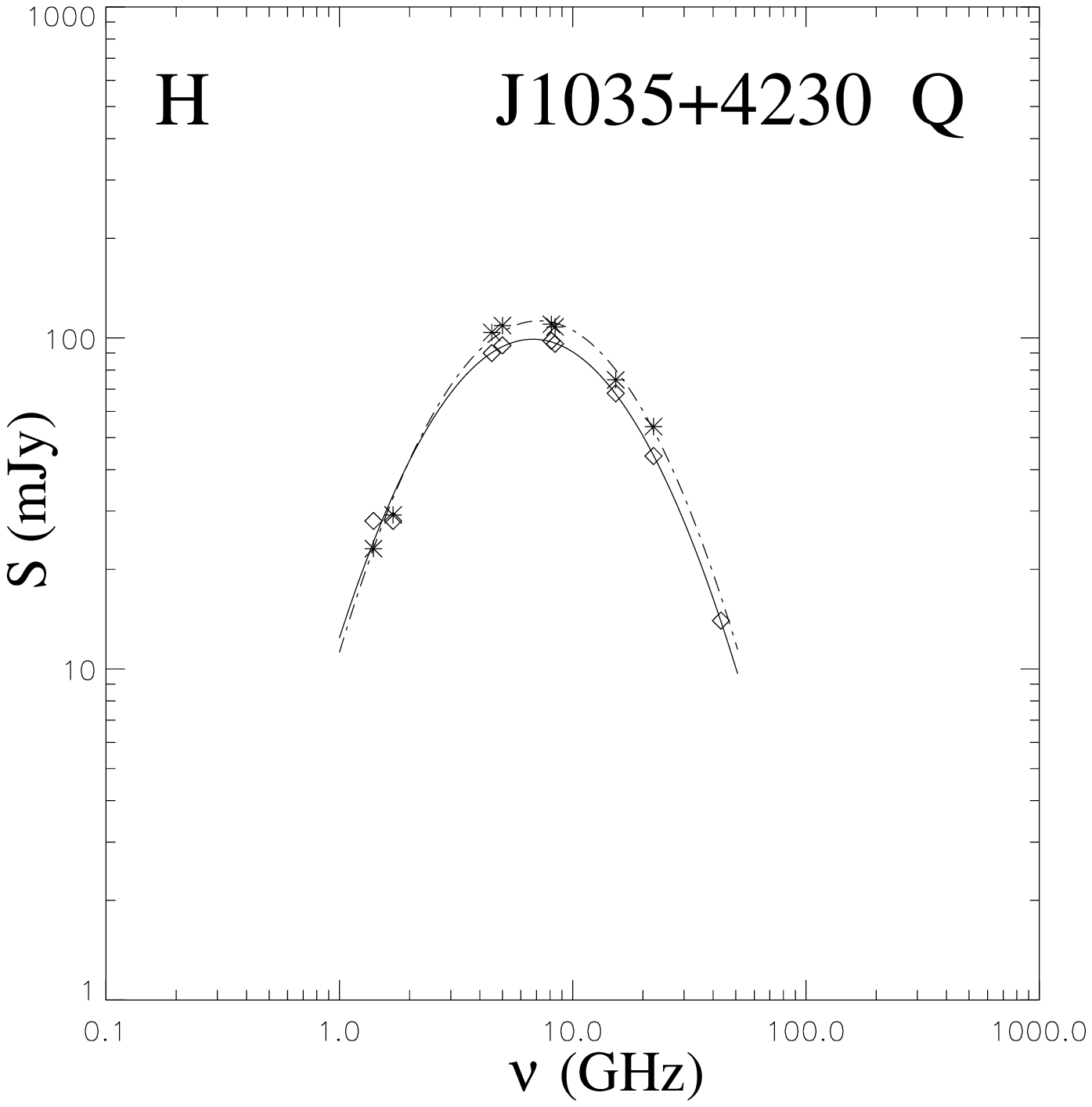}
\vspace{22cm}
\caption{Radio spectra of the 57 candidate HFPs from the ``faint'' HFP
  sample observed with the VLA during the observing runs presented in
  this paper. Asterisks and a dash-dotted line refer to the first
  epoch observations \citep[1998-2000,][]{cs09}; crosses and a dashed
  line refer to epoch {\it a} (2003); diamonds and a solid line
  refer to epoch {\it b} (2004); squares and a dotted line refer to
  epochs {\it c,d} (2006-2007).}
\label{radio_spettri}
\end{center}
\end{figure*}

\addtocounter{figure}{-1}
\begin{figure*}
\begin{center}
\includegraphics{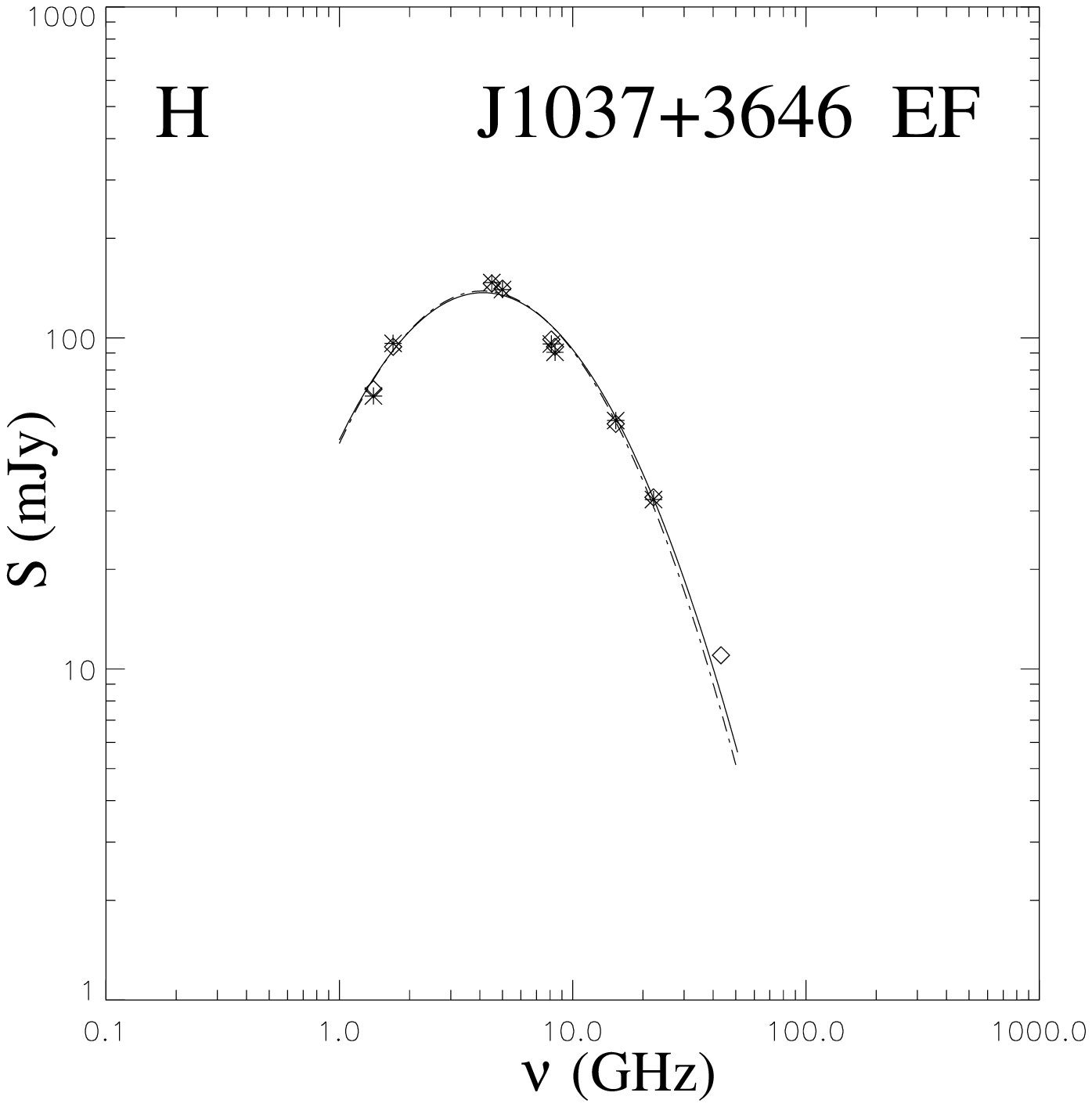}
\includegraphics{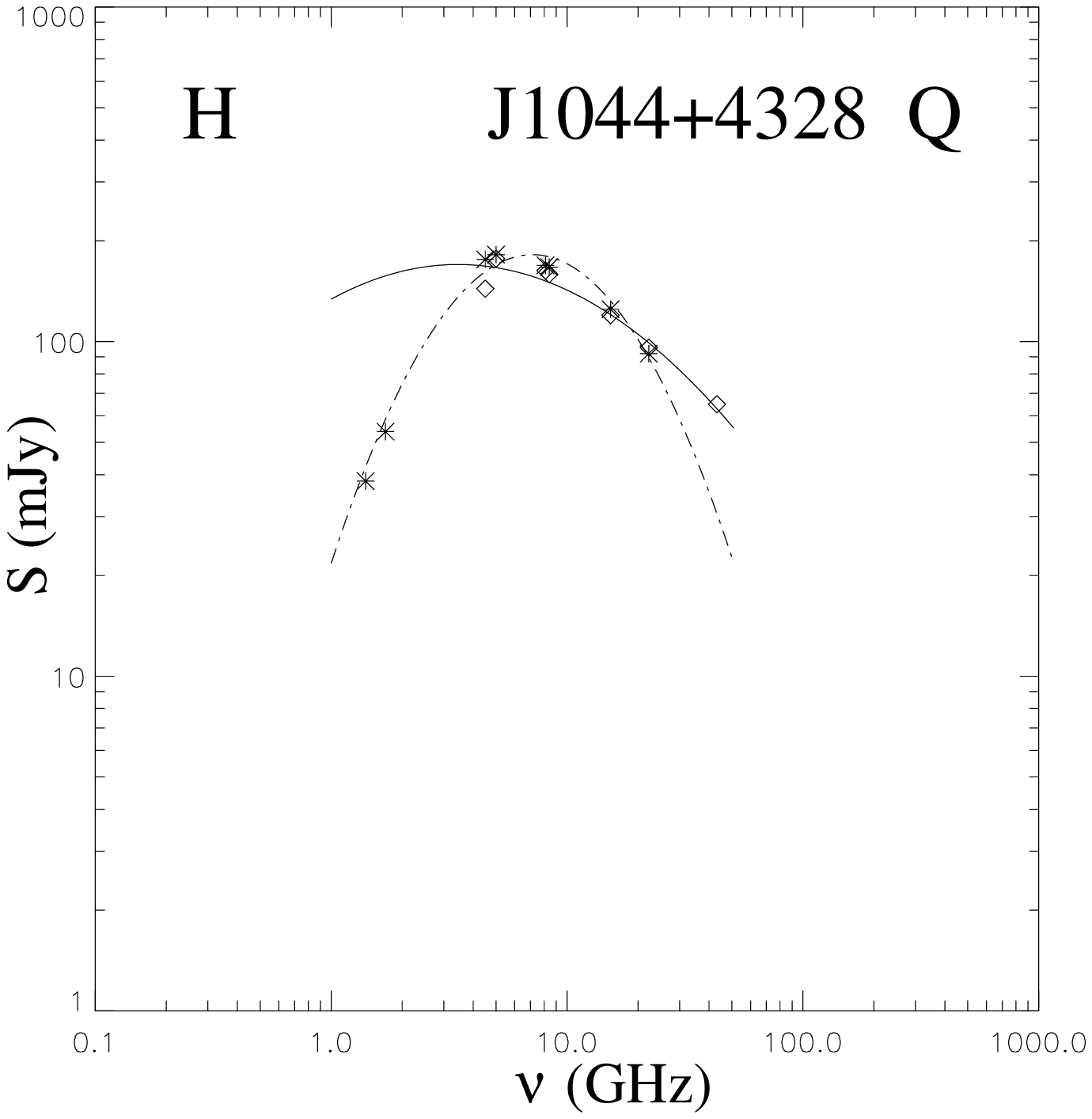}
\includegraphics{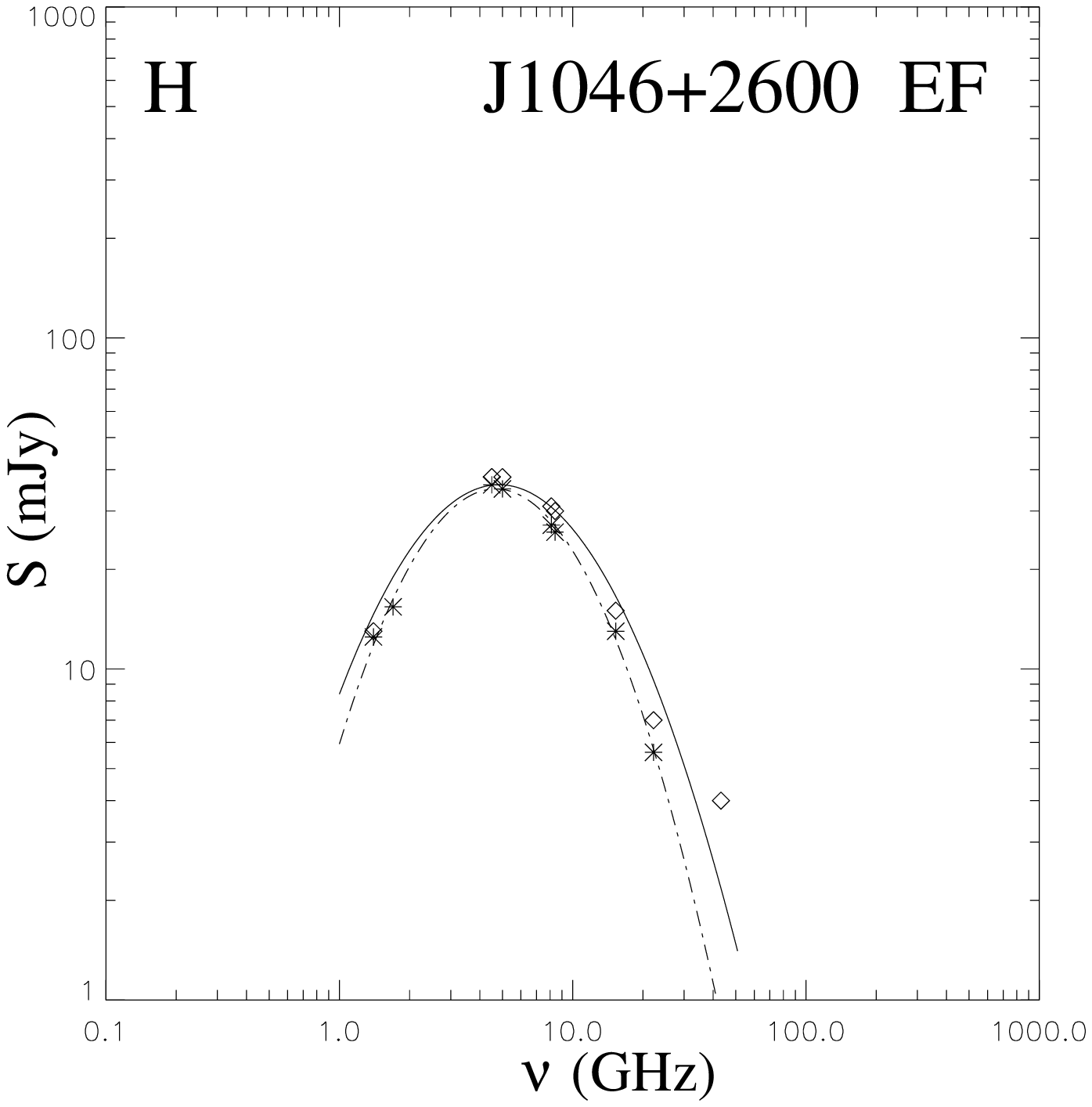}
\includegraphics{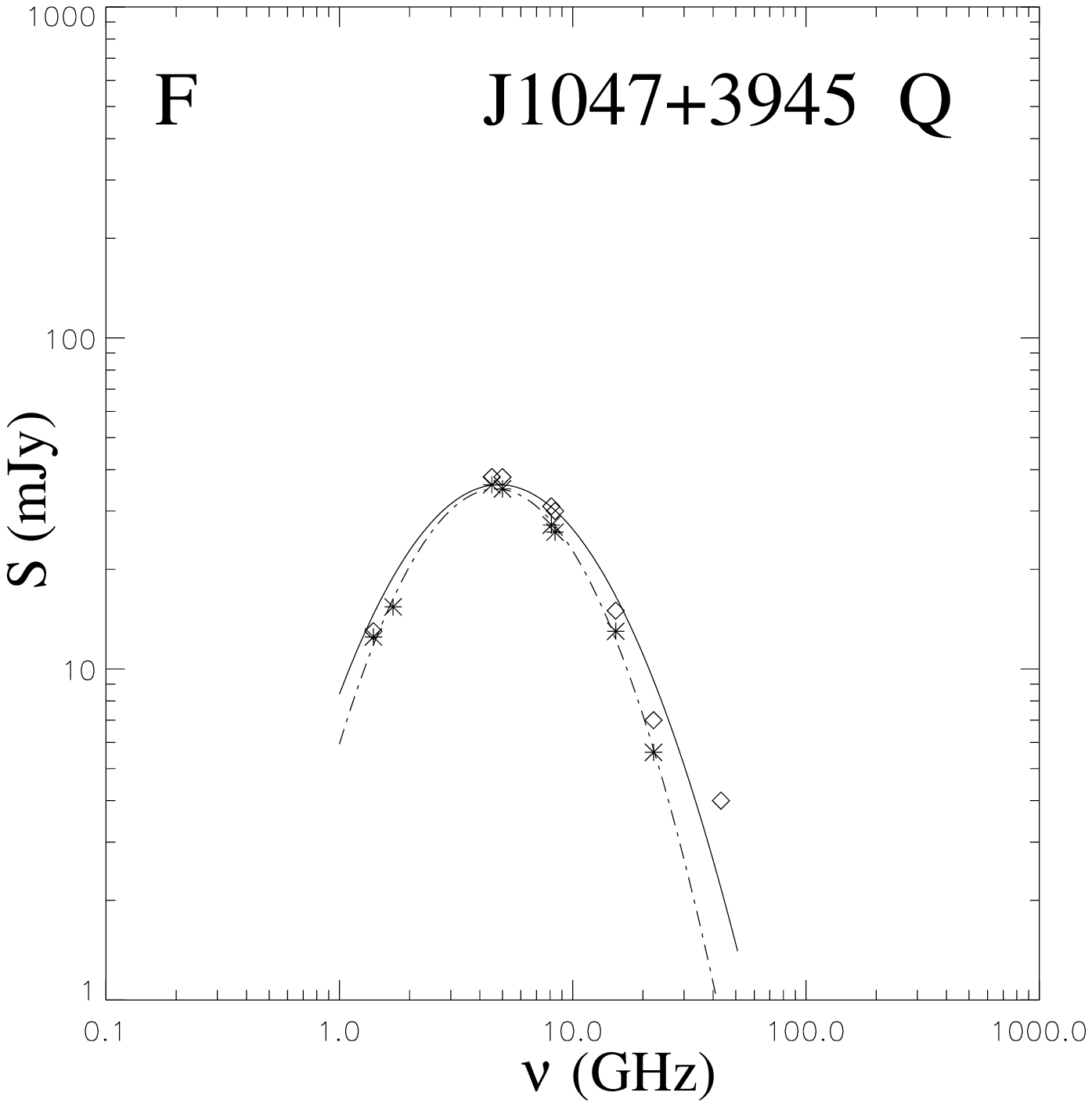}
\includegraphics{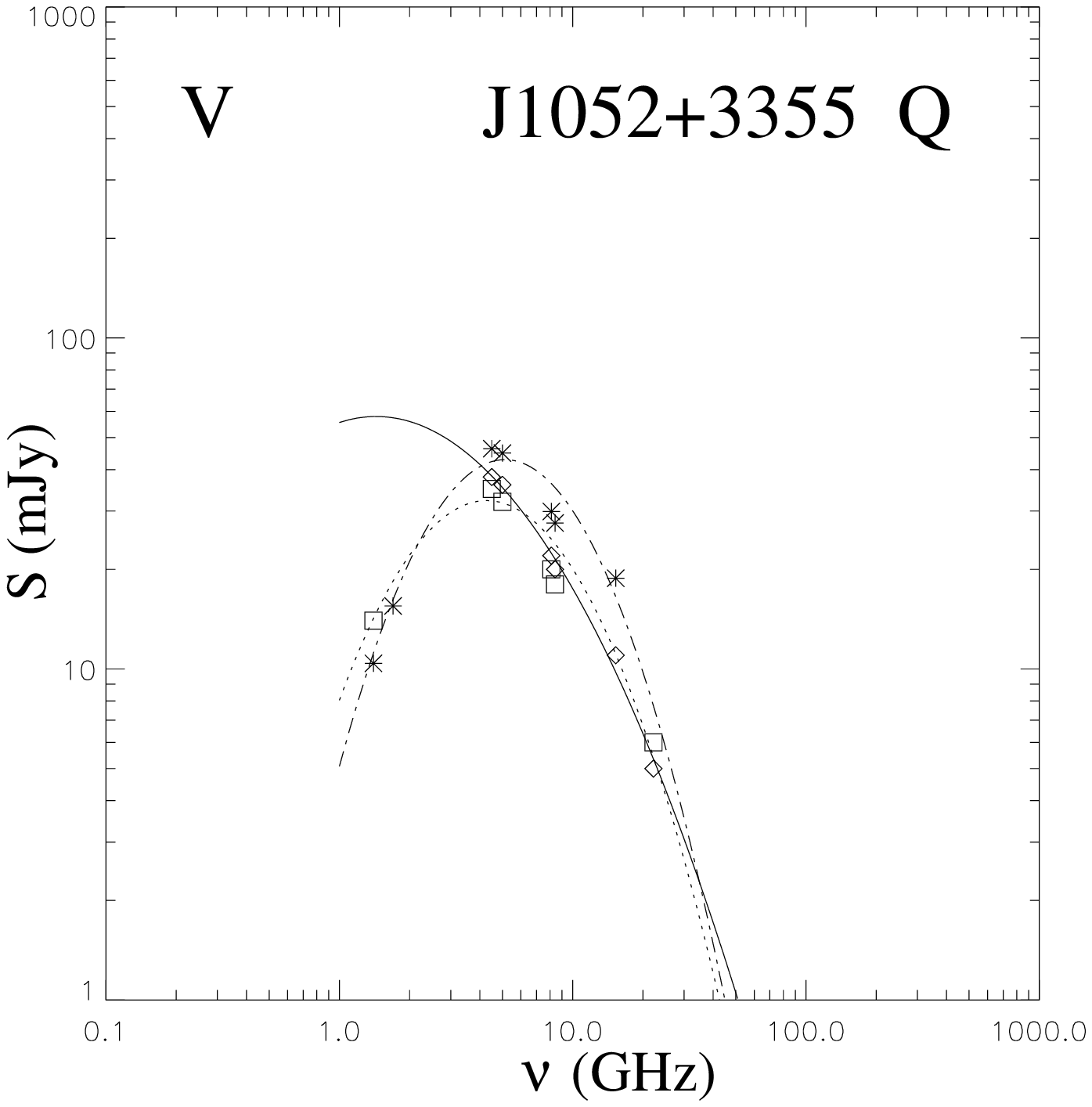}
\includegraphics{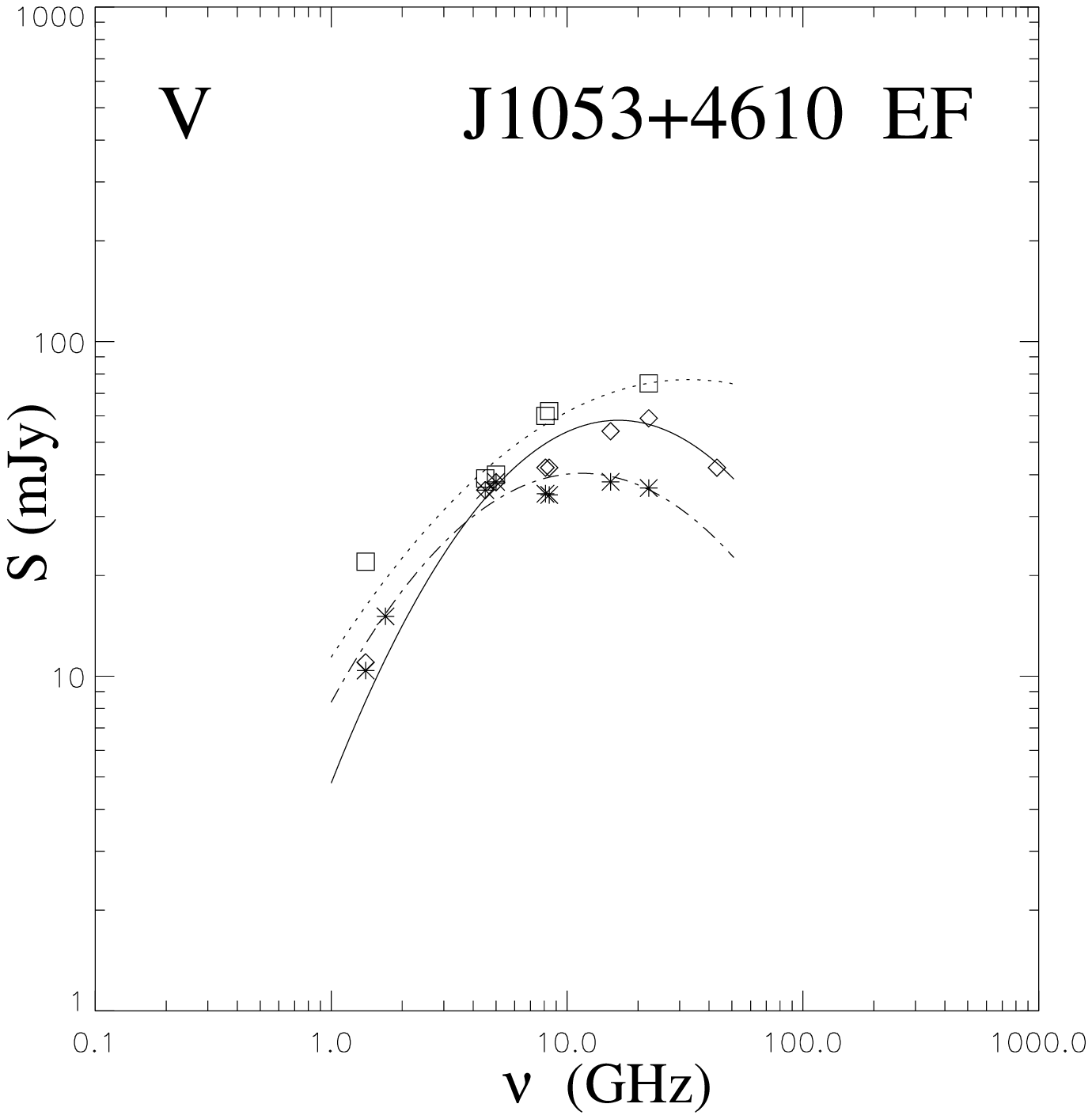}
\includegraphics{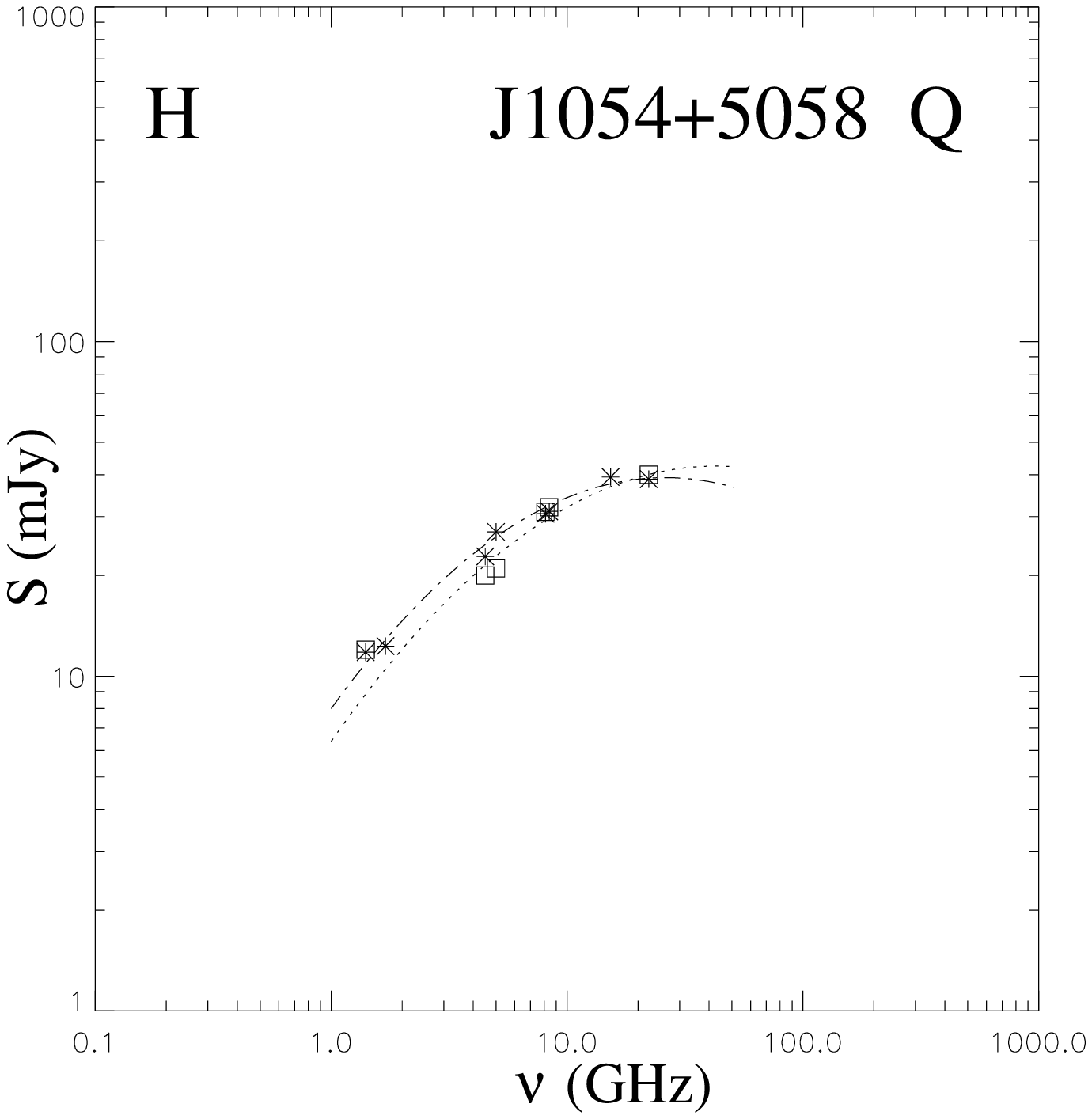}
\includegraphics{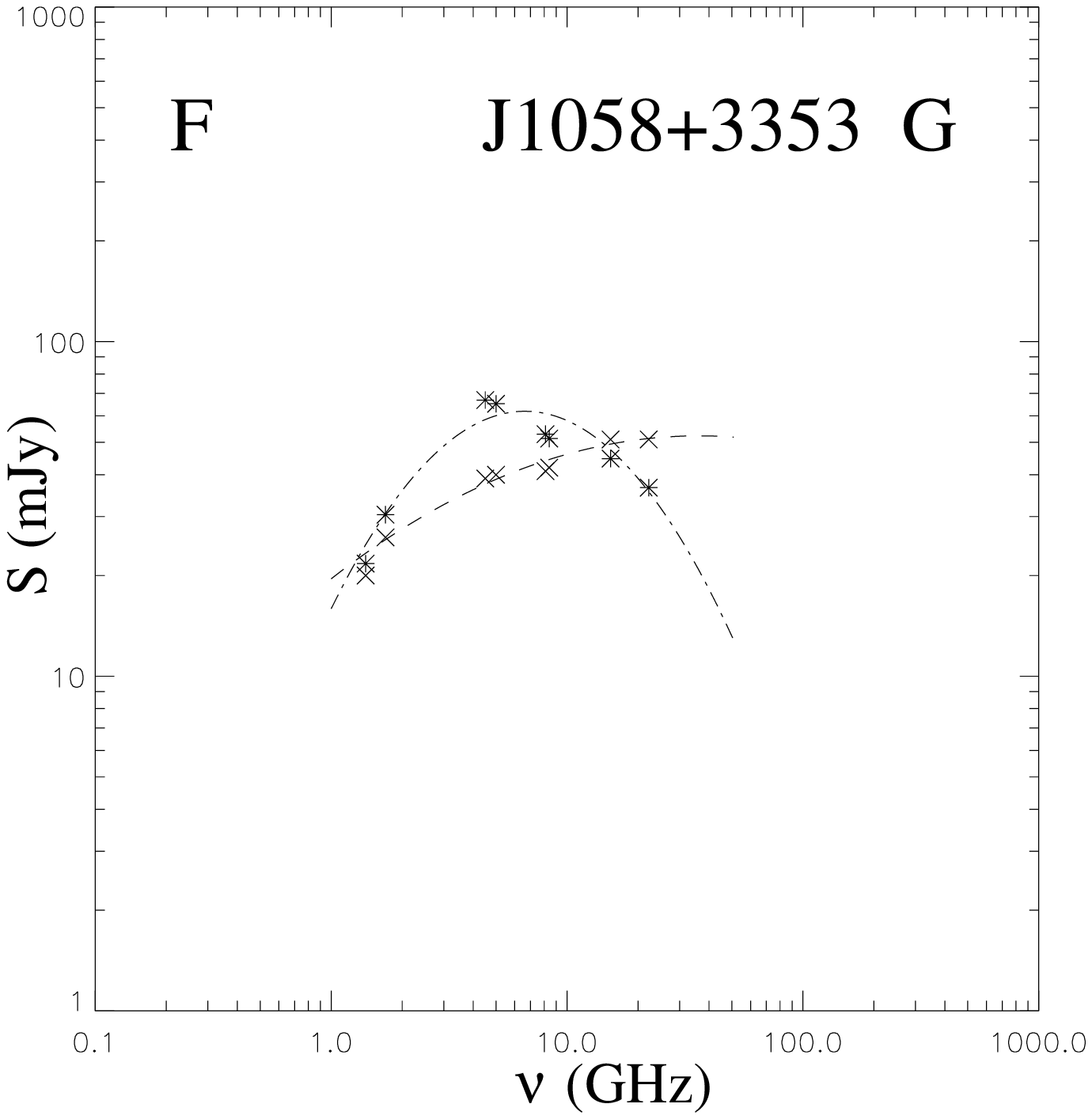}
\includegraphics{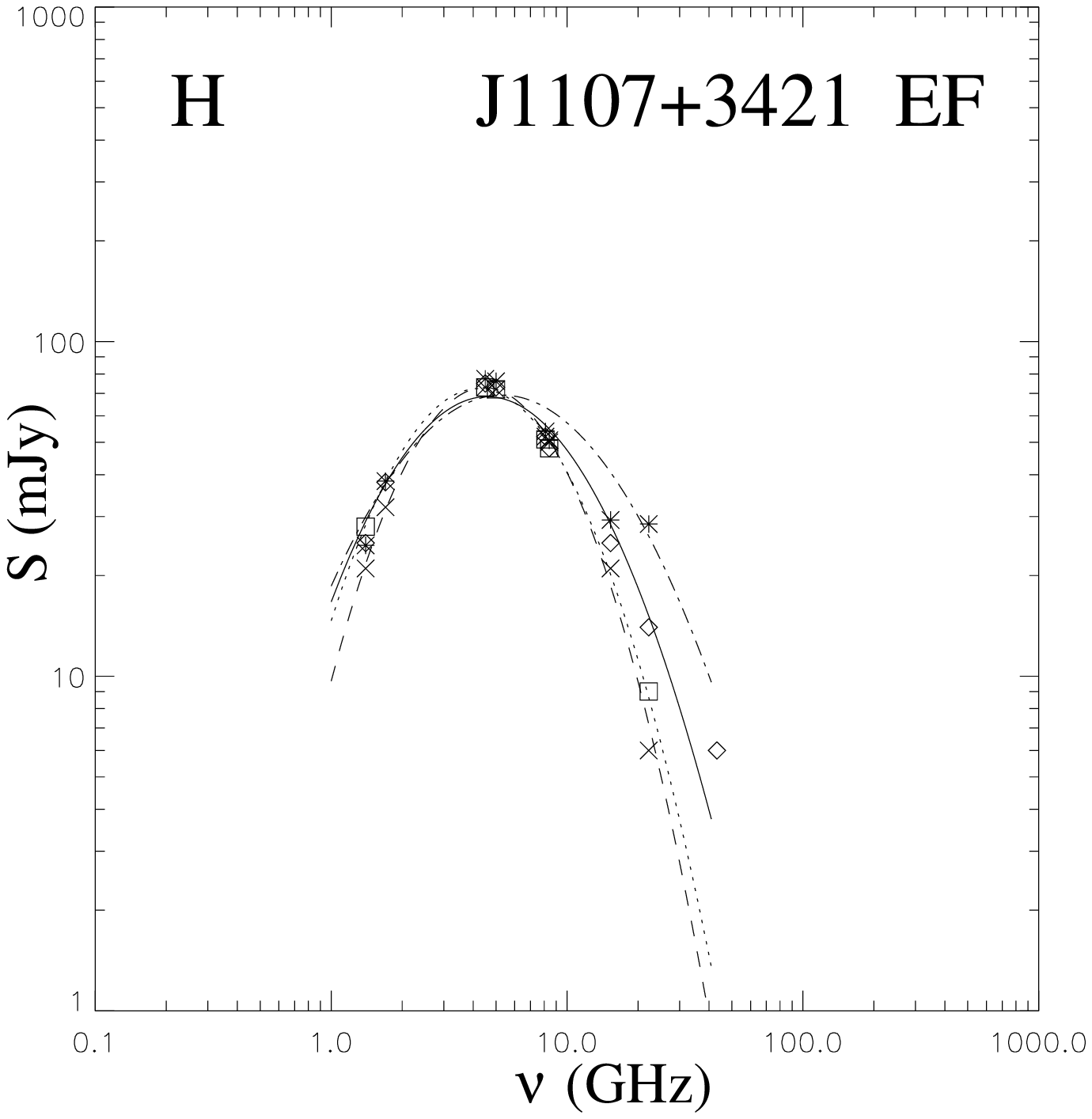}
\includegraphics{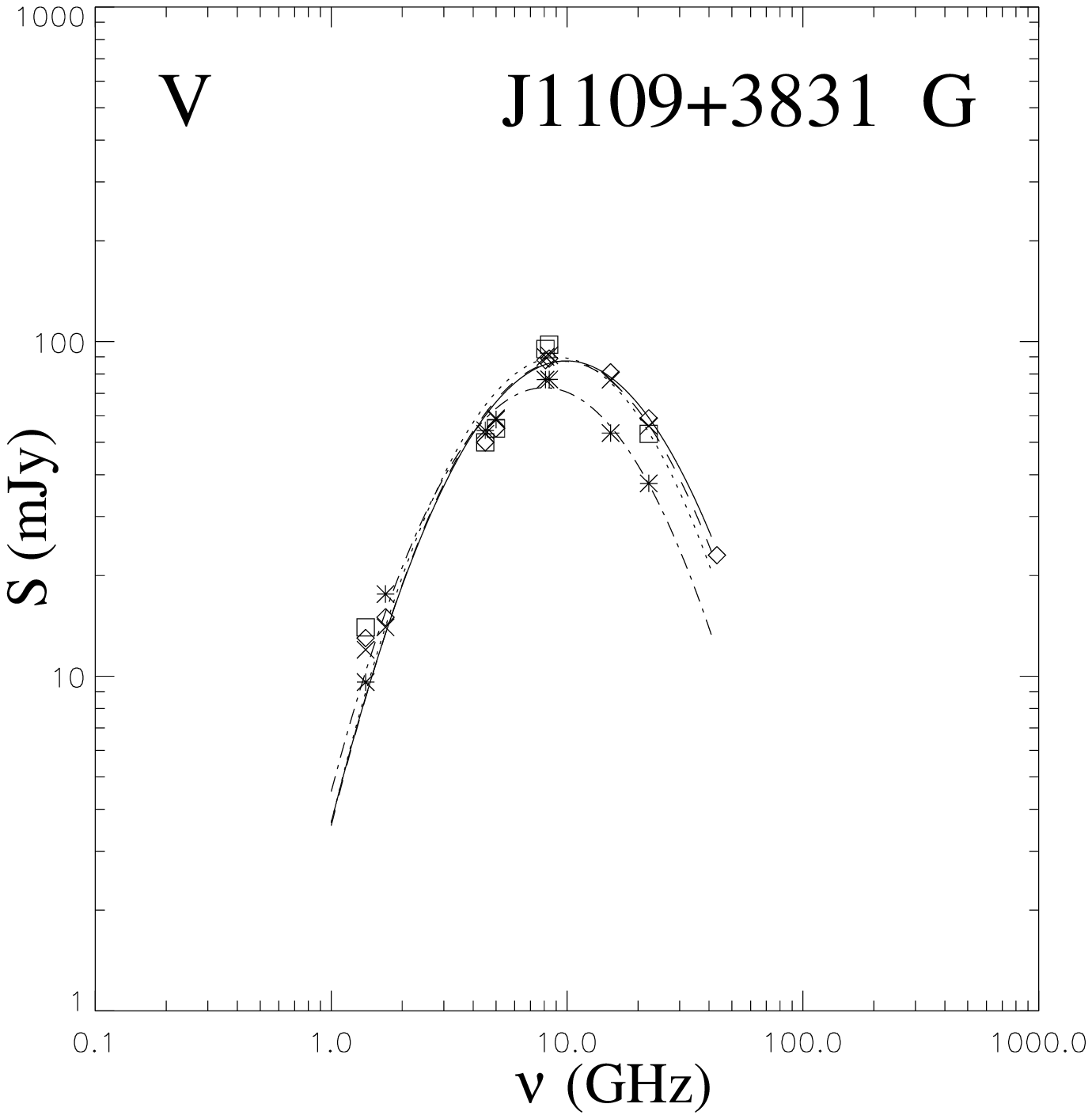}
\includegraphics{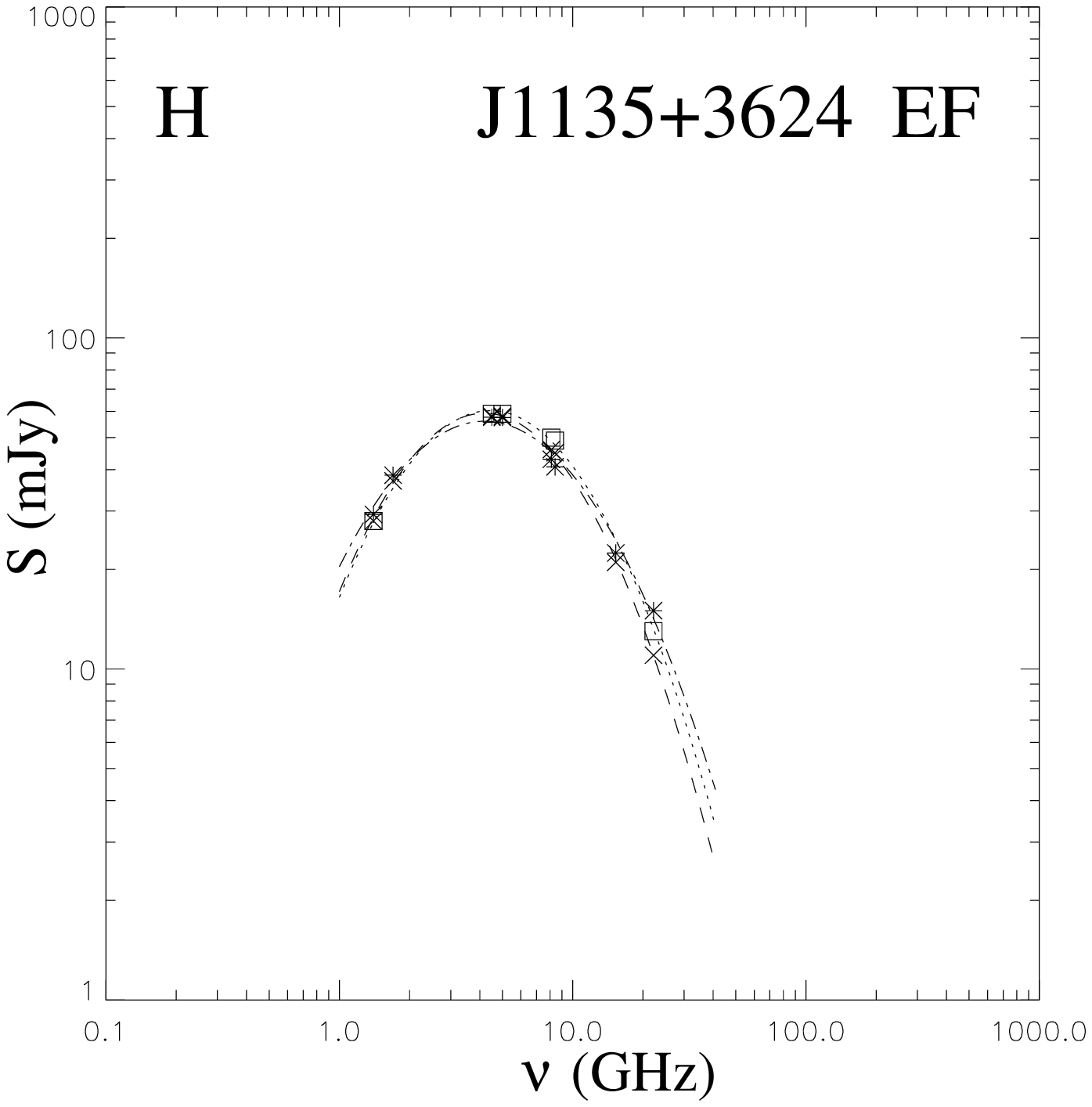}
\includegraphics{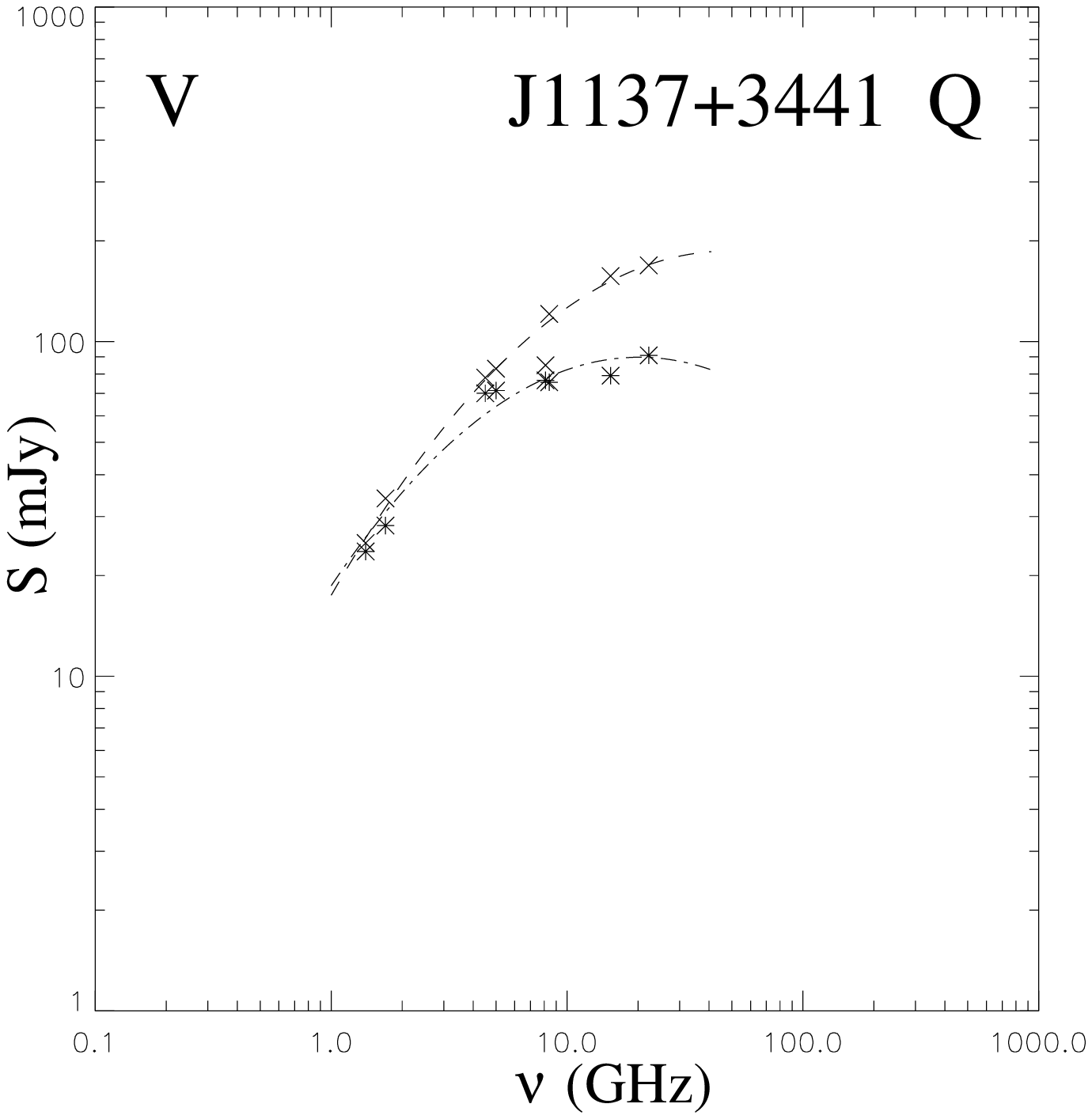}
\includegraphics{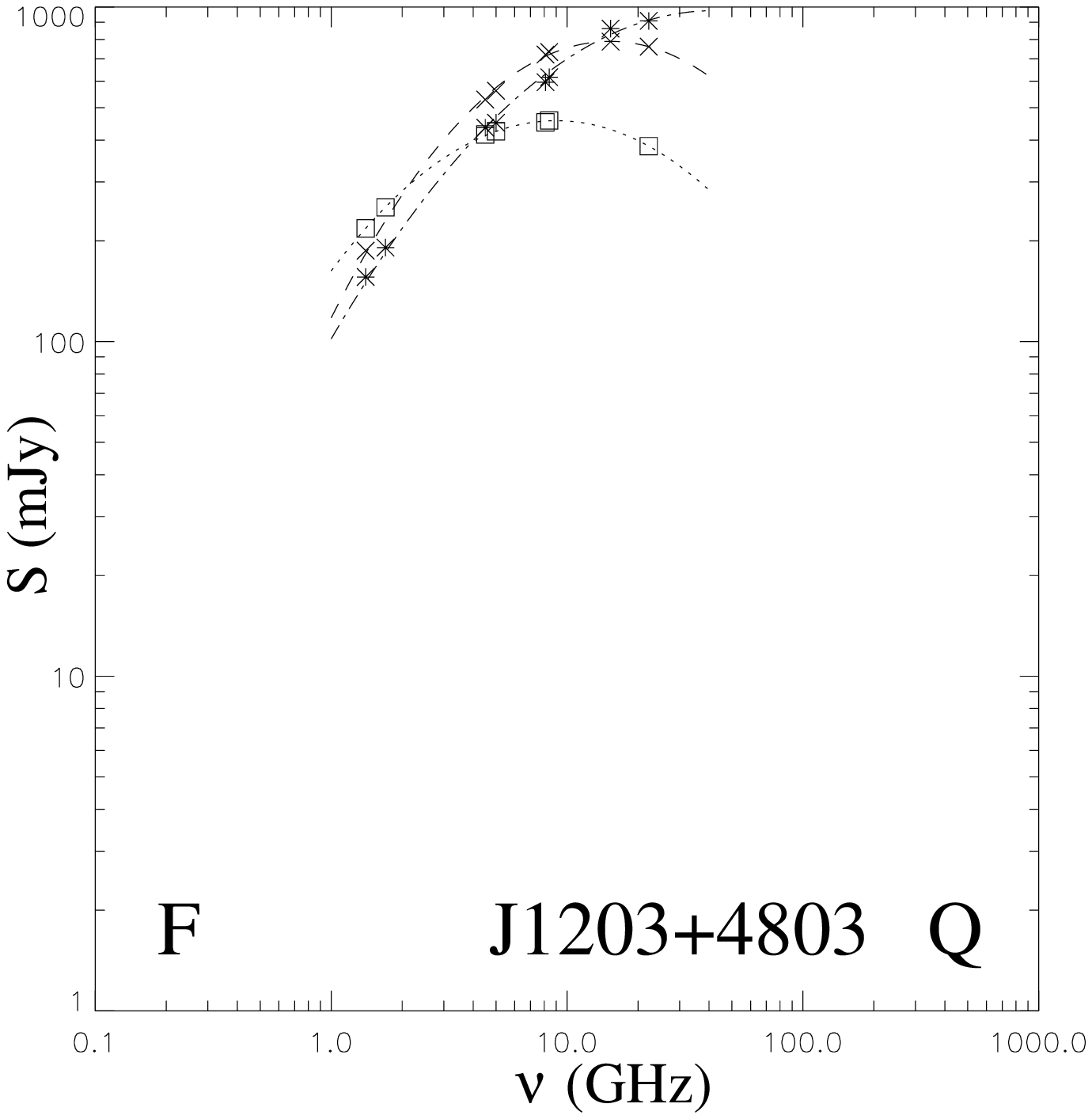}
\includegraphics{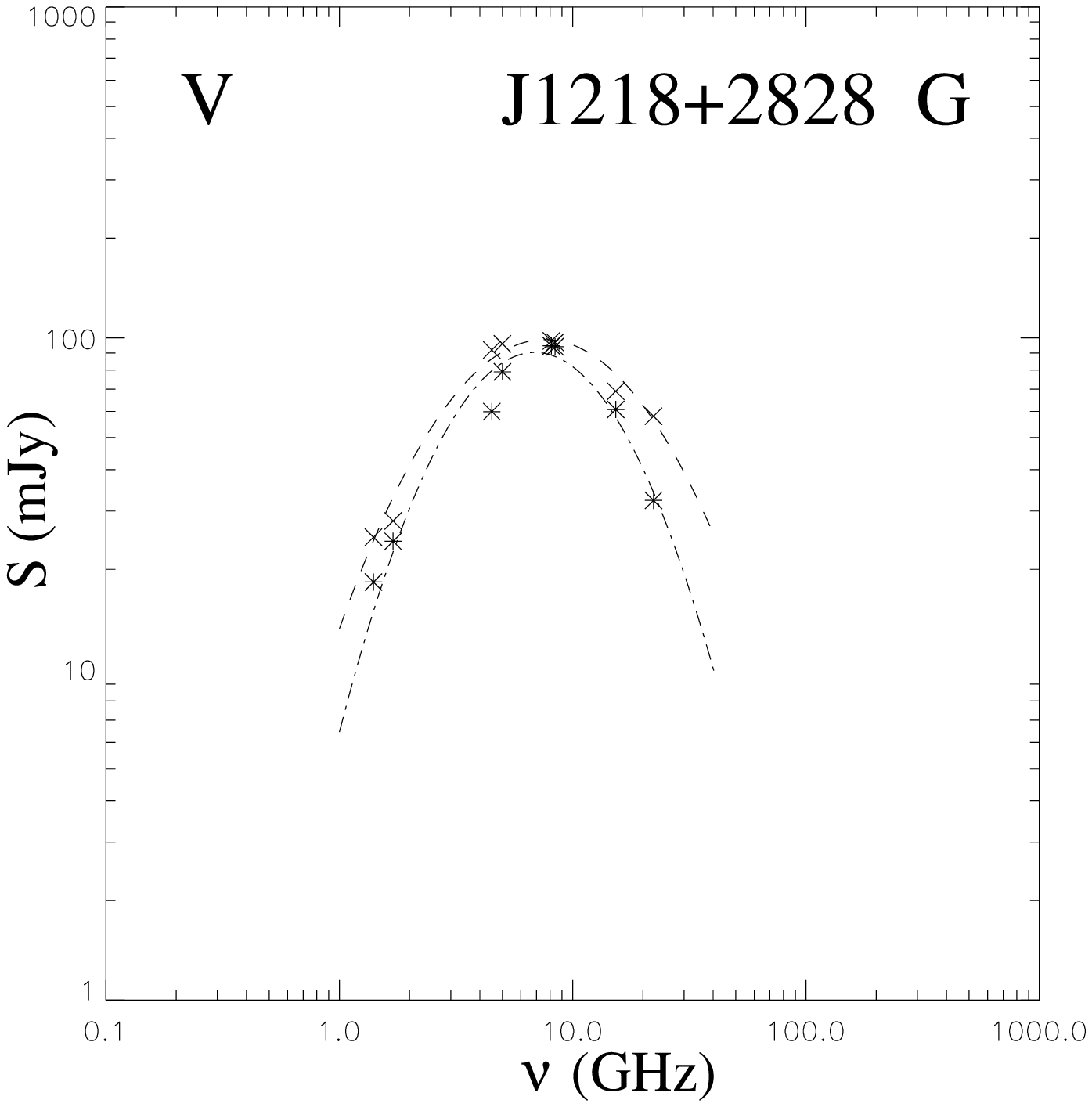}
\includegraphics{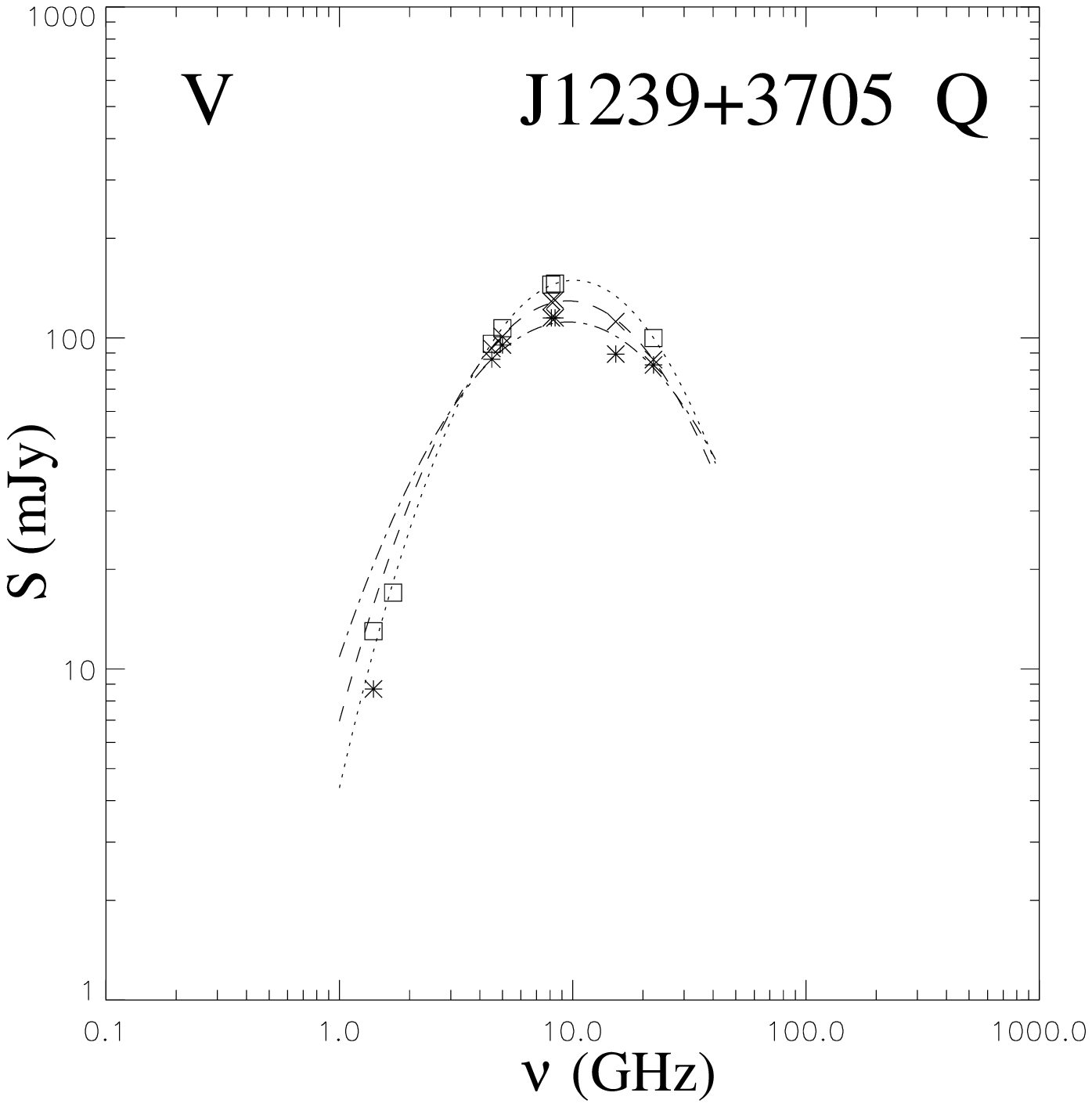}
\includegraphics{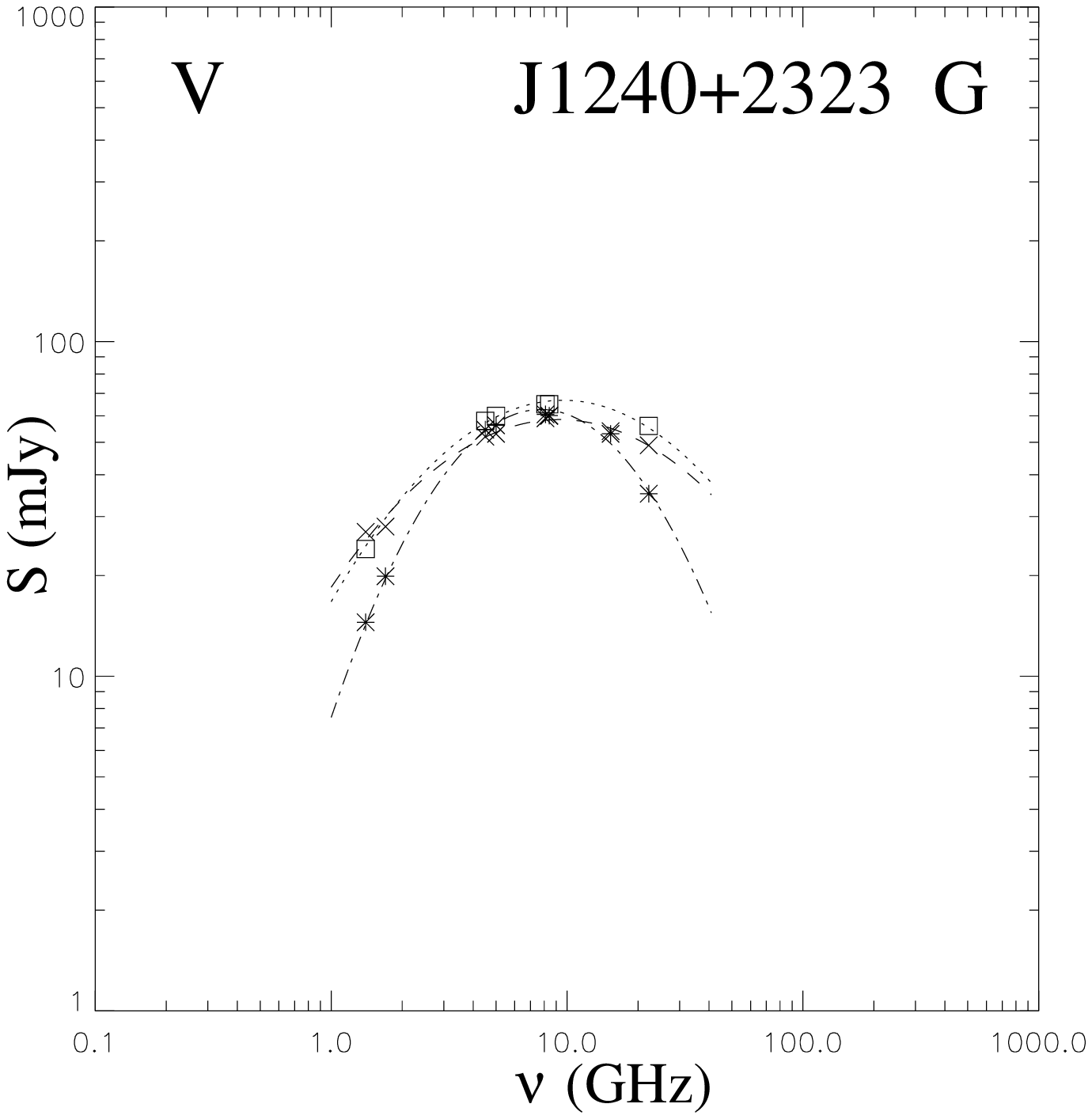}
\vspace{22cm}
\end{center}
\caption{Continued.}
\end{figure*}

\addtocounter{figure}{-1}
\begin{figure*}
\begin{center}
\includegraphics{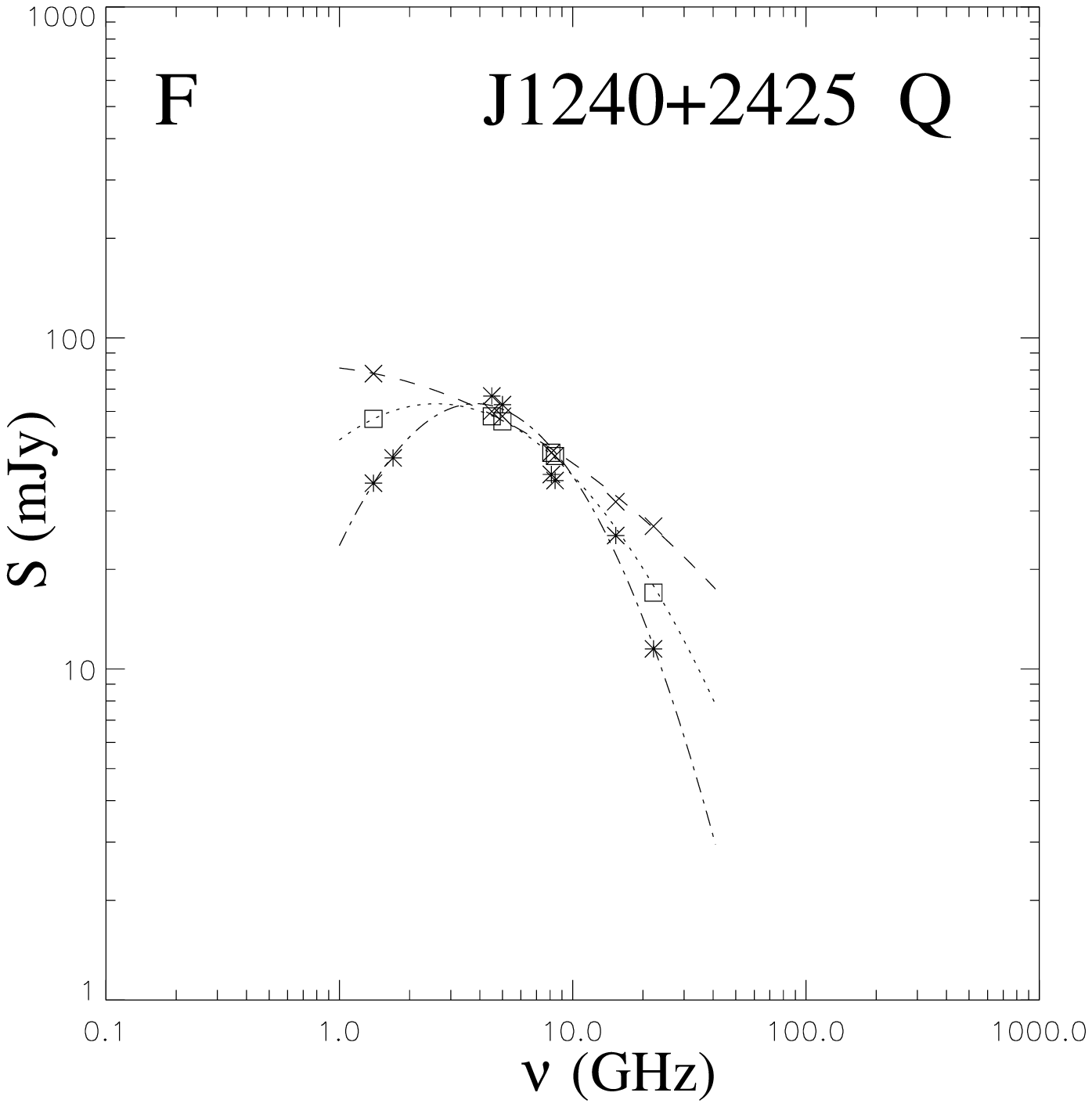}
\includegraphics{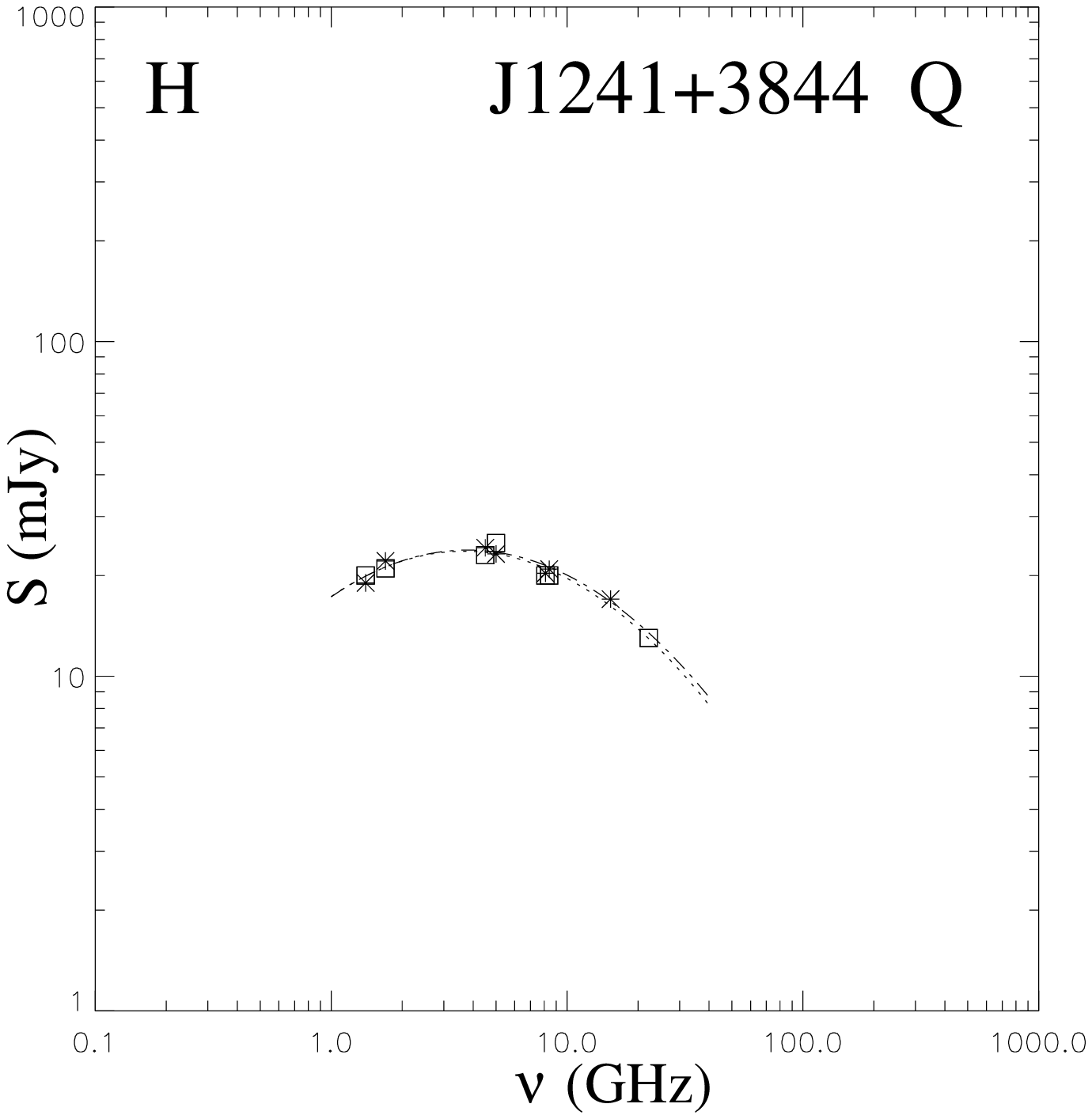}
\includegraphics{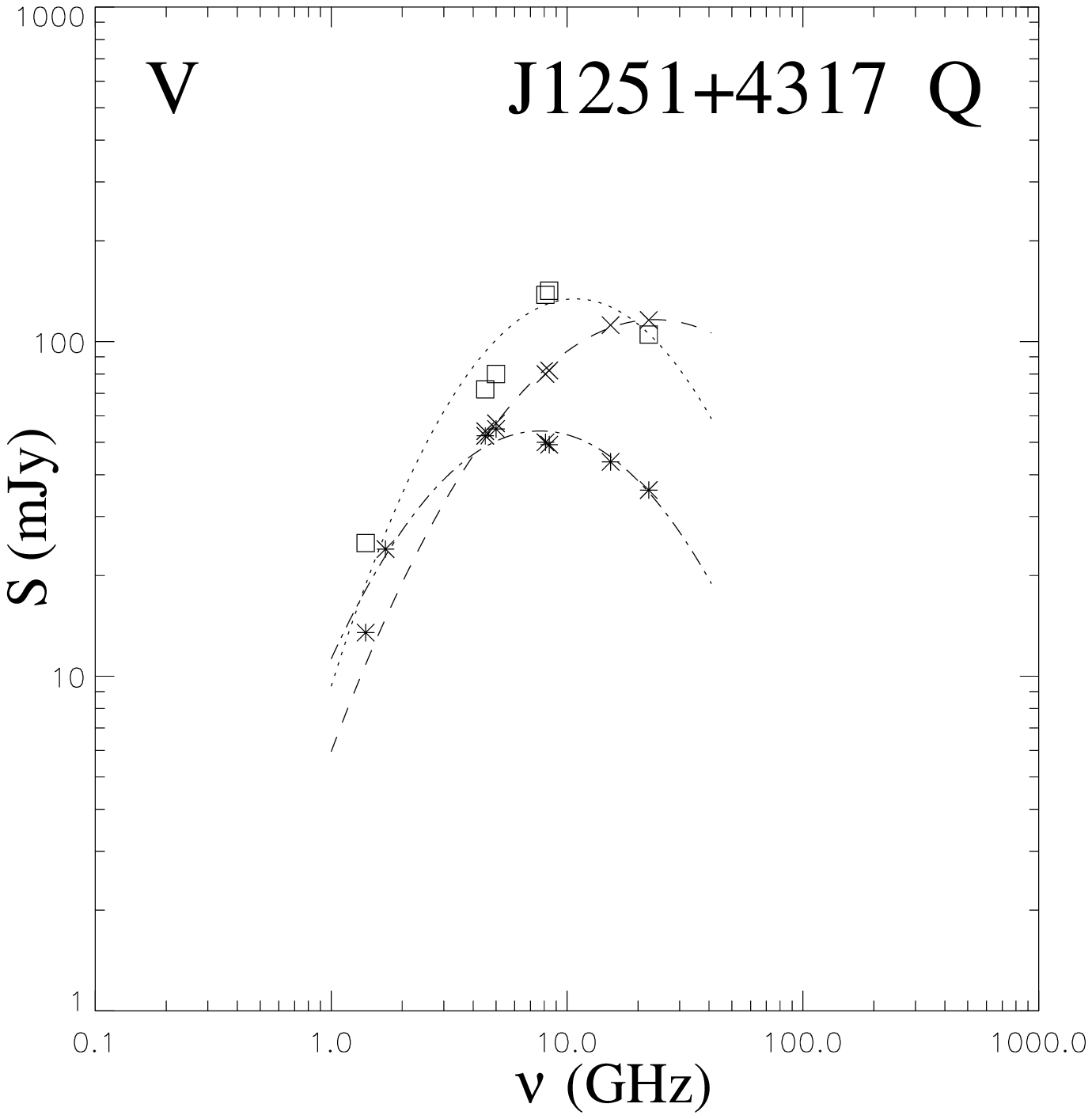}
\includegraphics{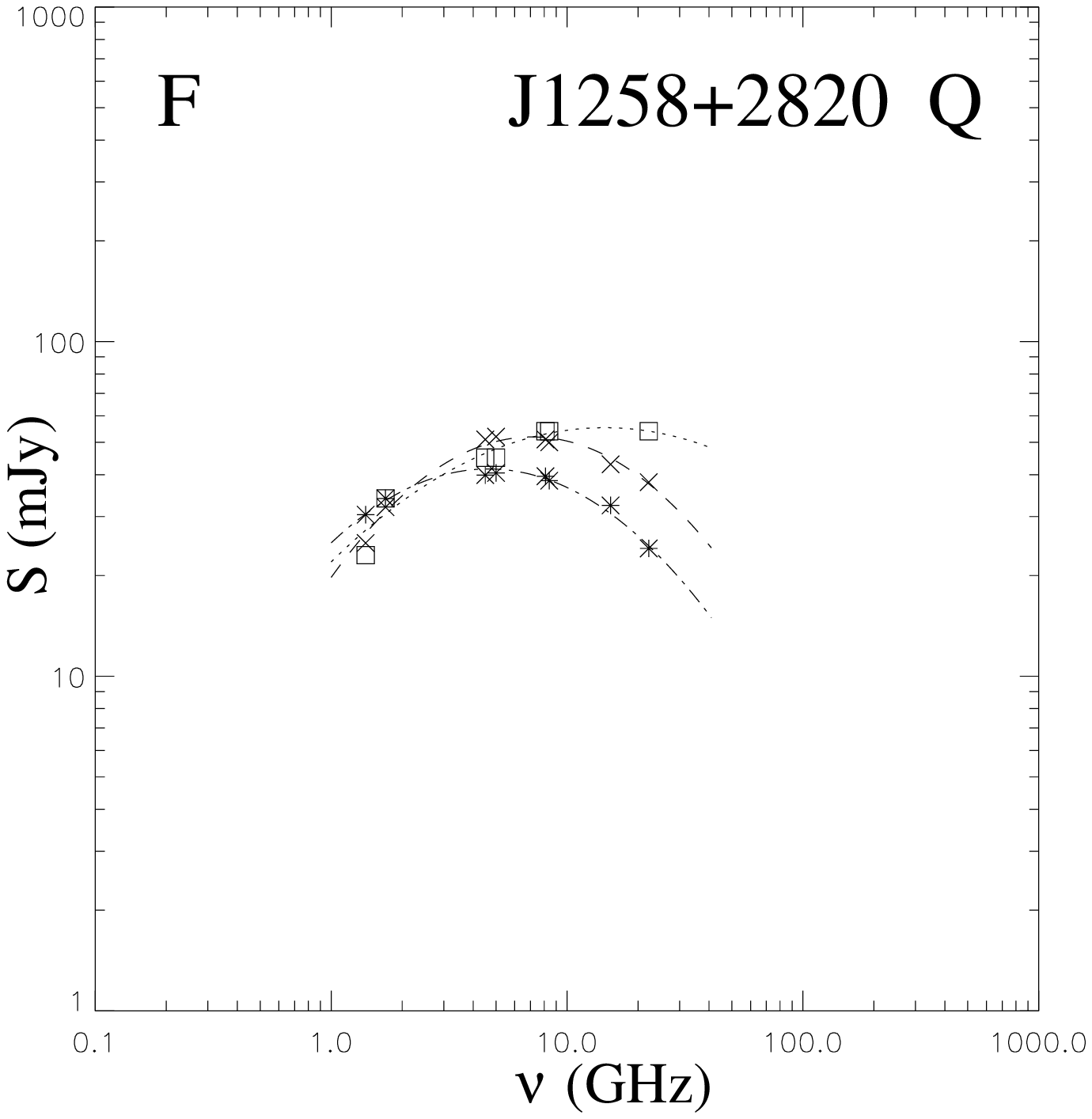}
\includegraphics{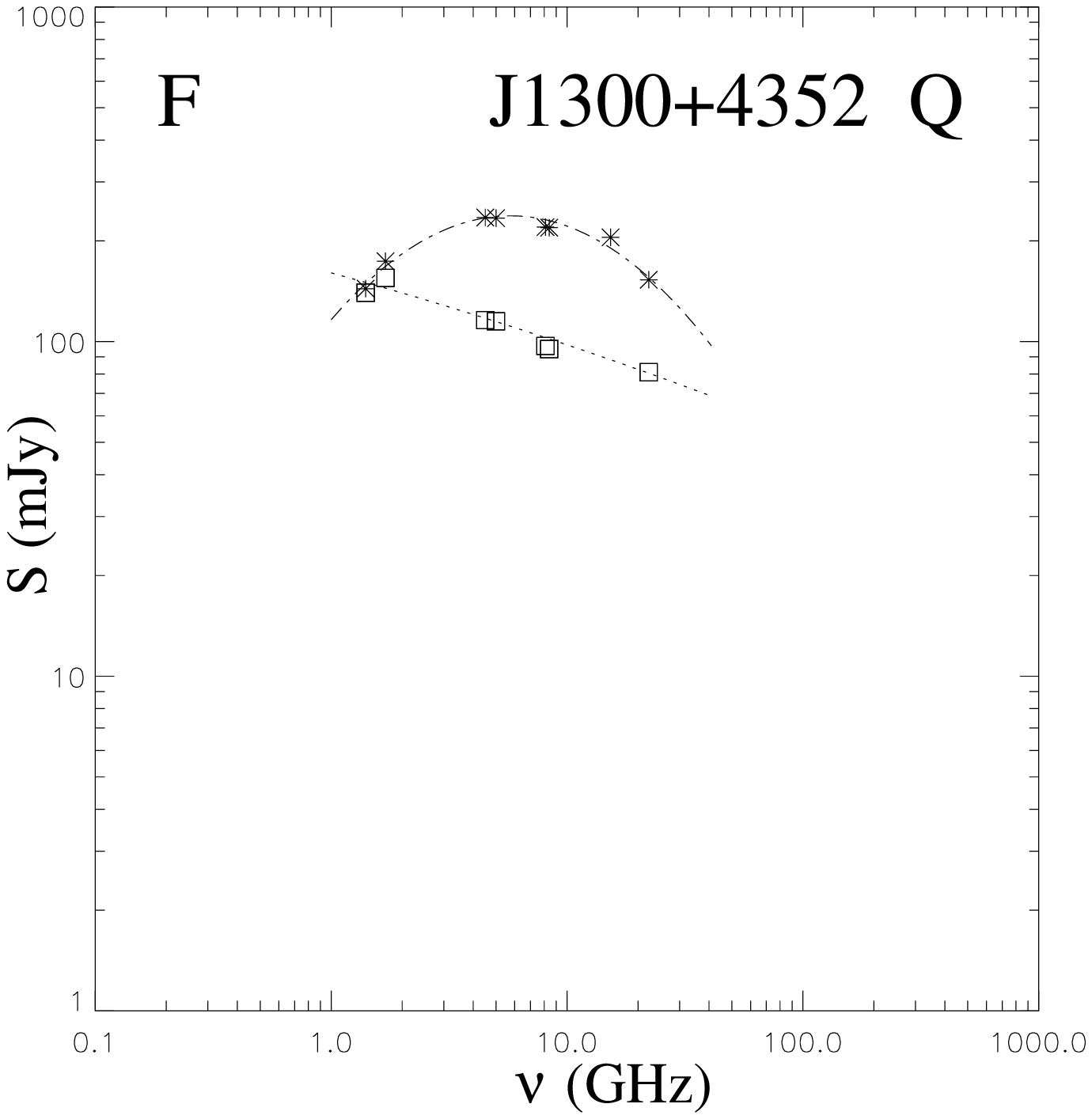}
\includegraphics{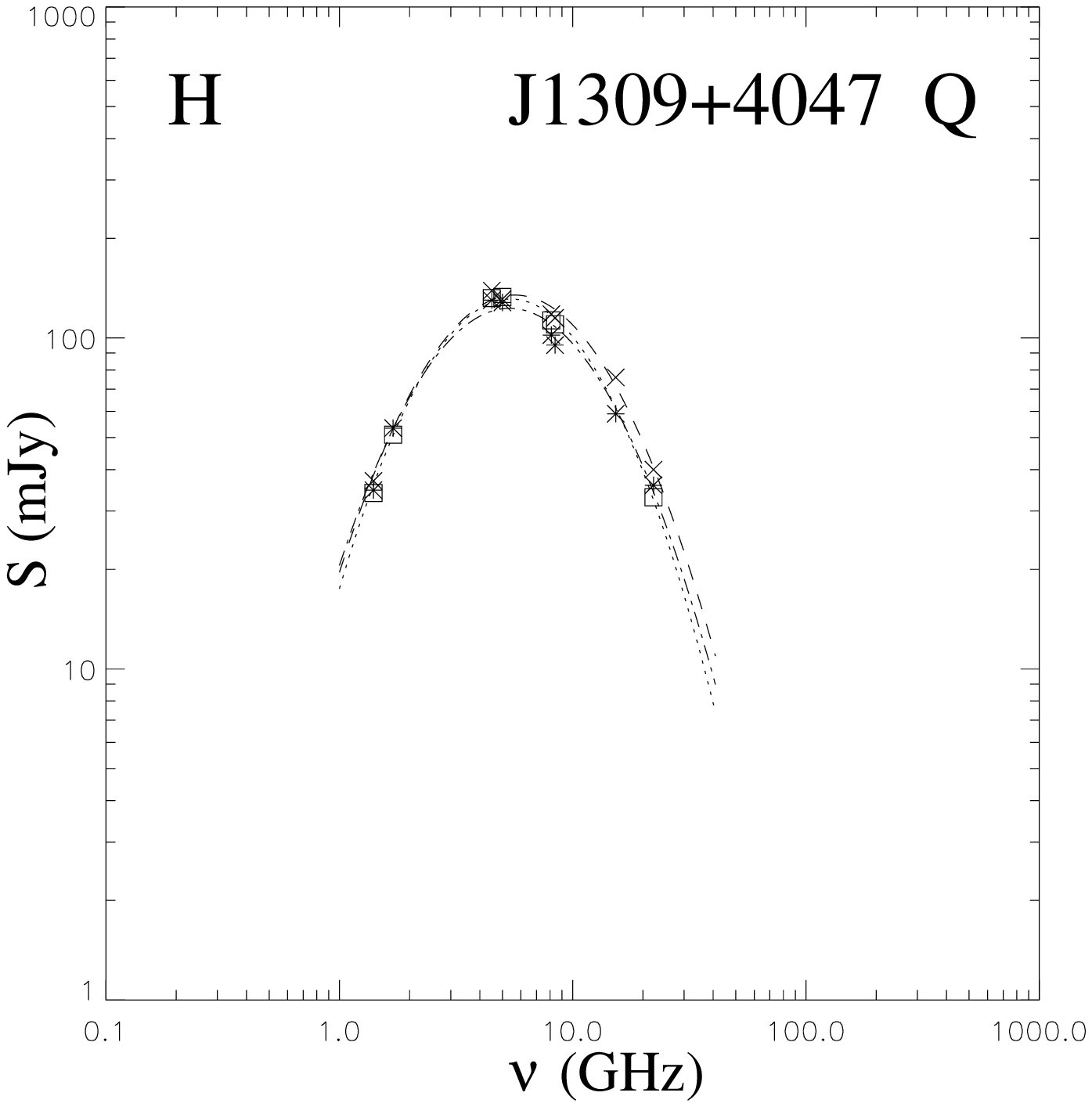}
\includegraphics{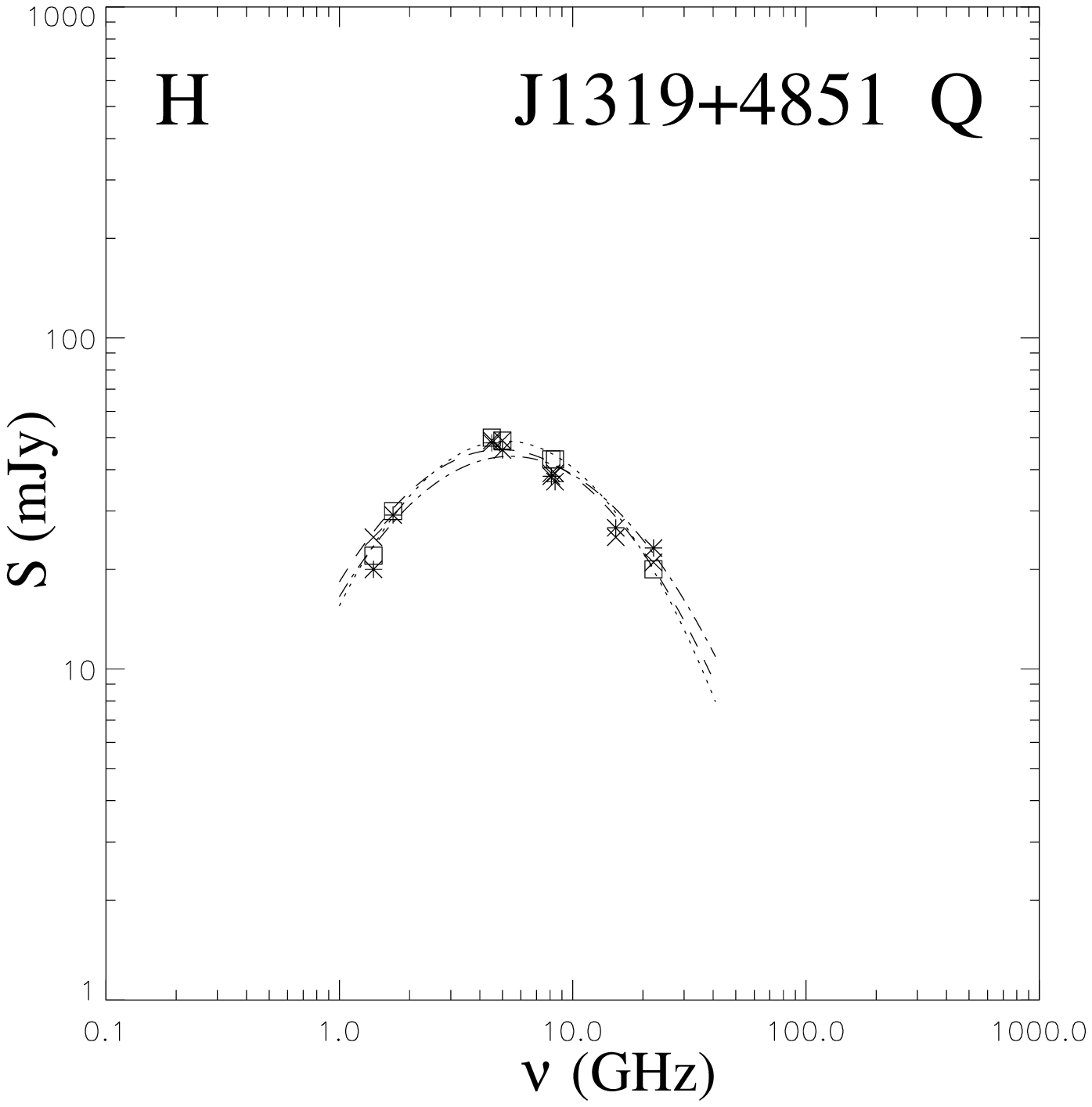}
\includegraphics{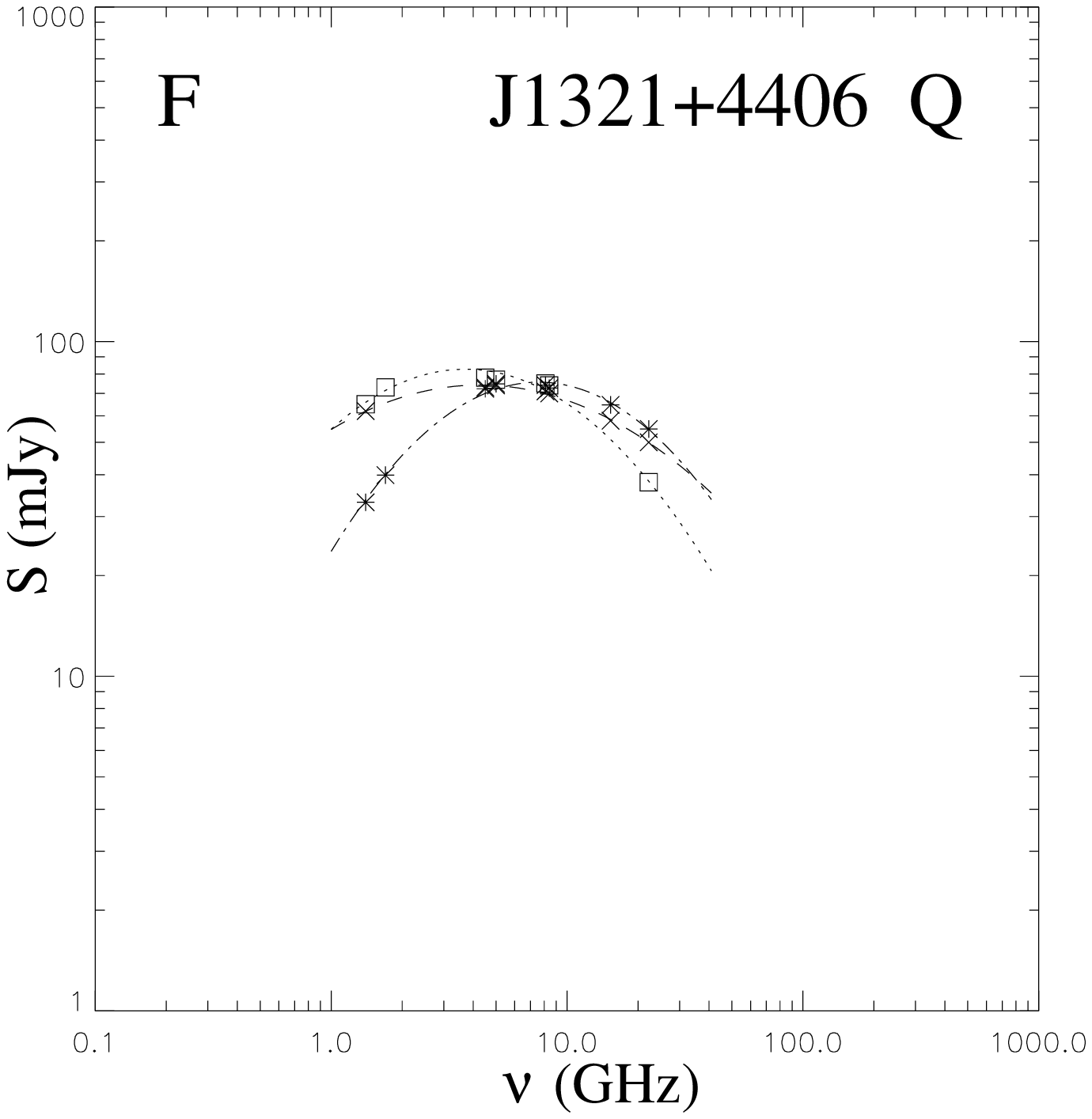}
\includegraphics{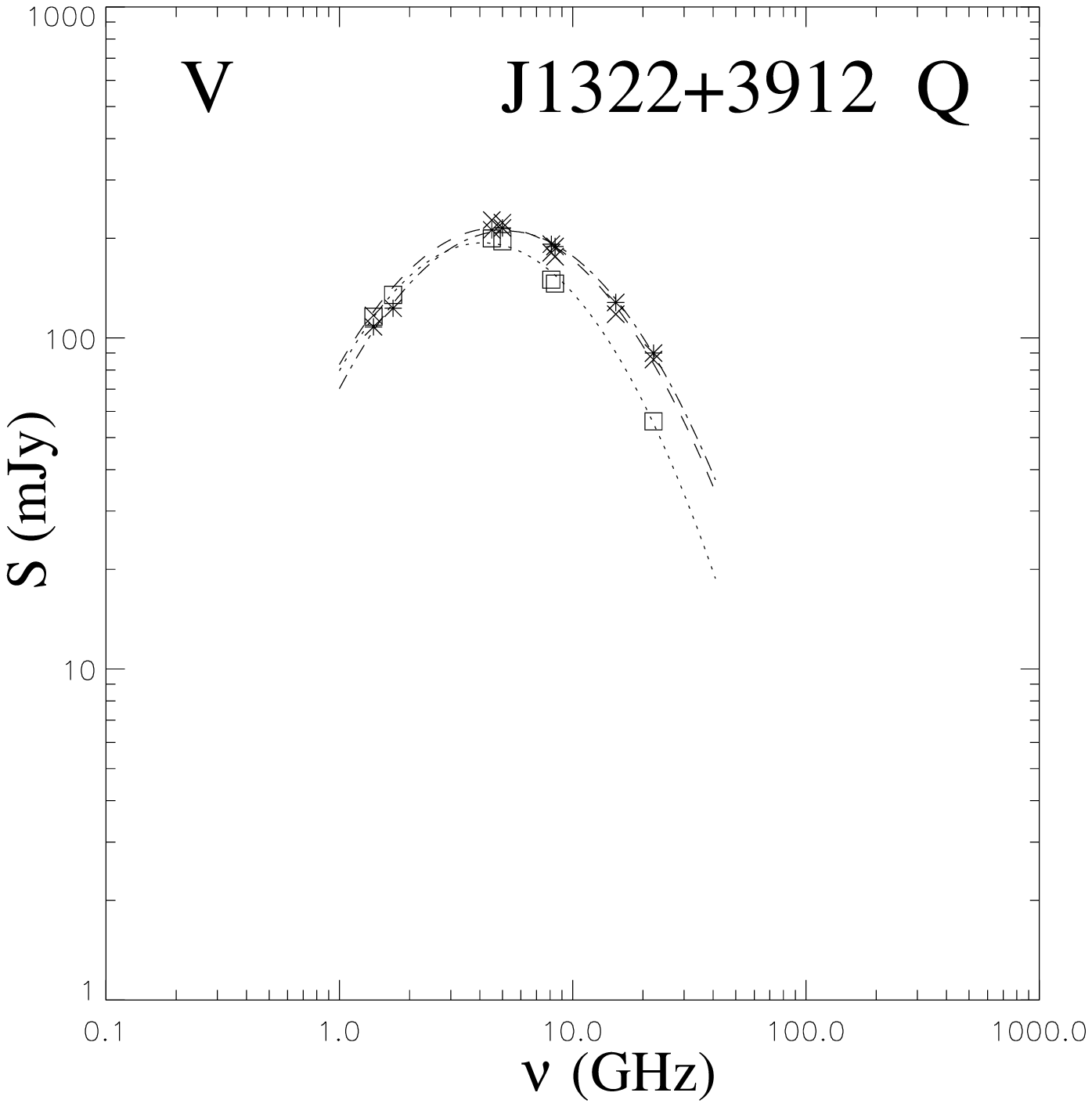}
\includegraphics{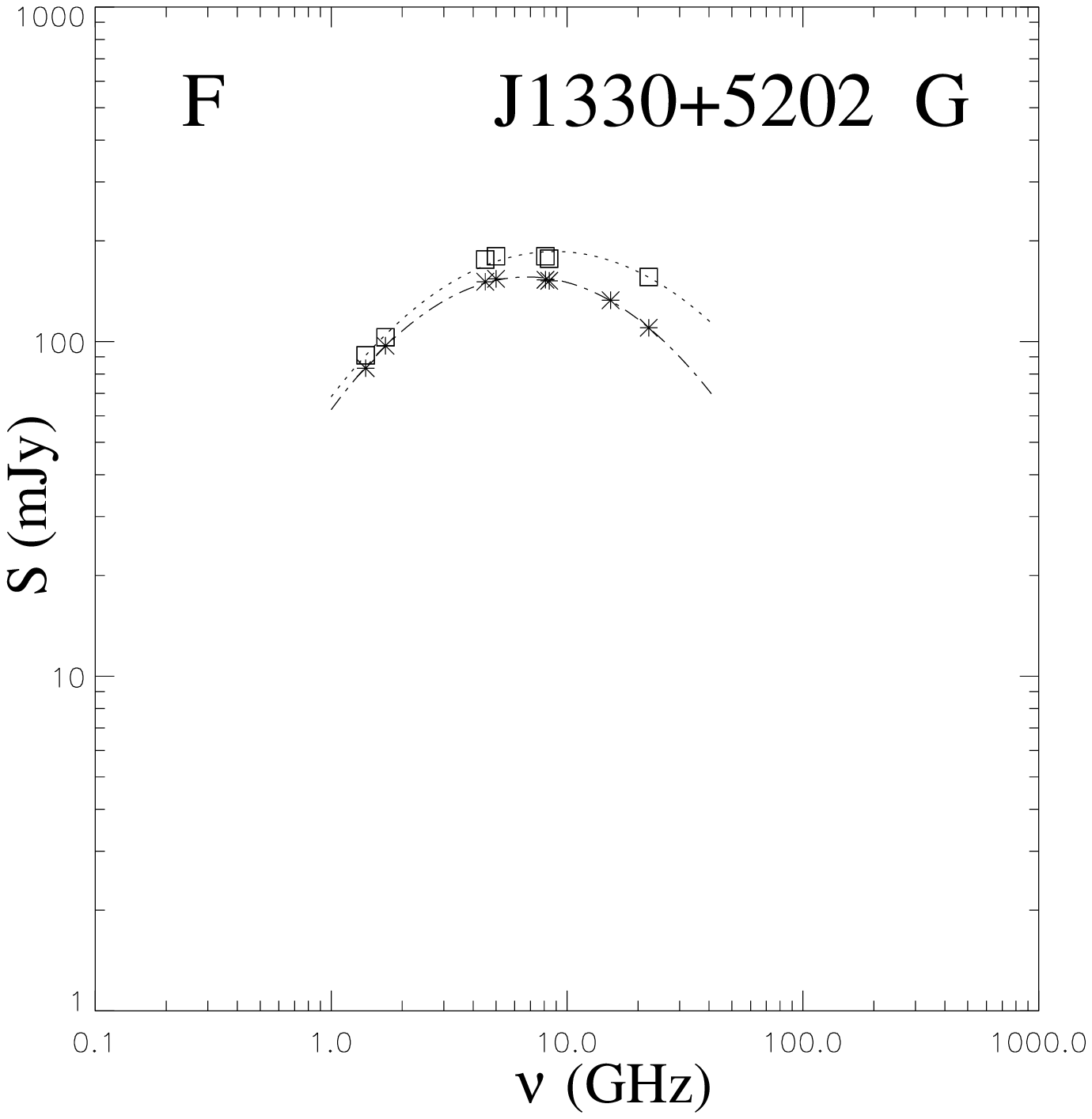}
\includegraphics{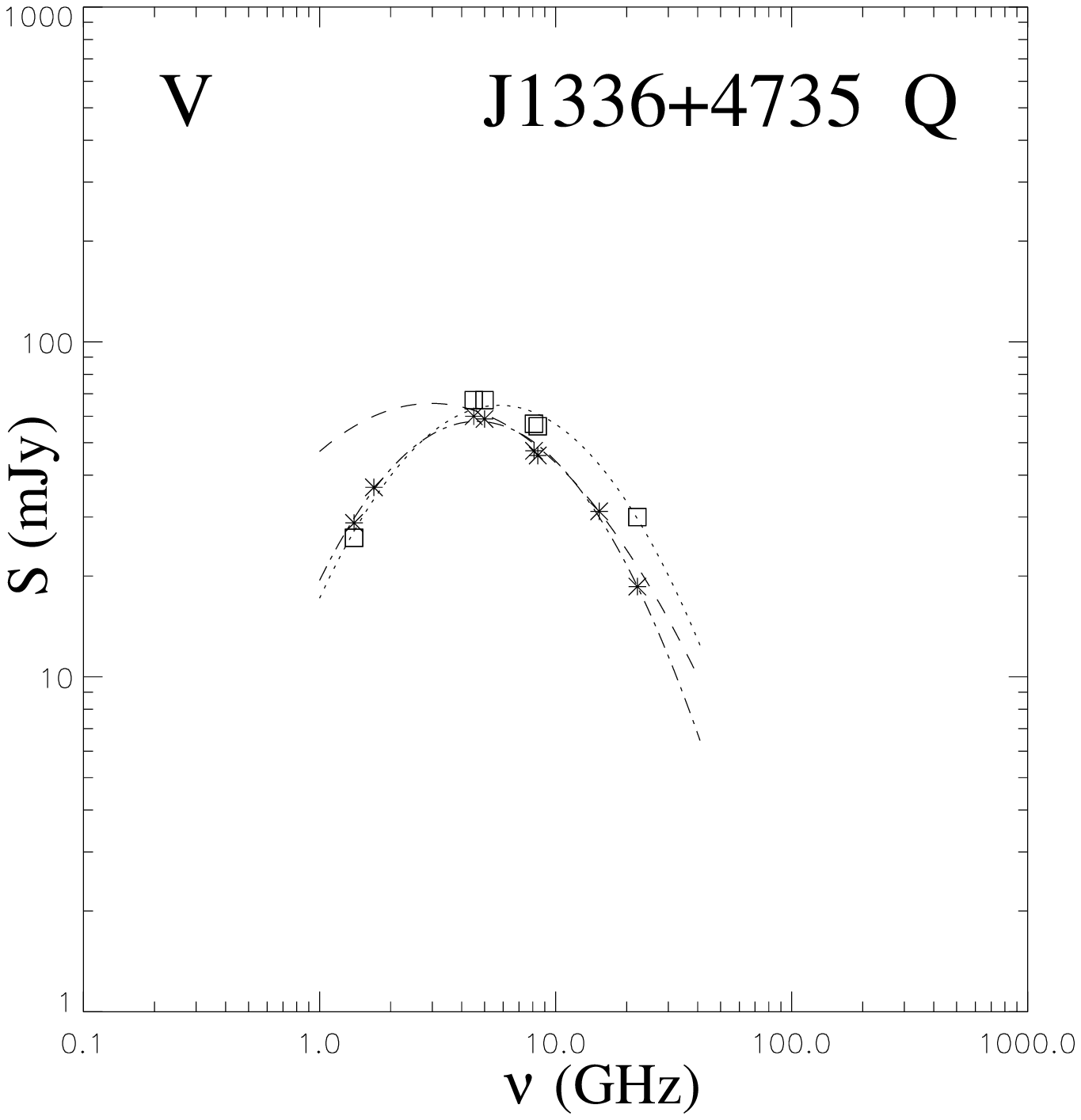}
\includegraphics{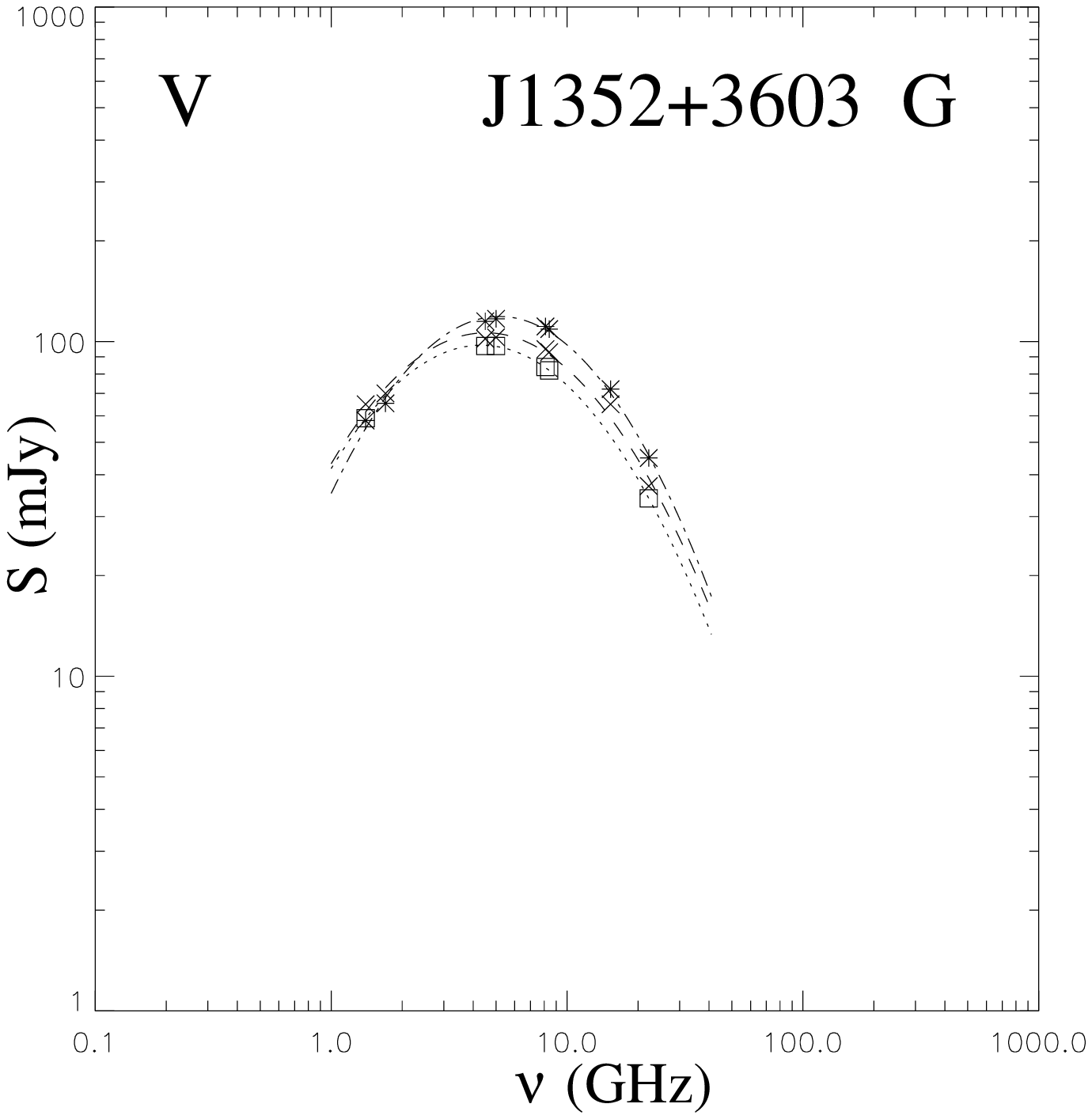}
\includegraphics{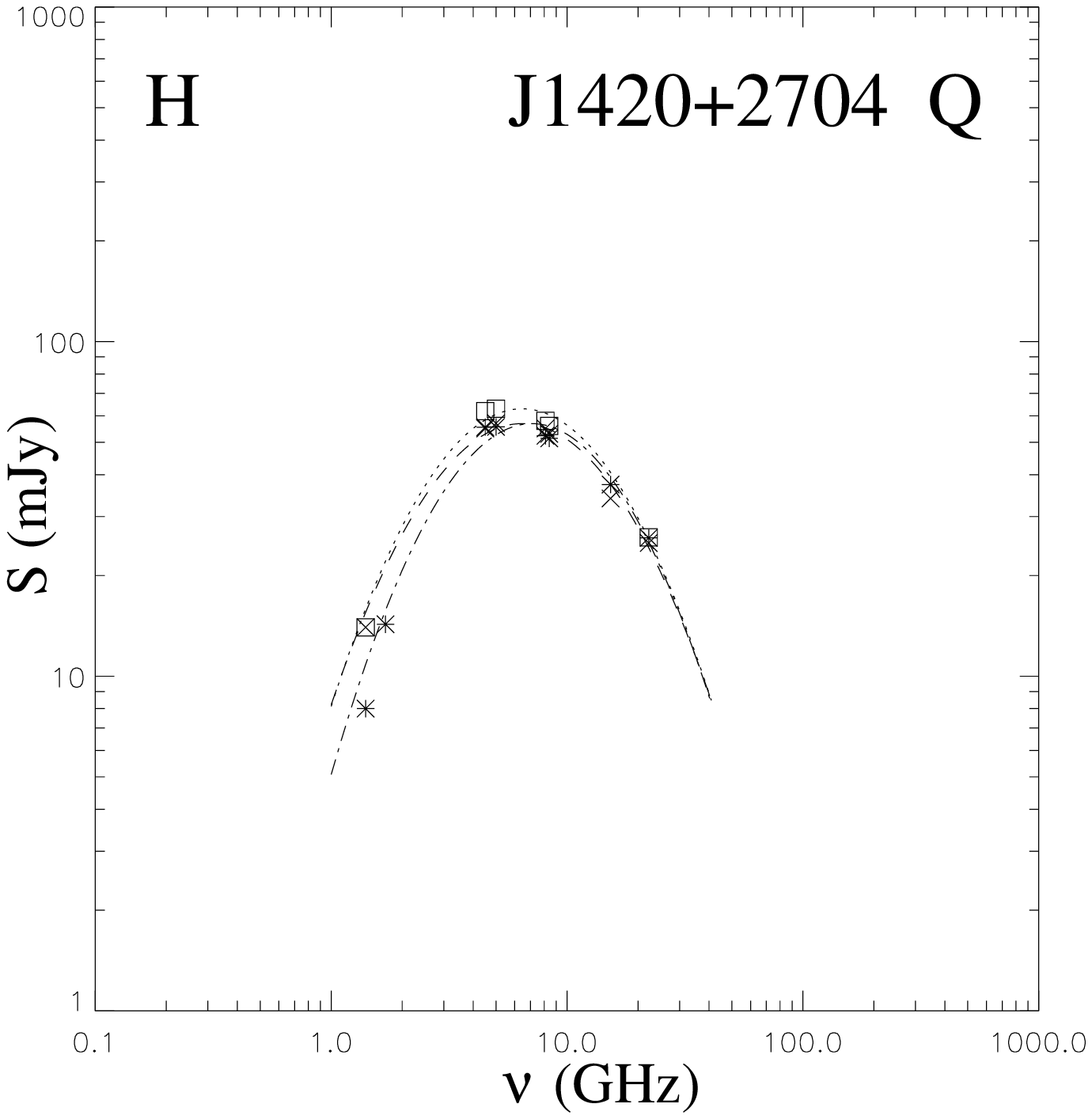}
\includegraphics{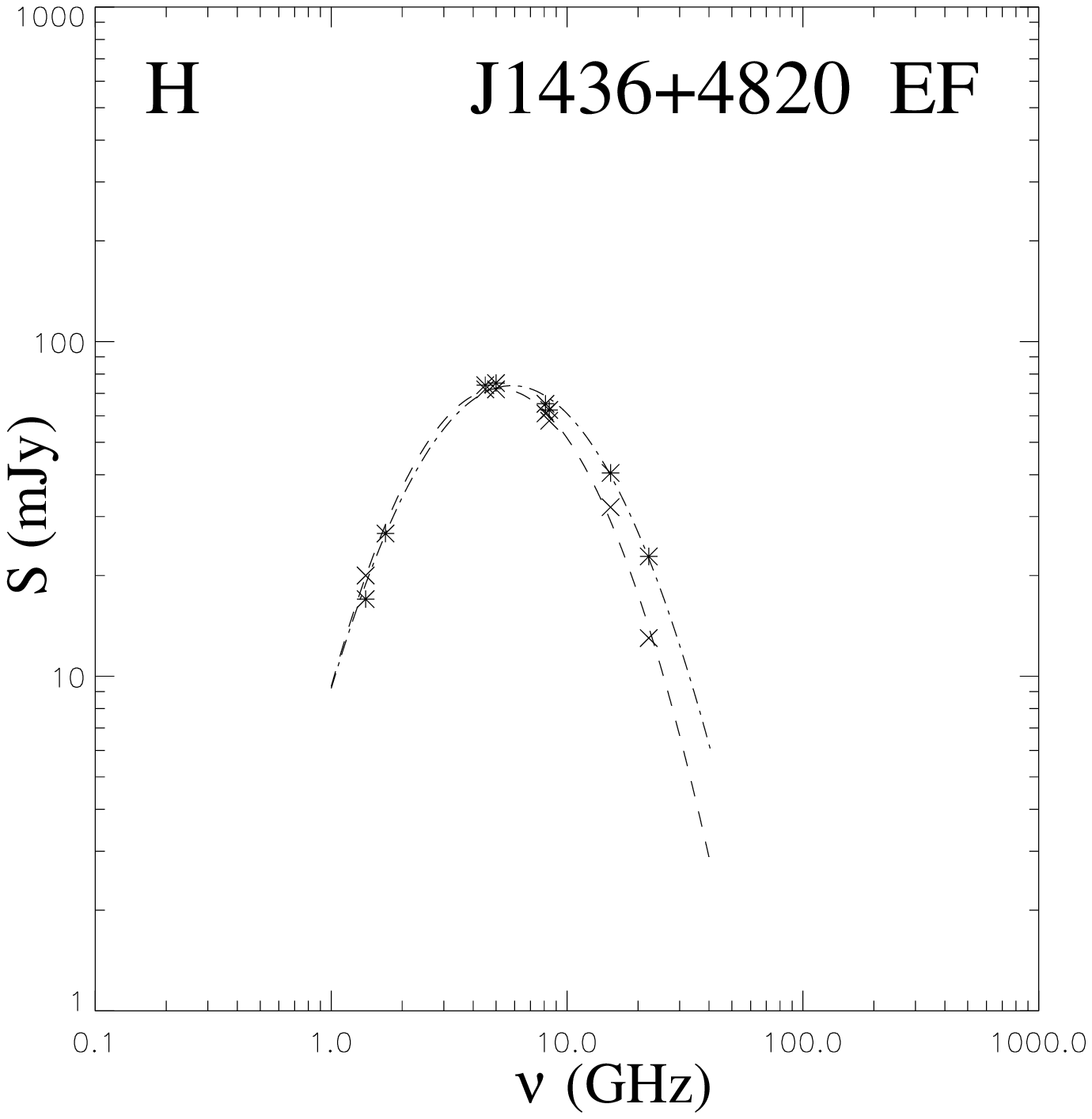}
\includegraphics{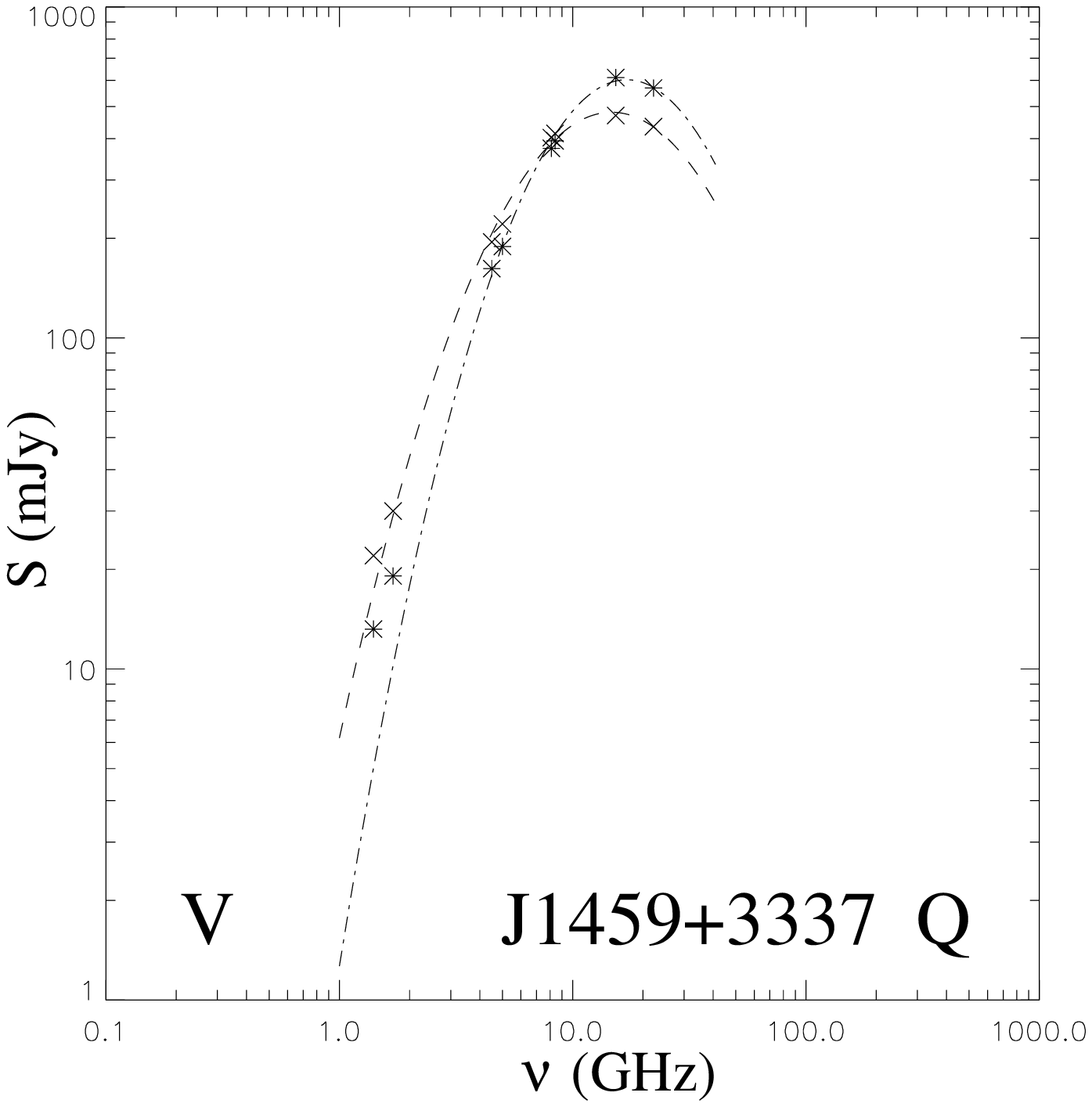}
\includegraphics{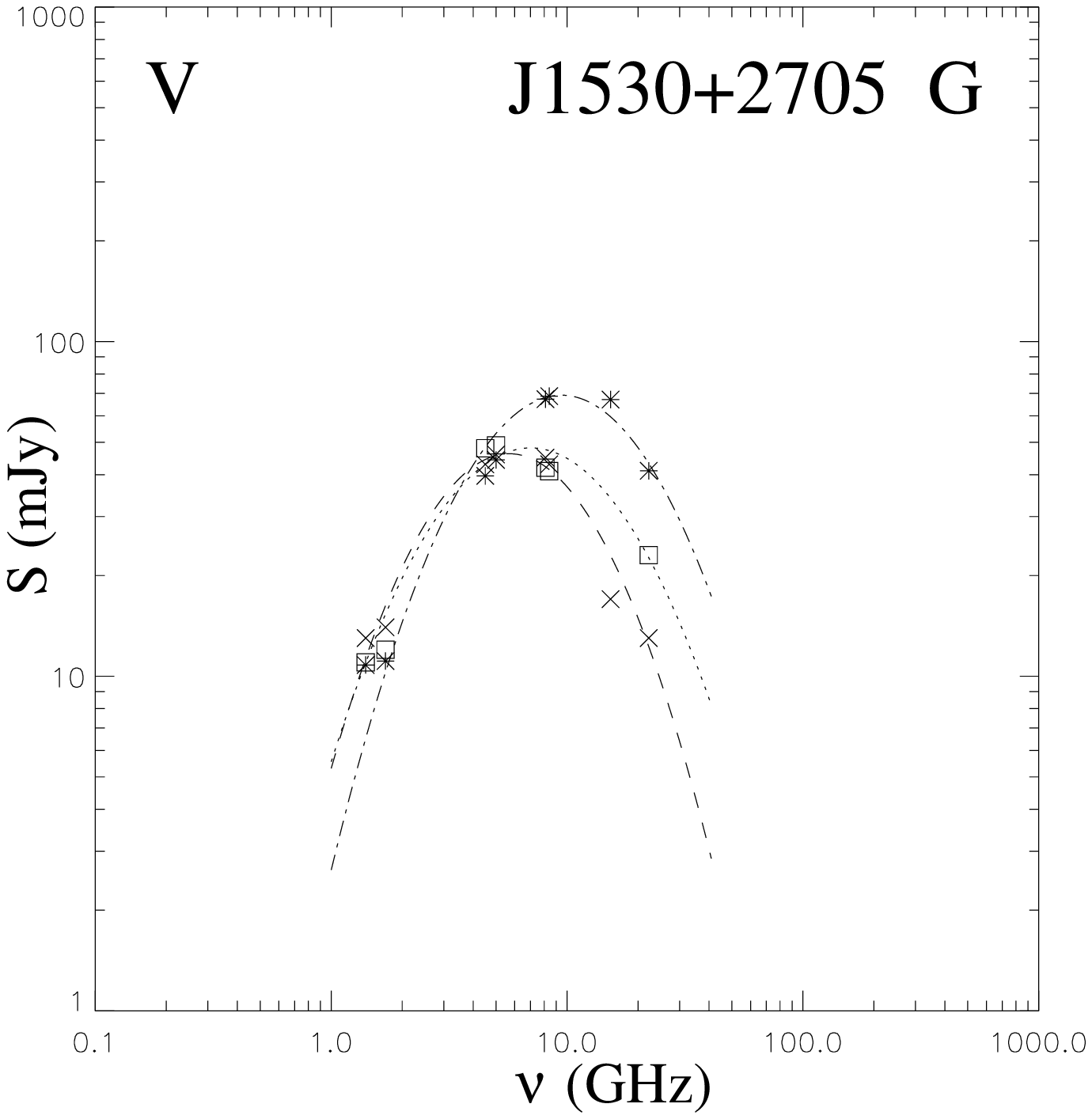}
\vspace{22cm}
\end{center}
\caption{Continued.}
\end{figure*}

\addtocounter{figure}{-1}
\begin{figure*}
\begin{center}
\includegraphics{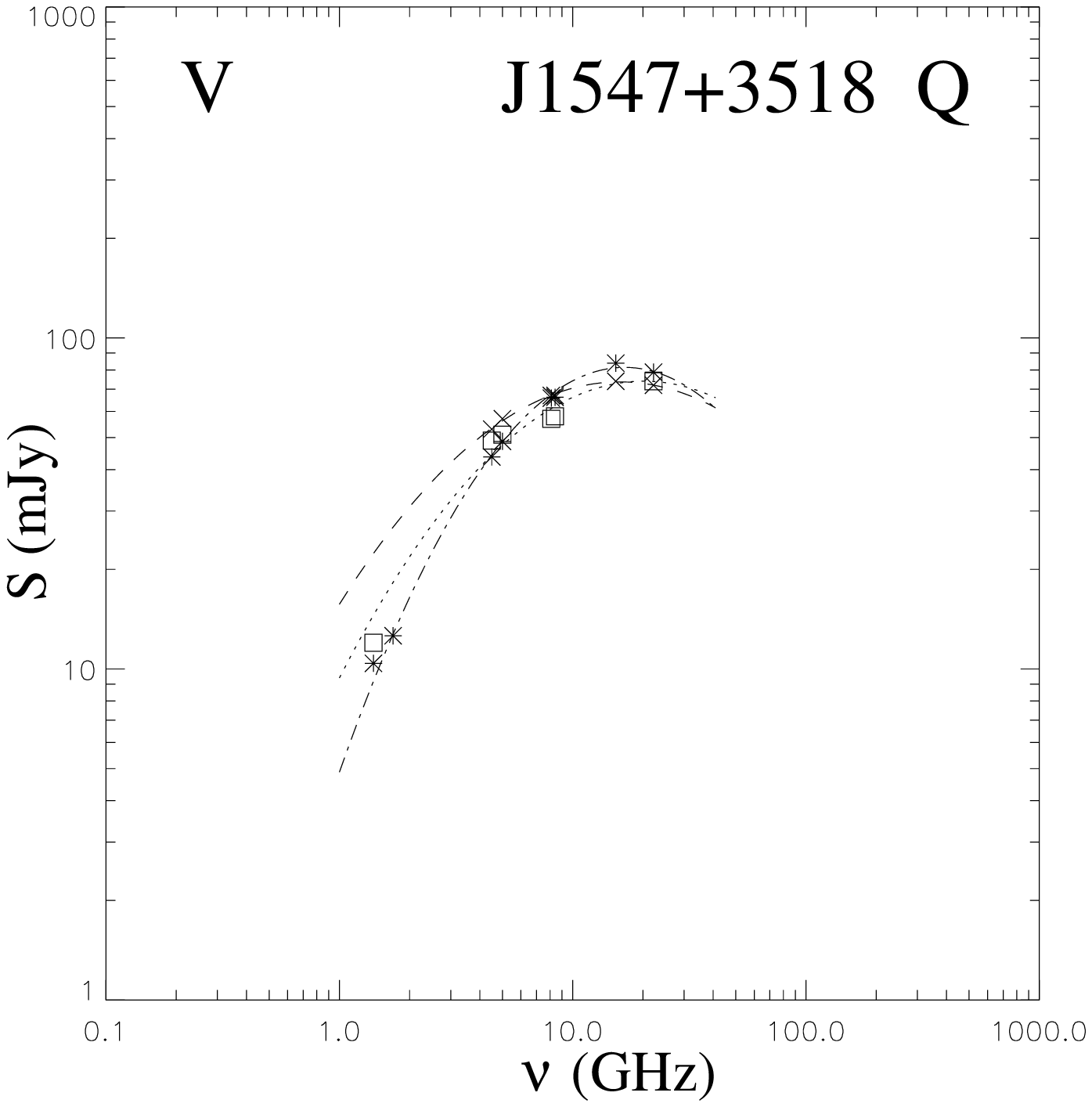}
\includegraphics{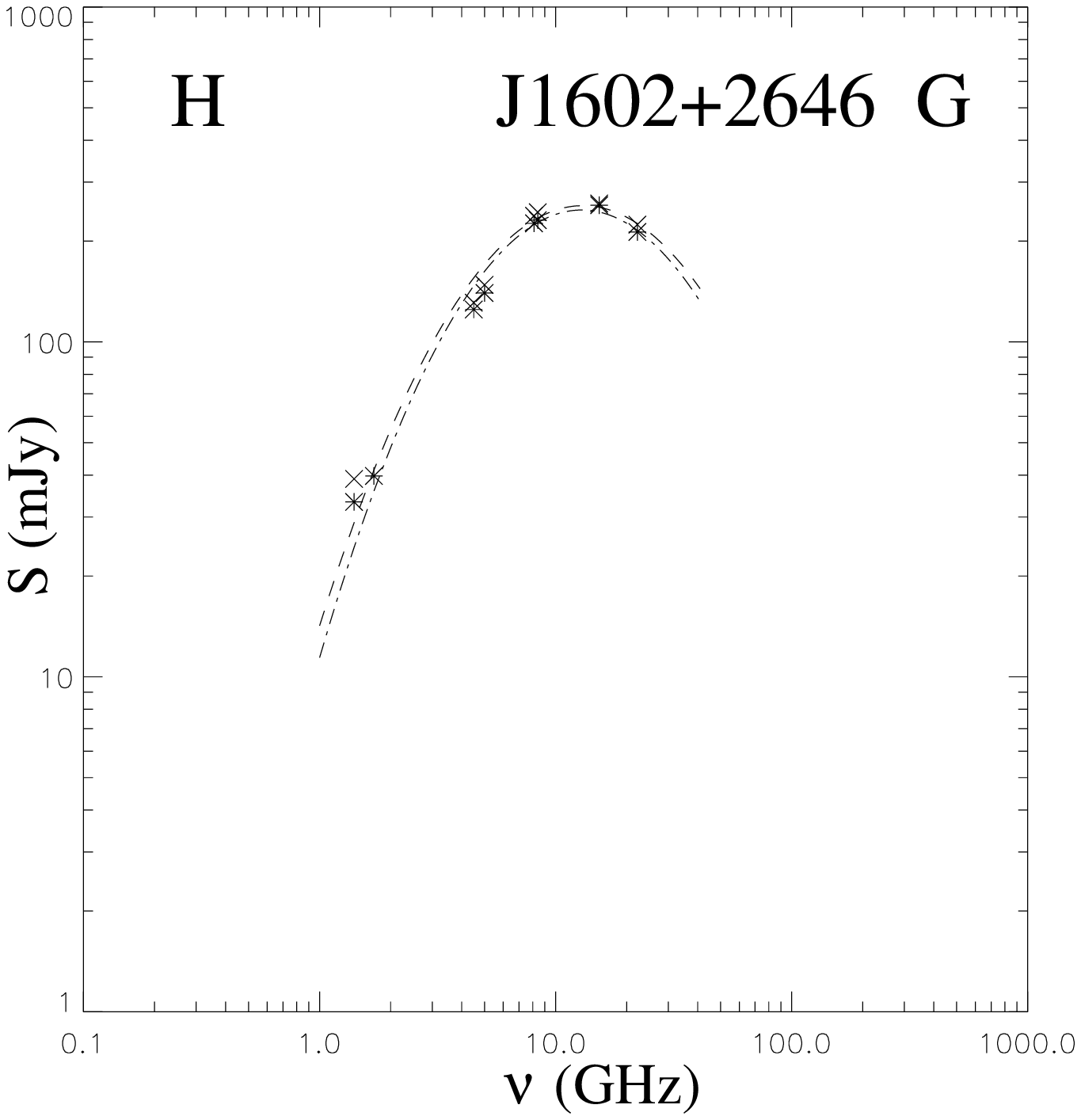}
\includegraphics{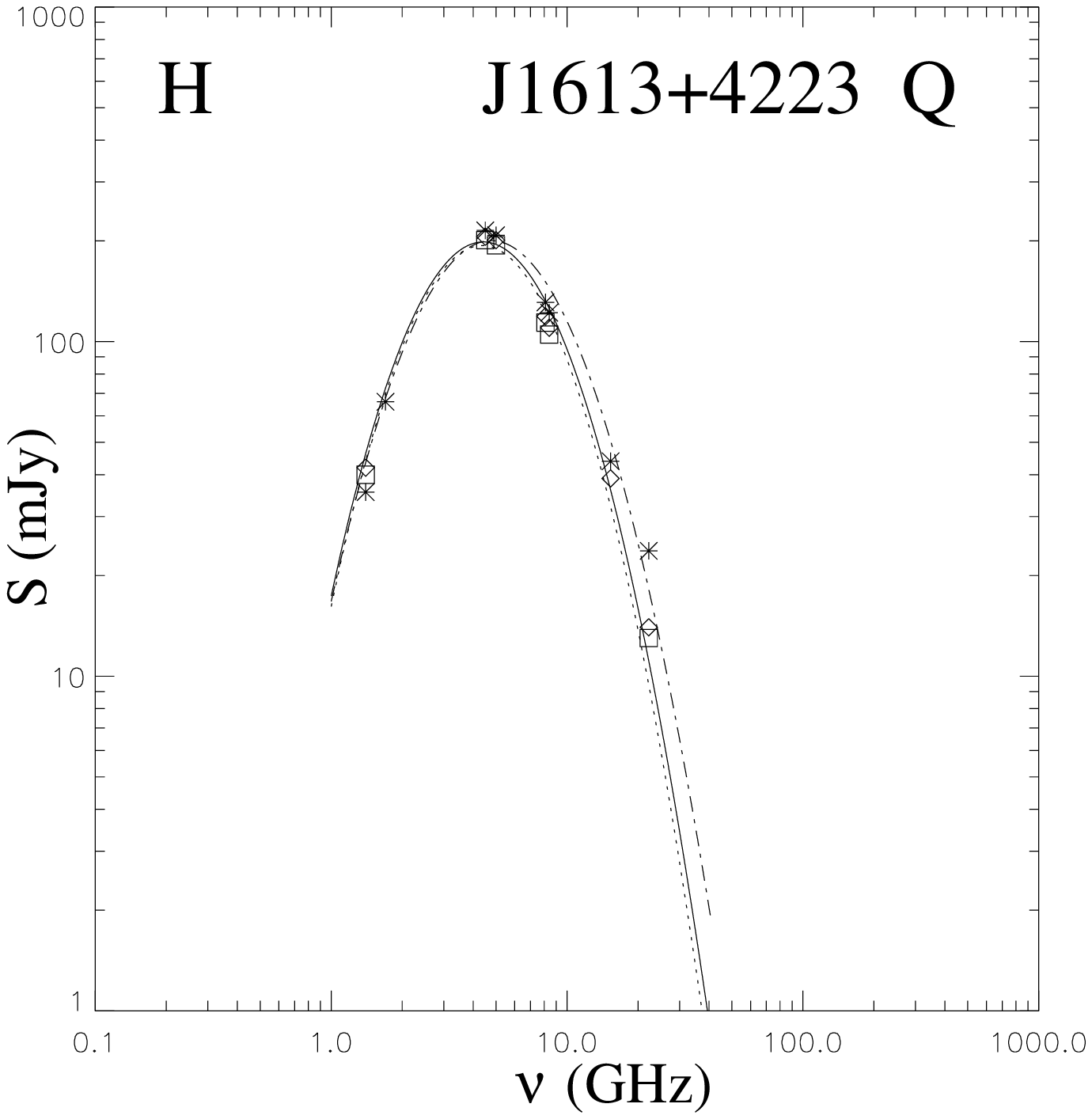}
\includegraphics{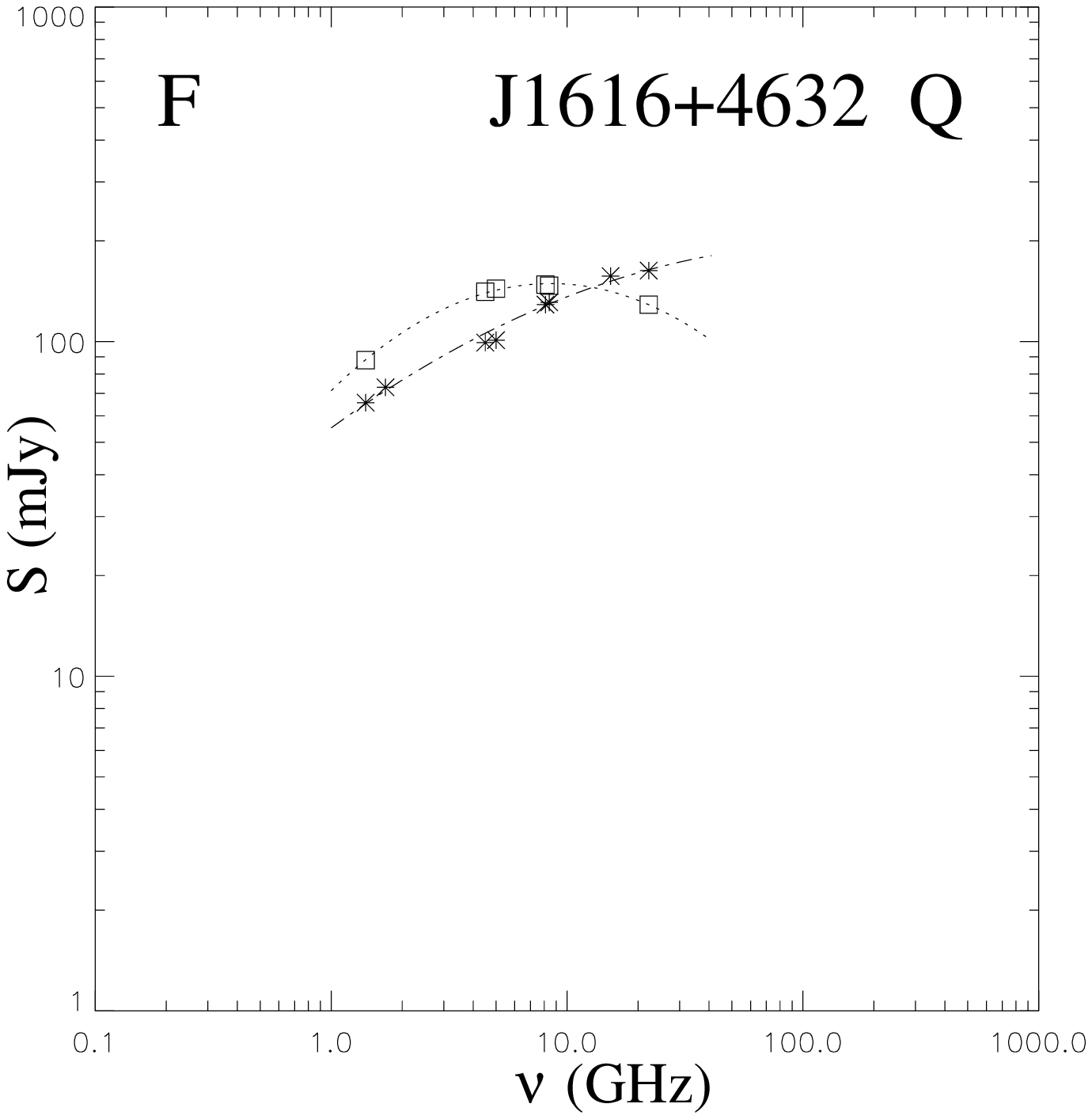}
\includegraphics{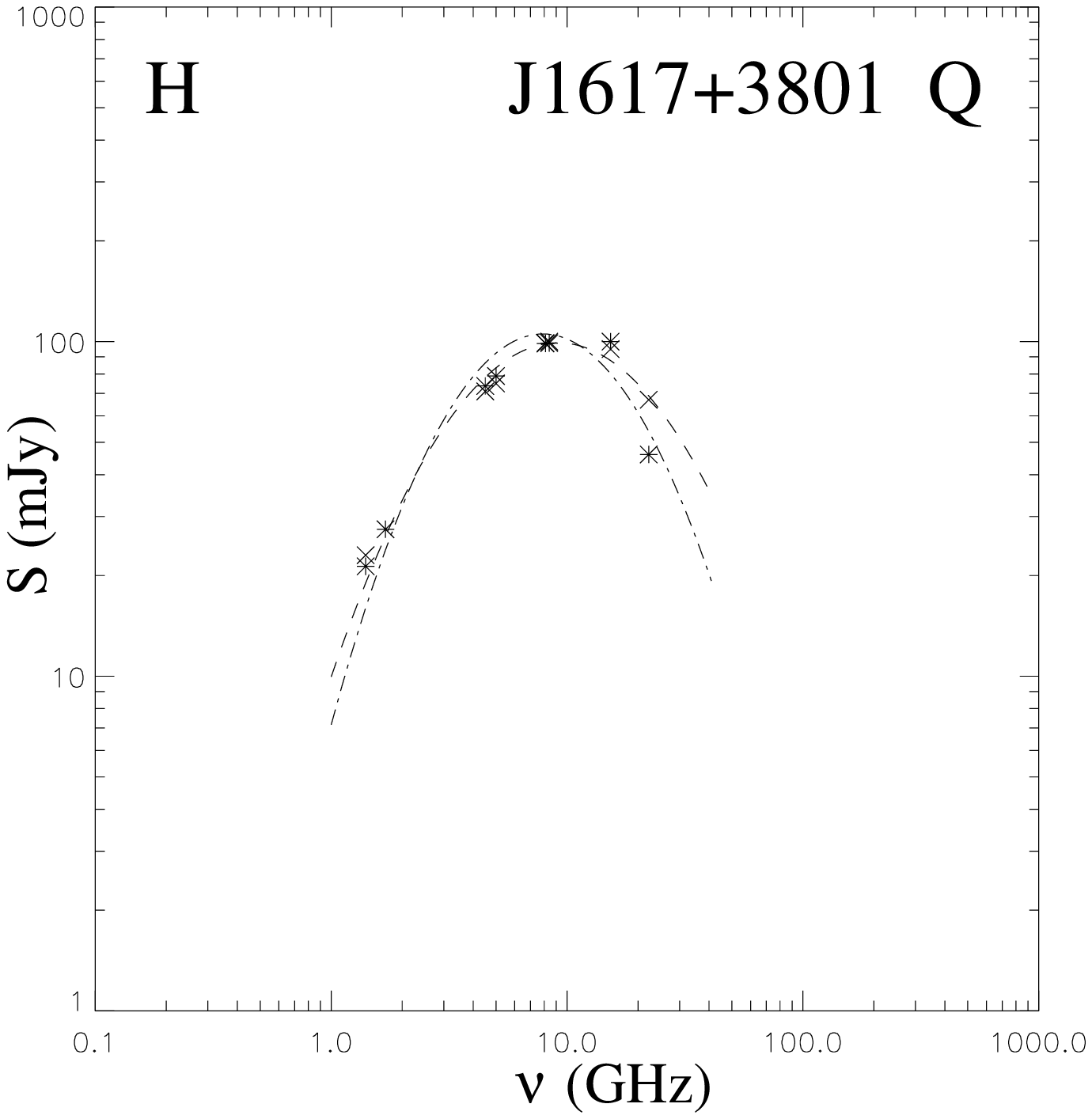}
\includegraphics{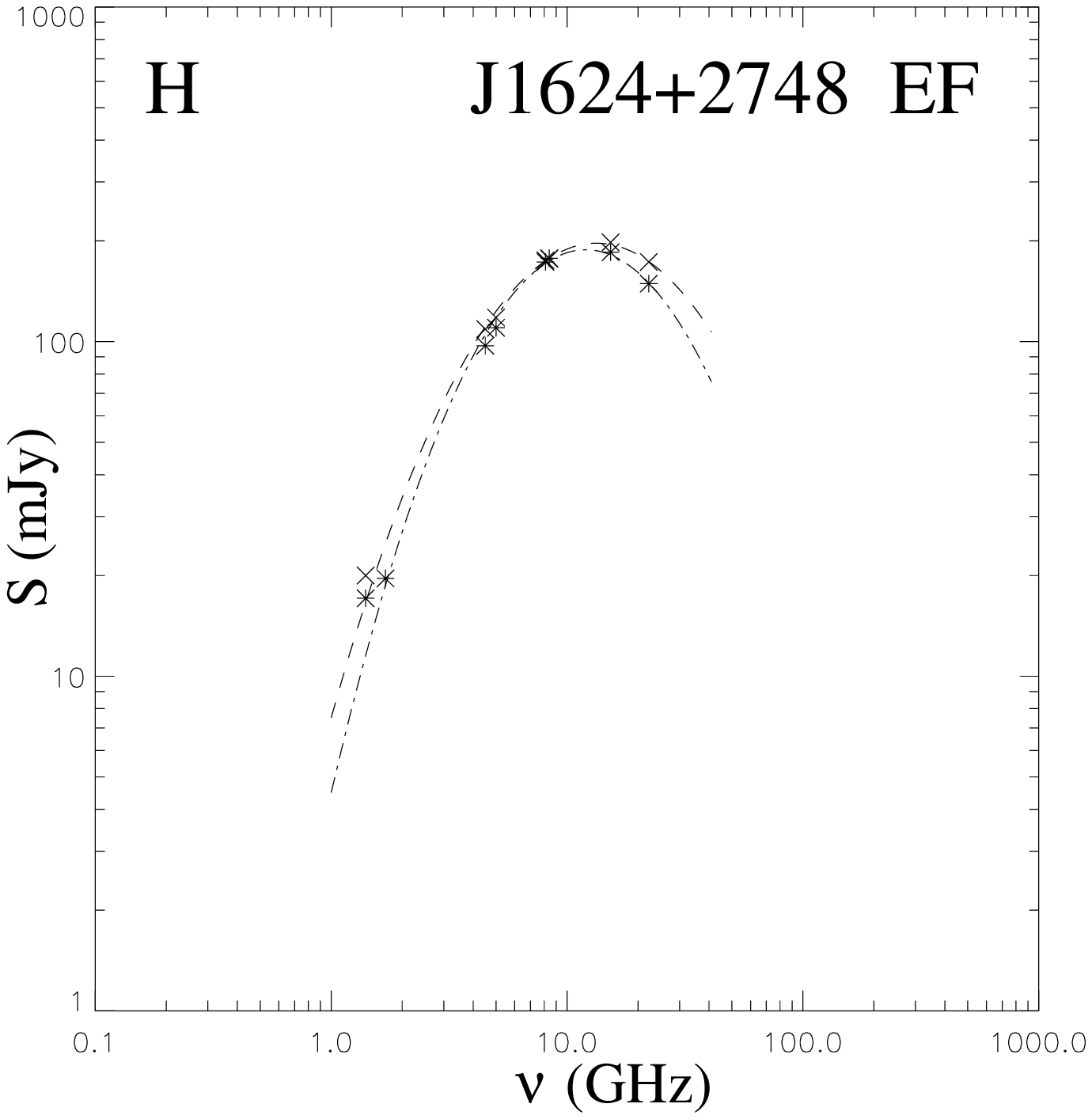}
\includegraphics{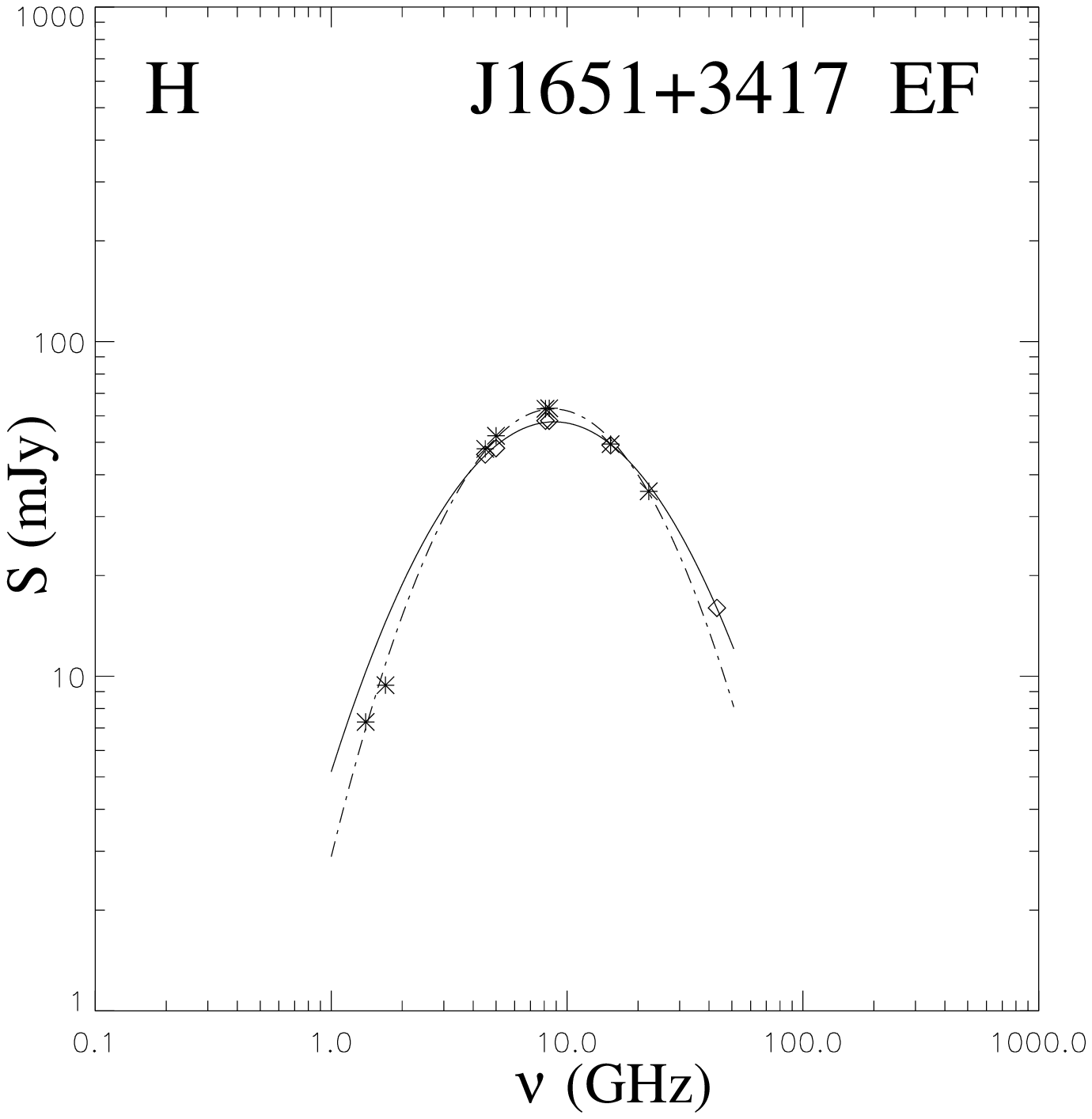}
\includegraphics{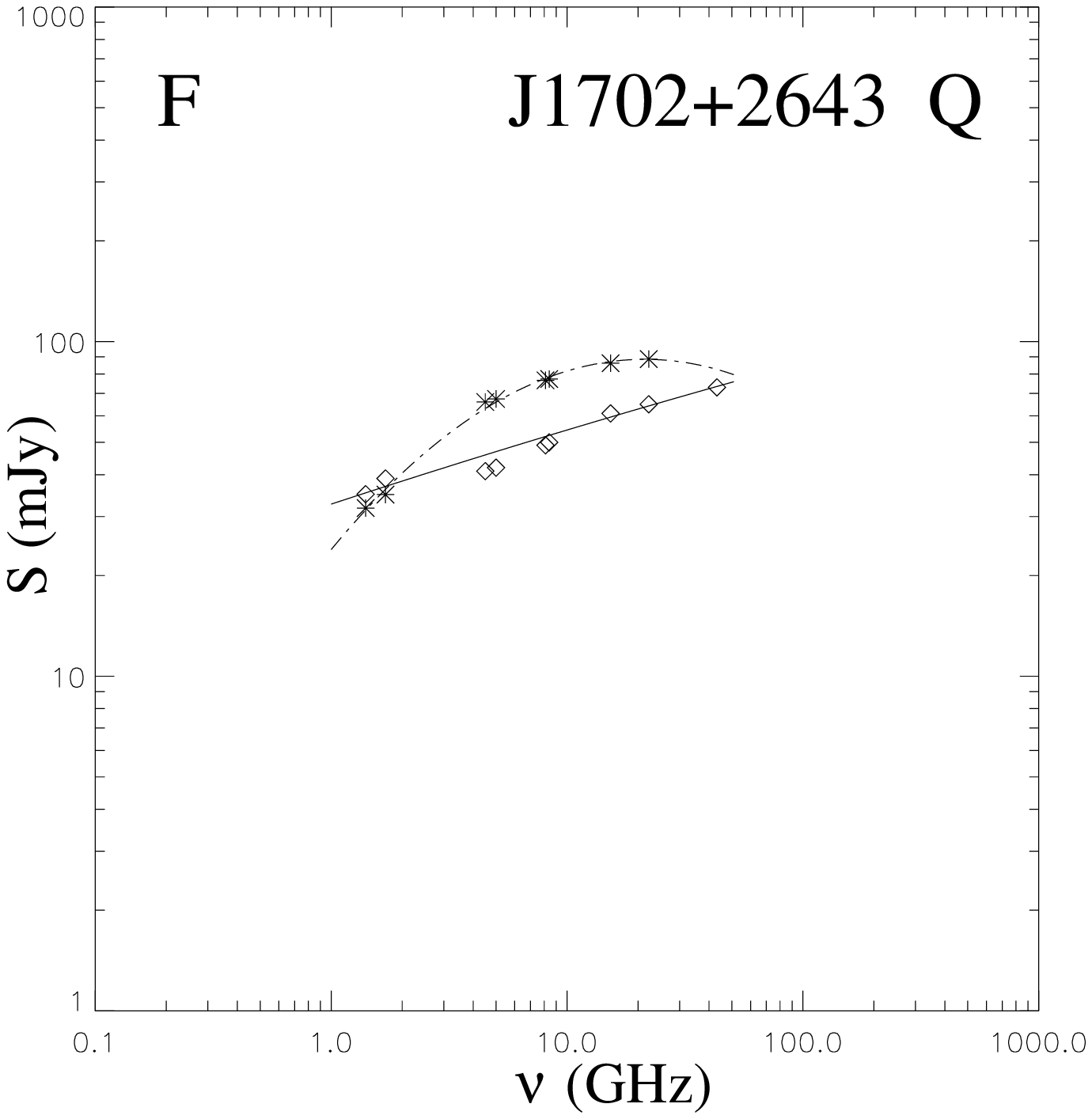}
\includegraphics{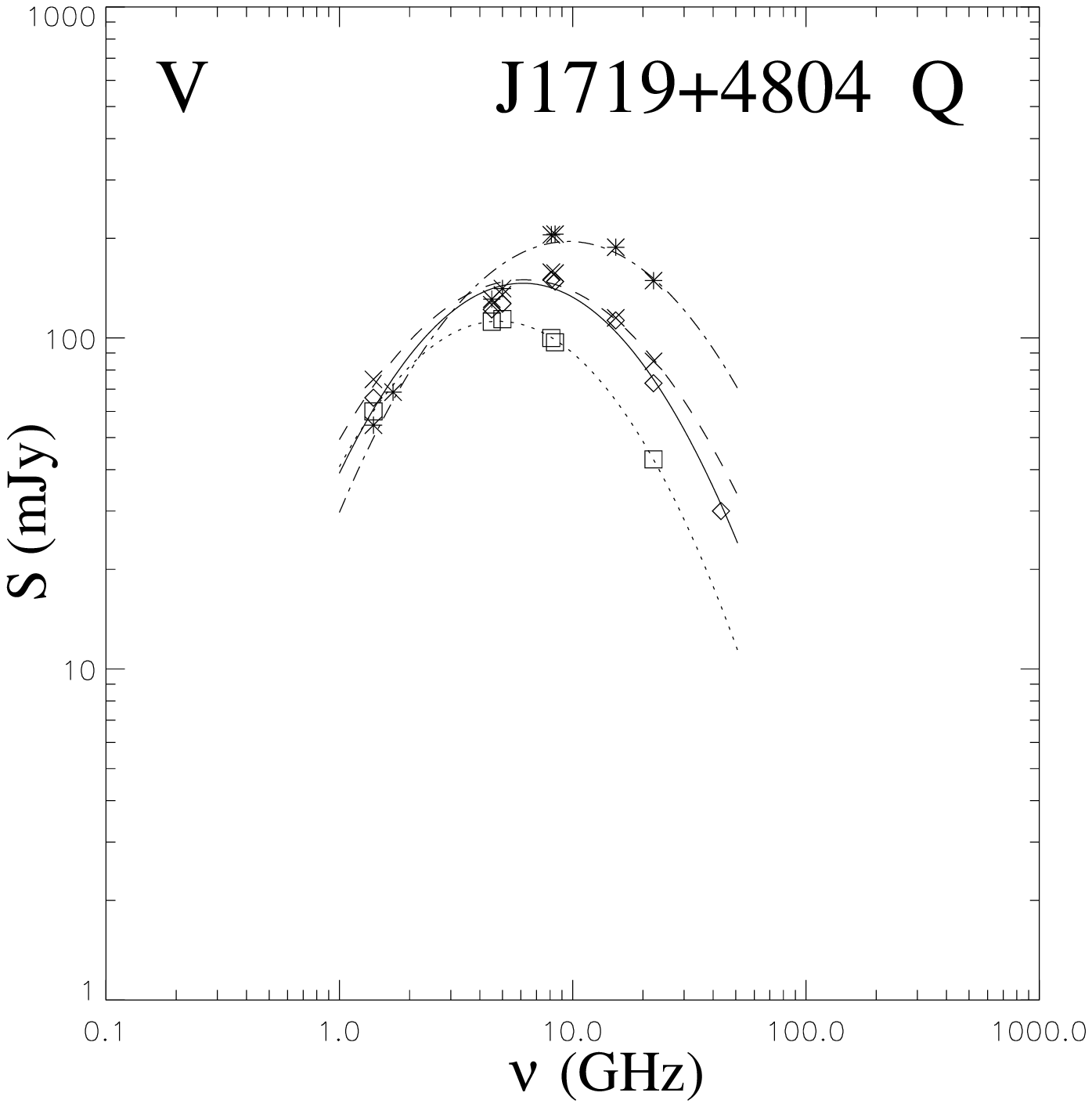}
\vspace{16cm}
\end{center}
\caption{Continued.}
\end{figure*}

\subsection{Variability index}

Following the approach by \citet{tinti05} and \citet{mo07} we
investigate the presence of flux density variability, by computing the
variability index $V$:

\begin{equation}
V= \frac{1}{m}\sum_{i=1}^m \frac{(S_{i}- \overline S_{i})^2}{\sigma^2_{i}}
\label{parameter_var} 
\end{equation}

\noindent where $S_{i}$ is the flux density at the {\it i}-th
frequency measured at one epoch, $\overline S_{i}$ is the mean value
of the flux density 
computed averaging the flux density at the {\it i}-th frequency measured at 
all the available epochs, $\sigma_{i}$ is the rms on $S_{i}-
\overline S_{i}$, and $m$ is the number of sampled frequencies. We
prefer to compute the variability index for each new epoch (see Table
\ref{variability}) instead of considering all the epochs together in order
to better detect the presence of a possible burst. As in
\citet{tinti05}, we consider variable those sources with a variability
index $V > 9$.\\

From the comparison of the multi-epoch spectral properties and
variability we find:\\

\noindent - 24 objects maintain the convex spectrum without showing
  significant flux density variability ($V<9$). They are labelled
  ``H'' in Column 9 of Table \ref{variability};\\

\noindent - 21 sources preserve the convex spectrum at the various
  epochs, although with some amount of flux density variability
  ($V>9$). They are labelled ``V'' in column 9 of Table
  \ref{variability};\\

\noindent - 12 sources with a convex spectrum in the first epoch
  \citep{cs09} show a flat spectrum at least in one of the subsequent
  epochs. They are marked with ``F'' in Column 9 of Table
  \ref{variability}.\\

\subsection{Radio properties versus optical identification}

\begin{figure}
\begin{center}
\includegraphics{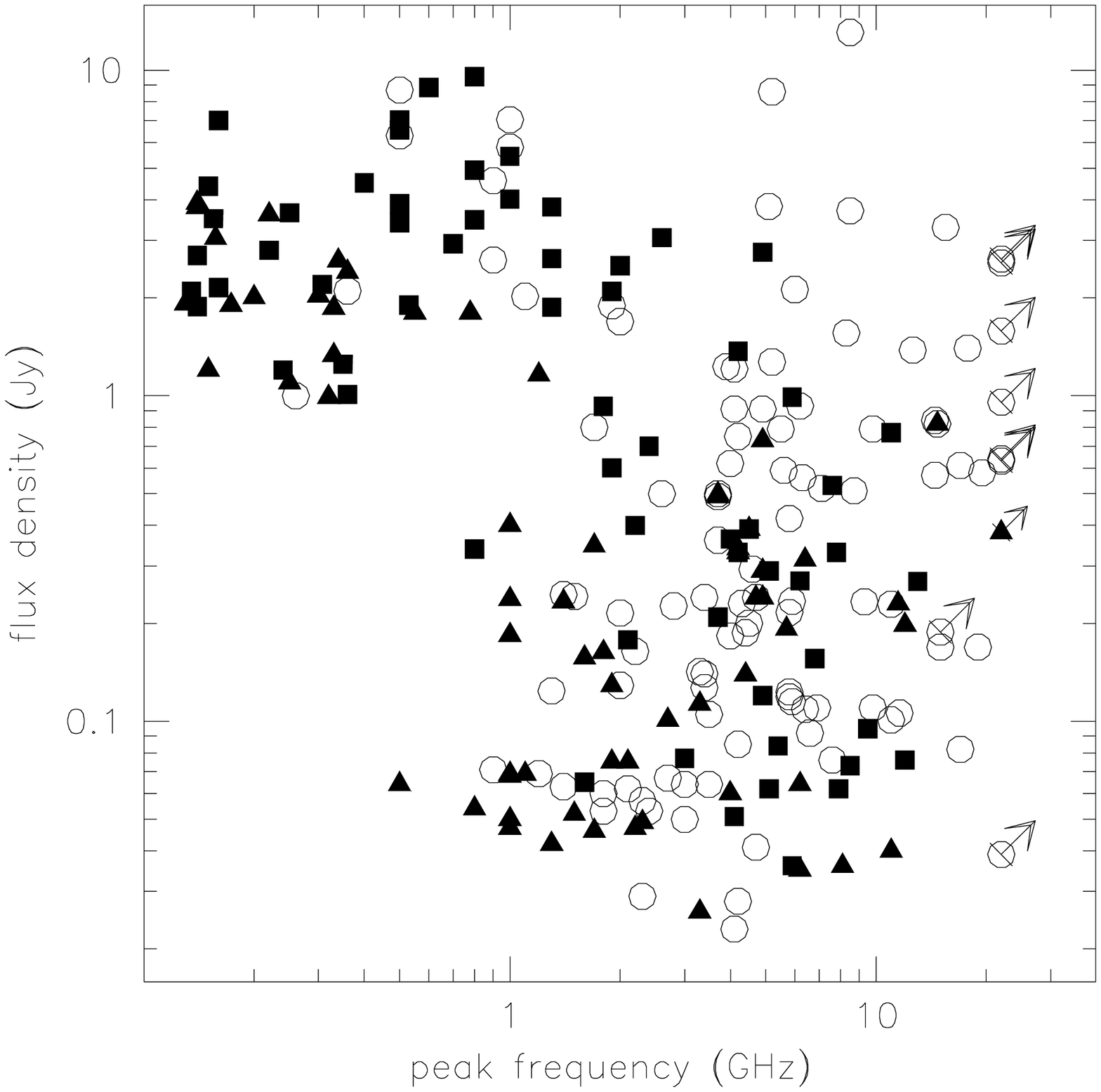}
\vspace{7cm}
\caption{The peak frequency versus the peak flux density for the radio
sources from the B3-VLA CSS sample \citep{cf01}, the faint GPS sample
\citep{sn98}, the half-Jansky GPS
sample \citep{sn02}, the GPS sample from \citet{cs98}, the bright
\citep{dd00} and the faint HFP samples \citep{cs09}. Circles, squares
and triangles refer to quasars, galaxies and empty field,
respectively.}
\label{plottato}
\end{center}
\end{figure}

The optical identification of the radio sources in samples of CSS and
GPS objects \citep[e.g.][]{rf90,cs98,sn98,cf01} showed that
in ``faint'' samples there is a higher fraction of objects identified with
galaxies with respect to ``bright'' samples, 
and the fraction of galaxies seems to be anti-correlated 
with the peak frequency, i.e. in CSS samples a higher percentage of
radio sources are hosted in galaxies than in GPS and HFP samples. 
This is easily seen in Fig. \ref{plottato}, where we
  have plotted the peak frequency versus the peak flux density for the
sources from the bright HFP sample
\citep{dd00}, the faint HFP sample \citep{cs09}, 
the GPS sample \citep{cs98}, the faint GPS sample \citep{sn98}; the
half-Jansky GPS sample \citep{sn02}, and 
the B3-VLA CSS sample \citep{cf01}. In the latter sample only objects
with spectra peaking above 100 MHz have been considered, since
for sources with a peak at lower frequency, the turnover frequency
could not be reliably constrained.
In this plot, 
radio sources identified with galaxies (squares) mainly occupy both
the top left part of the panel, i.e. low peak frequency and high peak
flux density, and the bottom right panel, i.e. high peak frequency
(usually below 10 GHz) but low peak flux density. On the other hand,
quasars (circles)
are mostly found in the left part of the plot, with peak
frequencies above a few GHz, and peak flux densities that span
almost three orders of magnitude. Radio sources without an
optical identification (triangles) are found in the same regions as
galaxies, suggesting that the majority of these objects share the same
properties of the galaxies, but they are fainter probably because
they are at
higher redshift. It is worth noting that the criteria that have been used
to date to select radio source catalogues cannot pick up objects with
both low peak frequency and low peak flux density (bottom left panel),
indicating that we miss the population of faint CSS objects.\\
Statistical studies on the correlation between the radio
characteristics and the optical counterpart indicate that those
objects hosted in a galaxy have the typical properties of young
radio sources (i.e. symmetric radio structure and no spectral
variability), whereas those identified with quasars are more similar
to flat-spectrum radio objects. This suggests that there is a
dichotomy between the optical identification and the radio properties:
quasars are more likely part of the flat-spectrum blazar population,
while galaxies are likely associated with genuinely young radio
sources.
In the case of the sources from the faint HFP sample studied in this
paper, the comparison between their optical identification and the radio
variability showed that the majority of quasars have a variable
spectrum, while a smaller fraction maintain the convex spectrum
without variability (33\% H, 39\% V, and 28\% F). 
In the case of galaxies, we found that the
fraction of variable objects is still larger than those without
spectral variability (34\% H, 50\% V, and 16\% F). For the sources
still lacking an optical identification we found that the majority
do not show significant spectral variability (75\% H, 17\% V, 8\%
F).\\
Among the 12 sources with a flat radio spectrum,
we found that two objects (J1058+3353 and J1330+5202) are identified
with a galaxy, and one (J1020+2910) lacks an optical
identification. When we compare the variability properties of sources
with different optical identification by means of the Student's
t-statistic we find that there is a significant difference ($>$90\%) between
galaxies and quasars, indicating that the majority of galaxies and
quasars are associated with different radio source populations.\\

\begin{table*}
\caption{Peak frequency and flux density variability. Column 1: source
name (J2000); Col. 2: peak frequency in GHz of the first epoch
(1998-2000, Stanghellini et al. 2009); Cols 3, 4, and 5: peak
frequency in GHz of epoch {\it a } (2003), {\it b} (2004), and {\it c,
d} (2006-2007); Cols. 6, 7, and 8: variability index V computed for
the epoch {\it a}, {\it b}, {\it c,d} respectively; Col. 9: the
classification of the source spectrum (V=variable, H=genuine HFP,
F=flat; see Section 3). An asterisk
indicates that the peak frequency is not reliable due to poor
frequency coverage. $^{1}$: for the source J1008+2533 we report the peak
of the lowest part of the spectrum.}
\begin{center}
\begin{tabular}{ccccccccc}
\hline
Source&$\nu_{\rm ep1}$&$\nu_{\rm ep2}$&$\nu_{\rm ep3}$&$\nu_{\rm
  ep4}$&V$_{a}$&V$_{b}$&V$_{c,d}$&Var.\\
(1)&(2)&(3)&(4)&(5)&(6)&(7)&(8)&(9)\\
\hline
&&&&&&&&\\
J0736+4744& 3.6&    & 3.9& 4.6&  6.3&     & 22.2& V \\
J0754+3033& 8.8& 7.9& 8.1& 7.3& 27.5& 22.4& 19.9& V \\
J0804+5431& 5.4&    &    & 5.5&     &     &  2.3& H \\
J0819+3823& 5.8& 5.6& 6.2& 6.1&  6.7&  3.0&  4.0& H \\
J0821+3107& 3.3&    &    & 2.6&     &     &166.6& V \\
J0905+3742& 3.9& 3.8& 3.7&    & 13.6&  1.5&     & H \\
J0943+5113& 3.7& 3.4&    & 3.7& 14.7&     & 12.6& H \\
J0951+3451& 6.0& 6.1&    & 5.6&  1.1&     &  4.3& H \\
J0955+3335& 5.8& 5.2&    & 4.5&  7.4&     & 57.9& V \\
J1002+5701& 4.6& 4.1&    &    & 14.4&     &     & H \\
J1004+4328& 8.0&    & 5.5& 6.5&     & 10.3& 10.9& V \\
J1008+2533$^{1}$& 5.8&    & 5.0& 5.4&& 64.6& 27.2& F\\
J1020+2910& 3.3&    & 1.5& 2.6&     & 14.2& 22.5& F \\
J1020+4320& 4.5&    & 4.5&    &     & 10.9&     & H \\
J1025+2541& 4.1&    & 3.6&    &     & 14.3&     & V \\
J1035+4230& 7.0&    & 6.7&    &     & 18.2&     & H \\
J1037+3646& 4.0&    & 4.0&    &     &  1.0&     & H \\
J1044+4328& 6.9&    &3.4$^{*}$&  &  & 20.3&     & H \\
J1046+2600& 4.7&    & 4.3&    &     &  4.1&     & H \\
J1047+3945& 3.7&    & 1.6&    &     & 62.2&     & F \\
J1052+3355& 5.0&    &2.3$^{*}$&4.1&  & 11.0&28.0& V \\ 
J1053+4610&11.5&    &17.6&$>$22&   &110.0&352.4& V \\
J1054+5058&$>$22&   &  &$>$22&    &    &   5.3 & H \\
J1058+3353& 6.4&$>$22& & &71.6&    &     &       F \\
J1107+3421& 4.6& 4.5& 5.2& 4.2& 12.8&  0.9&  4.0& H \\
J1109+3831& 8.1& 9.3& 9.2& 9.3&  5.1&  8.4& 14.1& V \\
J1135+3624& 4.1& 4.2&    & 4.4&  0.9&     &  5.5& H \\
J1137+3441&23.0&$>$22&  &     &146.8&     &     & V \\
J1203+4803&$>$22&15.2&  &  9.0&600.0&    &1058.1& F \\
J1218+2828& 7.1& 7.5&    &    & 50.7&     &     & V \\
J1239+3705& 9.5& 9.5&    &10.0& 23.6&     & 35.2& V \\
J1240+2323& 7.8& 9.0&    & 9.8& 16.0&     & 12.2& V \\
J1240+2425& 3.8& 0.7&    & 2.6& 55.7&     &  5.3& F \\
J1241+3844& 3.7&    &    & 3.6&     &     &  1.0& H \\
J1251+4317& 7.5&$>$22&   &11.4&150.3&     &430.3& V \\
J1258+2820& 4.8& 7.0&    &14.6& 22.2&     & 40.2& F \\
J1300+4352& 5.8&    &  &$<$1.4&     &     &883.0& F \\
J1309+4047& 5.2& 5.6&    & 5.4& 16.5&     &  2.6& H \\
J1319+4851& 5.3& 5.7&    & 5.2&  2.7&     &  4.0& H \\
J1321+4406& 7.5& 4.3&    & 3.8& 11.9&     & 51.9& F \\
J1322+3912& 5.2& 4.6&    & 4.1& 18.0&     & 83.9& V \\
J1330+5202& 6.8&    &    & 8.8&     &     & 52.0& F \\
J1336+4735& 4.6& 3.0&    & 5.7&  2.5&     & 17.6& V \\
J1352+3603& 5.2& 4.6&    & 4.3& 34.6&     & 55.7& V \\
J1420+2704& 6.8& 6.2&    & 6.5&  3.8&     &  9.7& H \\
J1436+4820& 5.8& 5.1&    &    & 12.9&     &     & H \\
J1459+3337&16.9&14.6&    &    &109.4&     &     & V \\
J1530+2705& 9.7& 5.7&    & 7.0&106.3&     & 33.4& V \\
J1547+3518&16.5&15.9&&$>$22$^{*}$&12.2&   & 10.9& V \\
&&&&&&&&\\
\hline
\end{tabular}
\label{variability}
\end{center}
\end{table*}

\addtocounter{table}{-1}
\begin{table*}
\caption{Continued.}
\begin{center}
\begin{tabular}{ccccccccc}
\hline
Source&$\nu_{\rm ep1}$&$\nu_{\rm ep2}$&$\nu_{\rm ep3}$&$\nu_{\rm
  ep4}$&V$_{a}$&V$_{b}$&V$_{c,d}$&Var.\\
(1)&(2)&(3)&(4)&(5)&(6)&(7)&(8)&(9)\\
\hline
&&&&&&&&\\
J1602+2646&12.8&13.0&    &    &  7.5&     &     & H \\
J1613+4223& 4.6&    & 4.3& 4.4&     &  5.9& 12.4& H \\
J1616+4632&$>$22&   &    & 8.5&     &     &119.1& F \\
J1617+3801&11.6& 9.4&    &    &  1.7&     &     & H \\
J1624+2748&12.0&13.3&    &    &  9.5&     &     & H \\
J1651+3417& 8.6&    & 8.4&    &     &  9.0&     & H \\
J1702+2643&21.5&   &$>$22&    &     &115.5&     & F \\
J1719+4804& 9.7& 6.0& 7.5& 4.8& 37.6& 33.8&472.3& V \\  
&&&&&&&&\\
\hline
\end{tabular}
%\label{variability}
\end{center}
\end{table*}

\section{Discussion}

The anti-correlation found between the projected linear size and the
peak frequency \citep{odea97,bicknell97,snellen00} 
implies that the sources with the
spectral peak occurring above a few GHz should represent the
population of the smallest radio sources whose radio emission has
recently turned on. Samples of high-frequency peaking objects have
been selected by choosing sources with an inverted radio spectrum up
to 5 GHz \citep{dd00,torniainen05,cs09}, i.e. the highest observing
frequency where a large area survey is presently available. However,
due to the selection criteria these samples have been found to
comprise both young radio sources and flaring flat-spectrum objects
selected during particular phases of their spectral variability, for
example when their radio emission is dominated by a knot in the
jet. The study of flux density and spectral variability based on
repeated simultaneous multi-frequency observations has
proved to be an ideal tool in discriminating the different nature of
the sources. It was found that in samples of bright objects, where
there is a high incidence of sources optically identified with
quasars, flat-spectrum blazar objects represent the dominant
population \citep[e.g.][]{torniainen07,mo07,jauncey03}. 
A higher incidence of
genuinely young radio sources is expected in samples of faint objects
where the majority of radio sources should be hosted in galaxies and
boosting effects are supposed to play a minor role.\\
The optical identification of the sources studied in this paper by
means of the SDSS DR7 indicates that 21\% of objects are hosted in
galaxies, i.e. similar to the fraction of galaxies in the bright HFP
sample. However, in the faint sample, another 21\% of objects lack
optical identification and thus a reliable comparison between the two
samples cannot be done. The analysis of the optical images of the
galaxies hosting HFP pointed out the presence of companions around 6
HFP candidates, indicating that young radio galaxies, like powerful
extended radio sources, are in groups, 
as previously suggested by
\citet{cs93}, indicating a continuity between compact young objects
and the population of classical radio galaxies \citep{odea96}.\\
Although in 4 galaxies the presence of companions is
suggested by photometric information only, in J1530+2705 and
J1602+2646 the association is made by spectroscopic redshift. 
The companion galaxies are located within a projected distance of
about 150 - 200 kpc from the target which usually is hosted in the
brightest elliptical at the group centre. A peculiar case is
represented by J1109+3831, whose parent galaxy seems to be a
spiral that is interacting with an elliptical. 
Young radio sources are normally associated with ellipticals. The
case represented by J1109+3831 may be explained by the
possible interaction between the hosting spiral and the companion that
may have triggered the radio emission.
The small redshift of J1530+2705 enabled us
to identify the morphology of its brightest companions that turned out
to be barred spirals. This
group resembles that of the bright HFP J0655+4100  \citep{mo06b}, 
and in both cases the
HFP is hosted by the central elliptical galaxy at the group
centre. The presence of companion galaxies in the environment of
galaxies hosting young radio sources suggests that the onset of the 
radio emission may be triggered by merger or interaction events that
occurred not long ago. This scenario is supported by 
the proximity of the companions in J0804+5431
and in J1109+3831, although observations to establish a
physical interaction are needed to unambiguously verify
this idea. \\

From the analysis of the multi-epoch radio spectra of the sources in
the faint HFP sample, we find a high fraction of objects displaying
some level of variability. This result does not imply that all these
sources are part of the blazar population.  
In fact, changes in the radio
spectrum may be a direct consequence of the source expansion 
\citep[e.g.][]{tingay03}. In newly
born radio sources, the evolution timescales can be of the order of a few
tens of years. Changes in the radio spectrum of such young objects can
be appreciable after the short time (5-8 years) elapsed between the
first and last observing run. A clear example is represented by the
faint HFP J1459+3337 \citep{mo08}. This HFP showed a steadily increasing
flux density at 1.4 and 5 GHz, in the optically thick regime, and its
spectral peak shifted from 30 GHz down to 12 GHz in about 10 years
\citep{edge96}. This behaviour is consistent with the flux density and
spectral evolution of a young object,
with an age of about 50 years, undergoing adiabatic expansion.\\
In the presence of adiabatic expansion of a homogeneous synchrotron
source, the radio spectrum undergoes a shift towards low
frequencies. In the optically-thick regime this means that at a given
frequency the flux density $S$ increases with time (see Pacholczyk
1970, Orienti et al. 2007, and Orienti et al. 2008b 
for a detail analysis of the radio
spectrum evolution):

\begin{equation}
 S_{1} = S_{0} \left( \frac{t_{0} + \Delta t}{t_{0}} \right)^{3}
\label{flusso}
\end{equation}

\noindent where $S_{0}$ and $S_{1}$ are the flux densities at the time
$t_{0}$ and $t_{0} + \Delta t$, respectively. On the other hand, the
spectral peak $\nu_{p}$ moves to lower frequencies:

\begin{equation}
\nu_{p,1} = \nu_{p,0} \left( \frac{t_{0}}{t_{0} + \Delta t}
\right)^{4}
\label{turn}
\end{equation}

\noindent where $\nu_{p,0}$ and $\nu_{p,1}$ are the peak frequency at
the time $t_{0}$ and $t_{0} + \Delta t$ respectively.\\
In the optically-thin regime, the flux density at a given frequency
decreases with time:

\begin{equation}
S_{1} = S_{0} \left( \frac{t_{0} + \Delta t}{t_{0}} \right)^{-2
  \delta}
\label{thin}
\end{equation}

\noindent where $\delta$ is the spectral index of the electron energy
distribution $N(E) \propto E^{- \delta}$ that originates the radio
emission. It is clear from this relationship that substantial
variation in the optically-thick regime can be revealed in case that $\Delta
t$ is a non-negligible fraction of the total source age ($t_{0}$). In
our case, $\Delta t \sim 8$ years and it can produce detectable flux
density variation for sources with $t_{0} \sim$ 100 years. Among the
sources with some changes in the spectral properties compatible with
such a scenario, other 7 sources
(J0754+3033, J0955+3335, J1004+4328, J1025+2541, J1052+3355, J1322+3912, and
J1547+3518) in addition to J1459+3337 show a spectral and flux density
variability that may be explained in terms of source
expansion. Although in J1459+3337 there are several indicators
supporting this interpretation \citep{mo08}, in the case of the other
sources mentioned here this assumption is based on the flux density and peak
variation on a small time range only. Additional observations spanning a
longer time interval, together with information on the pc-scale
morphology are necessary in order to reliably constrain the source
nature.\\

\section{Conclusions}

We presented simultaneous multi-frequency VLA observations of 57
sources from the faint HFP sample, carried out at various epochs. From
the comparison of the spectral properties we found that 24 objects (4
galaxies, 11 quasars, and 9 empty fields) preserve their convex
spectrum without showing any evidence of flux density variability. Of
the remaining sources, 12 objects (2 galaxies, 9 quasars, and 1 empty
field), selected on the basis of their convex spectrum in the first epoch
by \citet{cs09}, turned out to show a flat spectrum in one of the
subsequent observing epochs. The remaining 21 sources (6 galaxies, 13
quasars, and 2 empty fields) possess high levels of variability,
although still displaying a convex spectrum. However, among these
variable sources we found that in 8 objects the changes in their
spectra are consistent with what expected if the source is
undergoing adiabatically expansion. This implies that out of the 57 sources
studied in this paper, 32 objects (56\%) can still be considered
young radio source candidates. The remaining 25 sources (44\%) are
part of the flat-spectrum blazar population, indicating that also in
samples of faint radio sources, where boosted effects are thought to
play a minor role, a large fraction of sources are represented by
flaring objects. \\
The analysis of the optical images of the HFPs hosted
by galaxies pointed out the presence of companion galaxies in the target
environment, supporting the idea that young radio sources reside in
groups. The parent galaxy
is usually the brightest elliptical at the group centre with the
exception of two sources. In J0804+5431 the galaxy hosting the HFP is
at the periphery of the group, and it seems interacting with a close
elliptical. A surprising result is represented by the HFP J1109+3831
that is hosted in a spiral that seems to be interacting with a close
elliptical. The fact that young radio sources reside in groups support
the idea that the interactions occurring between the galaxies are at
the origin of the radio emission.\\

\section*{Acknowledgments}

We thank S. Bardelli for his help on the analysis of the optical
spectra. The VLA and the VLBA are operated by the US 
National Radio Astronomy Observatory which is a facility of the National
Science Foundation operated under cooperative agreement by Associated
Universities, Inc. This work has made use of the NASA/IPAC
Extragalactic Database NED which is operated by the JPL, Californian
Institute of Technology, under contract with the National Aeronautics
and Space Administration.
Funding for the SDSS and SDSS-II has been provided by the Alfred
P. Sloan Foundation, the participating Institutions, the National
Science Foundation, the U.S. Department of Energy, the National
Aeronautics and Space Administration, the Japanese Monbukagakusho, the
Max Planck Society, and the Higher Education Funding Council for
England. The SDSS was managed by the Astrophysical Research Consortium
for the Participating Institutions.

\end{document}